\documentclass[aps,
prb,
reprint,
twocolumn,
groupedaddress,
superscriptaddress,
showpacs,
nobibnotes,
longbibliography,
notitlepage,
floatfix
]{revtex4-1}

\usepackage{graphicx}
\usepackage{amsmath}
\usepackage{amsfonts}
\usepackage[caption=false, position=top]{subfig}
\usepackage{color}
\usepackage{bbm}

\def\ket#1{\mathinner{|{#1}\rangle}}

\newcommand{\ketbra}[2]{|#1\rangle\langle#2|}

\def\Id{\mathbbm{1}}

\begin{document}

\title{Silicon and silicon-nitrogen impurities in graphene: structure, energetics and effects on electronic transport}

\author{Mikko M. Ervasti}
\affiliation{COMP Centre of Excellence, Department of Applied Physics, Aalto University, Helsinki, Finland}
\author{Zheyong Fan}
\affiliation{COMP Centre of Excellence, Department of Applied Physics, Aalto University, Helsinki, Finland}
\affiliation{School of Mathematics and Physics, Bohai University, Jinzhou, China}
\author{Andreas Uppstu}
\affiliation{COMP Centre of Excellence, Department of Applied Physics, Aalto University, Helsinki, Finland}
\author{Arkady Krasheninnikov}
\affiliation{COMP Centre of Excellence, Department of Applied Physics, Aalto University, Helsinki, Finland}
\author{Ari Harju}
\affiliation{COMP Centre of Excellence, Department of Applied Physics, Aalto University, Helsinki, Finland}

\date{\today}

\begin{abstract}

We  theoretically study the atomic structure and energetics of silicon
and silicon-nitrogen impurities in graphene. Using density-functional
theory, we get insight into the atomic structures of the impurities,
evaluate their formation energies and assess their abundance in
realistic samples. We find that nitrogen, as well as oxygen and
hydrogen, are trapped at silicon impurities, considerably altering the
electronic properties of the system. Furthermore, we show that nitrogen
doping can induce local magnetic moments resulting in spin-dependent transport
properties,even though neither silicon nor nitrogen impurities are magnetic
by themselves. To simulate large systems with many randomly
distributed impurities, we derive tight-binding models that describe the
effects of the impurities on graphene $\pi$ electron structure. Then by using the
linear-scaling real-space Kubo-Greenwood method, we evaluate the
transport properties of large-scale systems with random distribution of impurities,
and find the fingerprint-like scattering cross sections for each impurity type.
The transport properties vary widely, and our results indicate that some of
the impurities can even induce strong localization in realistic graphene
samples.

\end{abstract}


\maketitle

\section{Introduction}

Graphene is a remarkable two-dimensional (2D) material due to its unique
mechanical and electronic properties, which makes it a good candidate
for modern nanoscale devices and applications. While pristine graphene
is a semimetal with a high carrier mobility \cite{Novoselov_2005, Chen_2008,
Morozov_2008, Siegel_2011, Castro_Neto_2009}, 
for some applications it would be desirable to open a band gap in it, for example by cutting it to ribbons \cite{Son_2006_2, Yang_2007, Han_2007} or introducing impurities and defects\cite{Berashevich_2009, Balog_2010, Casolo_2011, Denis_2010}.
Moreover, impurities are
often the dominant scatterers that control the intrinsic electronic and
transport properties of realistic systems.

Silicon is commonly present in nature, so that Si impurities can appear in
synthetic graphene during the growth processes or be introduced later
when graphene is used together with
standard silicon-based electronics components. It is even preferred to
grow graphene on silicon wafers as one does not have to transfer it
elsewhere after growth. Specifically, graphene grown at high temperatures by chemical
vapor deposition can introduce silicon and oxygen impurities originating
from the quartz (SiO$_2$) substrate or the apparatus itself. Also using silicon carbide (SiC) \cite{Forbeaux_1998, Singh_2011} to grow graphene can produce silicon impurites.
Another avenue would be to deliberately introduce silicon impurities
by post-synthesis treatments \cite{Wang_2012, Wang_2014},
such as by low-energy ion irradiation, similar to direct ion implantation of N and B atoms into
single graphene sheets \cite{Bangert2013}, or deposition of Si atoms on
ion or electron-beam treated graphene with irradiation-induced vacancies.
Silicon atoms filling monovacancies and divacancies in
graphene have been explicitly identified in experiments \cite{Zhou_2012,
Ramasse_2012}. Their formation energies are expected to be fairly low
compared to other period 3 element substitution defects
\cite{Denis_2010}, and they are stable enough for the electron beam not
to break the atomic structure or the bonding easily \cite{Zhou_2012_2}. 

Si impurities can further pick-up various atoms from the environment.
Recent experimental studies have reported the presence of individual
defects formed by co-occurring silicon and nitrogen impurities
\cite{Zhou_2012, Zhou_2012_2}. Zhou et al. showed that surface plasmons
are locally enhanced at both silicon and a silicon-nitrogen
impurities \cite{Zhou_2012_2}. Therefore, such silicon impurities
occurring in graphene could in principle be used as plasmonic
waveguides. Additionally, they could be useful in optoelectronic devices \cite{Bonaccorso_2010}. 
Nitrogen impurities have been extensively studied,
as they dope their surroundings in graphene \cite{Wei_2009, Lambin_2012},
but the role of nitrogen binding to the
silicon impurities, forming silicon-nitrogen defects, is not fully
understood. Moreover, the presence of oxygen can also affect the formation of
silicon impurities. We answer these issues by finding the
defect geometries and formation energies for various silicon,
silicon-oxygen and silicon-nitrogen impurities through comprehensive
density-functional-theory calculations in a supercell geometry.

Such impurities are also interesting in the context of magnetism.
Local magnetic moments can be created in graphene in several ways. Examples
include graphene nanoribbon zigzag edges \cite{Son_2006}, monovacancies
and hydrogen adatoms \cite{Yazyev_2007}, and transition metal
substitutions \cite{Krasheninnikov_2009}. Local magnetic impurities
interact strongly with the  conduction electrons of graphene, as has
been shown by measuring the Kondo effect \cite{Chen_2011}. We demonstrate that
many of the nitrogen-doped silicon impurities exhibit finite spin
moments. Such defects can be important in graphene-based spintronics
applications, and we evaluate the spin-dependent electronic and
transport properties for the most stable defect types. 

Electronic transport in systems with silicon point-defects has been
studied in a ribbon geometry by Lopez-Bezanilla et al.
\cite{Lopez_Bezanilla_2014}. They found that the formation energies of
the silicon point-defects are smaller closer to the ribbon edges, and
the transmission functions for graphene nanoribbons with silicon defects
at the edges were computed. Furthermore, Cheng et al. \cite{Cheng_2014}
studied the electronic and transport properties of the SiN$_x$ defects
in armchair nanoribbons. However, it is unclear what are the transport
properties of realistic two-dimensional graphene systems with numerous
randomly positioned defects. We simulate transport in such a realistic
setting, and find the characteristic fingerprint-like scattering cross
sections for each defect type.

The paper is organized as follows. In Sections II A-D, we present first
principles results for the geometries and formation energies of
silicon, silicon-oxygen, and silicon-nitrogen impurities, and also
touch upon the effects of hydrogen adsorption on them. In section II E,
we evaluate the electronic band structures and density of states of
systems containing silicon and silicon-nitrogen impurities.
In Section III A-B we derive
tight-binding models to describe the effects of the impurities on the electronic structure.
In Section III C, we compare the
density of states of the system with impurities calculated
within the periodic supercell approach with that of many randomly
placed impurities in large realistic systems. In Section III D, we
report the results of a comprehensive real-space Kubo-Greenwood study of the
electronic transport of silicon and silicon-nitrogen impurities, where
we also discuss the localization effects in these systems.

\section{First principles calculations}

First principles calculations were performed by the all-electron
density-functional theory package FHI-aims \cite{Blum_2009}. The code
uses local basis functions specified for each atom, and we have chosen
the default \textit{tight} basis sets provided in the package. As the
exchange-correlation energy functional we used the generalized-gradient
approximation as parametrized by Perdew, Burke, and Ernzerhof (PBE)
\cite{Perdew_1996}.

The defects were placed in periodic supercells of $8 \times 8$ and $11
\times 11$ graphene unit cells. Without any defects they would contain
$2 \times 8 \times 8 = 128$ and $2 \times 11 \times 11 = 242$ carbon
atoms, respectively. The lattice vectors of the supercells were fixed to
those of pristine graphene (with a lattice constant of $2.467 $ \AA$ $
in a two-atom unit cell), even after the point-defects had been
introduced. The atom positions were relaxed until the forces between the
atoms were smaller than $10^{-3}$ eV/\AA. The self-consistency cycle was
considered converged if the change in the volume-integrated root-mean square of the charge
density, and changes in the sum of eigenvalues and total energy were
below $10^{-6}$ [electrons], $10^{-3}$ eV, and $10^{-6}$ eV,
respectively. A Monkhorst-Pack grid of $8 \times 8 \times 1$ k-points was
employed in the total energy calculations, and the band structure and
density of states (DOS) calculations utilized a $\Gamma$-centered grid
with at least $15 \times 15 \times 1$ k-points in order to achieve a
better coverage of the first Brillouin zone.

Performing the calculations with two distinct supercell sizes lets us
explicitly test whether the results have converged as a function of
system size. Systematic convergence studies have been done for the
silicon substitution defect in graphene in Ref.
[\onlinecite{Denis_2010}], where convergent results were obtained
already with a rather small $4 \times 4$ supercell, as contrary
to vacancies\cite{Kra2011tca}, substitutional impurities and adatoms 
do not give rise to long-ranged strain fields.
Studying structural
point-defects in graphene, a $7 \times 7$ supercell is typically large
enough for geometry relaxation \cite{Lherbier_2012}.
However, to obtain convergent electronic properties and spectra,
a larger supercell is typically needed.
Specifically, for isolated nitrogen substitution defect,
or any other defect with large effective
on-site potential at the impurity site, the supercell needs to be much
larger\cite{Lambin_2012}. The same is true for the adequate description
of magnetic properties of some defects \cite{Wang2012prb}.

To estimate the likelihood of a defect to form, and to compare the
energies of different defects in graphene, we define the formation
energy of a defect as
\begin{equation}
E_\text{f} = E - \sum_i n_i \mu_i ,
\label{eq:def:formation_energy}
\end{equation}
where $E$ is the total energy of a supercell with the defect in
question, $i$ sums over atom types, $n_i$ is the number of atoms of type
$i$, and $\mu_i$ is the chemical potential of atom type $i$. Now, the
carbon chemical potential $\mu_{\text{C}}$ was chosen as the energy of
graphene per carbon atom, so that defect-free graphene has zero
formation energy. For the silicon chemical potential $\mu_{\text{Si}}$,
we chose the energy per atom of crystalline silicon in a diamond cubic
lattice. The hydrogen, nitrogen, and oxygen chemical potentials
$\mu_{\text{H}}$, $\mu_{\text{N}}$, and $\mu_{\text{O}}$ were chosen as
half the energies of the H$_2$, N$_2$, and O$_2$ molecules,
respectively. The formation energies can be related to binding energies
given the formation energies of isolated H, C, N, O, and Si atoms, that
are $2.27$ eV, $8.01$ eV, $5.28$ eV, $2.92$ eV, and $4.66$ eV,
respectively.

We would like to stress here that even though concentration $n$ 
of a particular type of defects at temperature  $T$ can be evaluated as 
$n \sim \exp( - \beta E_{\text{f}} )$
for the system in the thermodynamic equilibrium only,
Eq.~\eqref{eq:def:formation_energy} makes it possible to qualitatively 
assess relative concentrations of different types of defects in the 
system subjected to mild chemical treatment, and even though in some 
cases after harsh treatment like irradiation, provided that the system 
was annealed after that. 
All defects were assumed to be neutral, as graphene is a semimetal.

\begin{figure*}[thb!]
\centering
\includegraphics[width=0.195\linewidth]{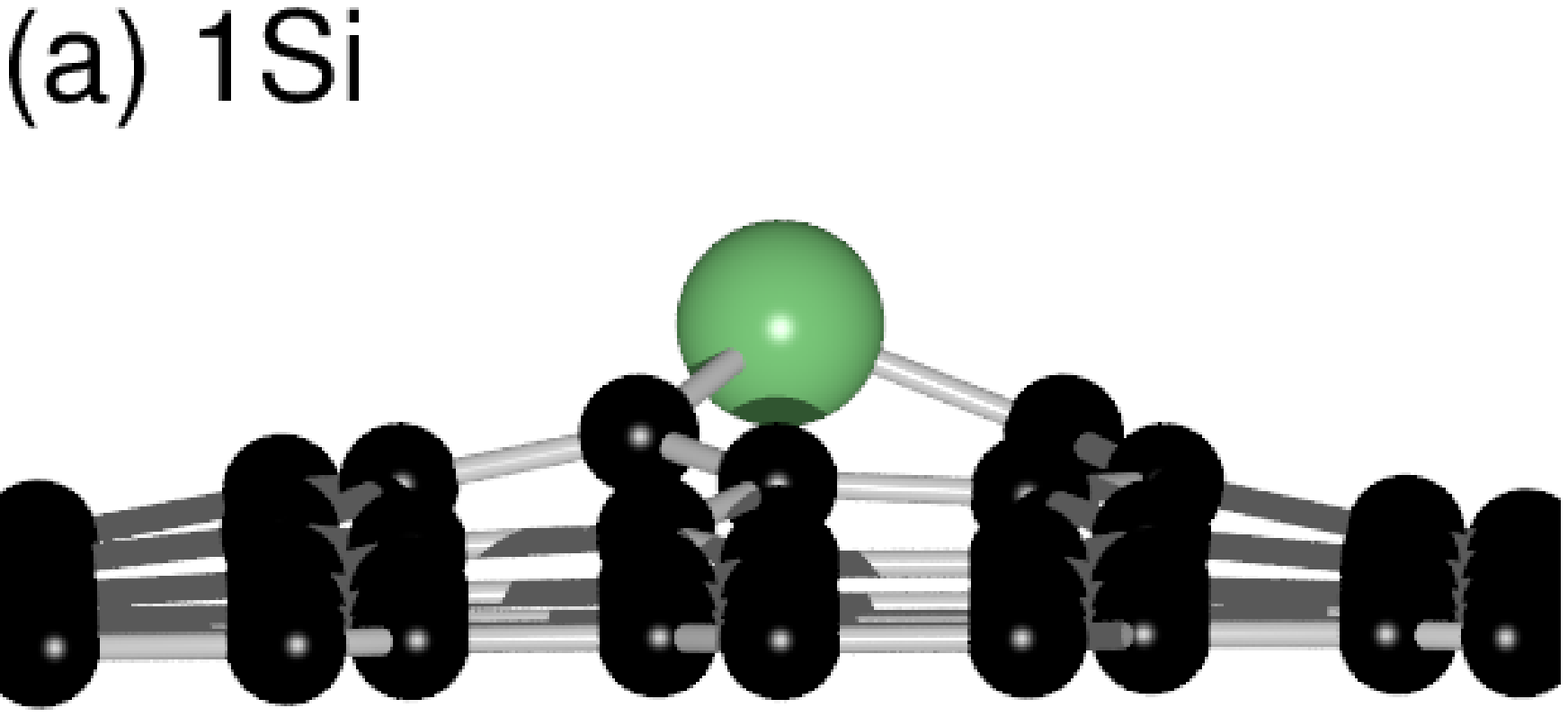}
\includegraphics[width=0.195\linewidth]{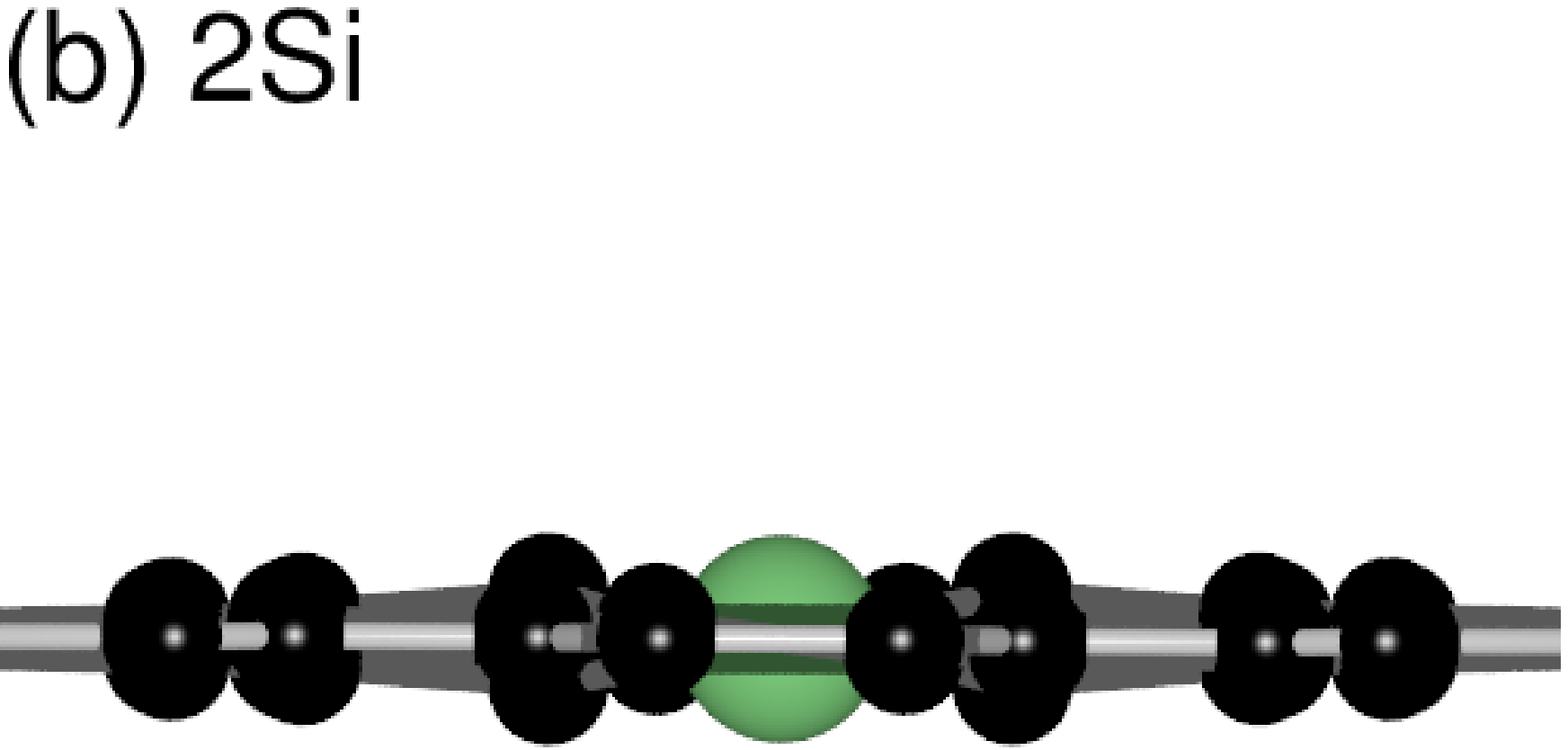}
\includegraphics[width=0.195\linewidth]{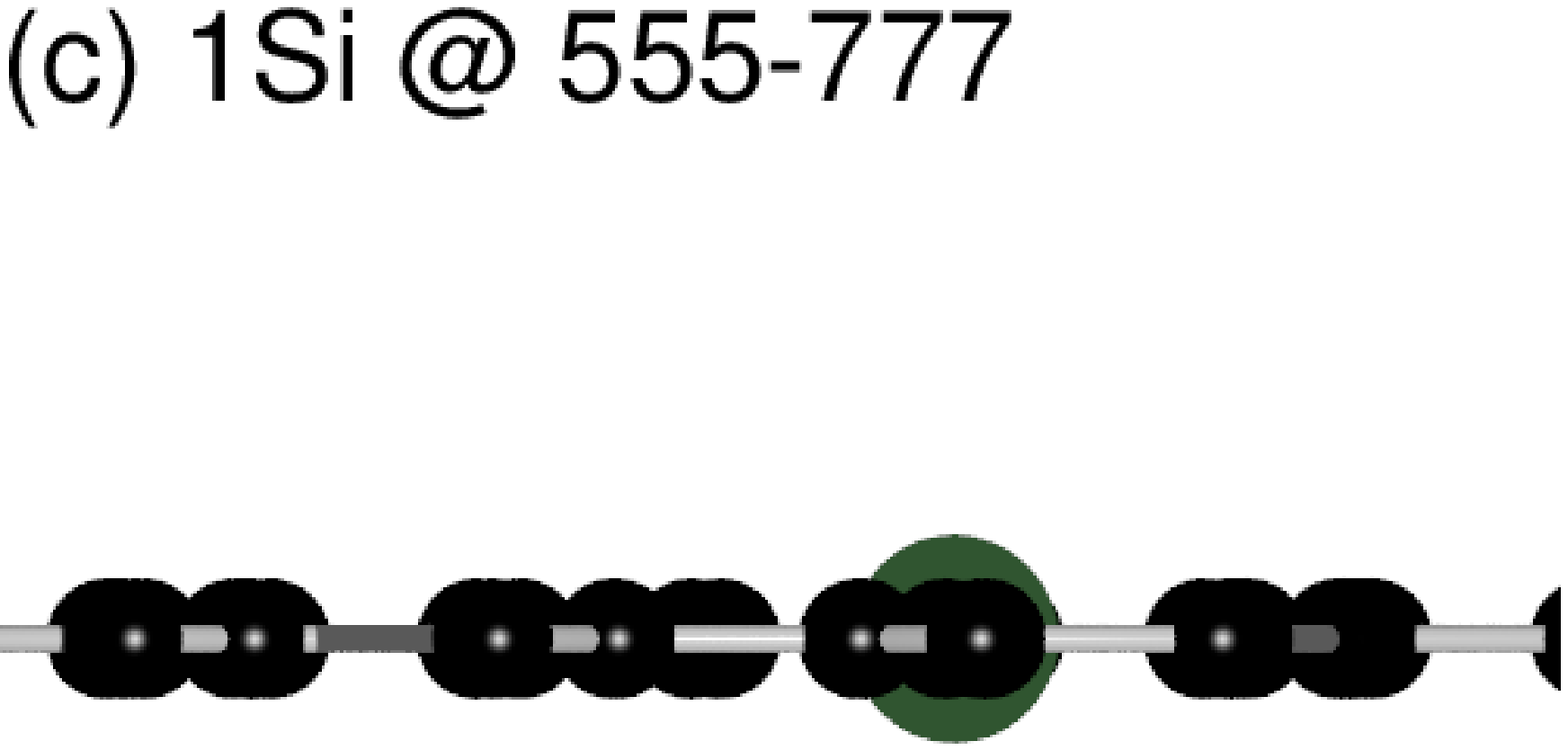}
\includegraphics[width=0.195\linewidth]{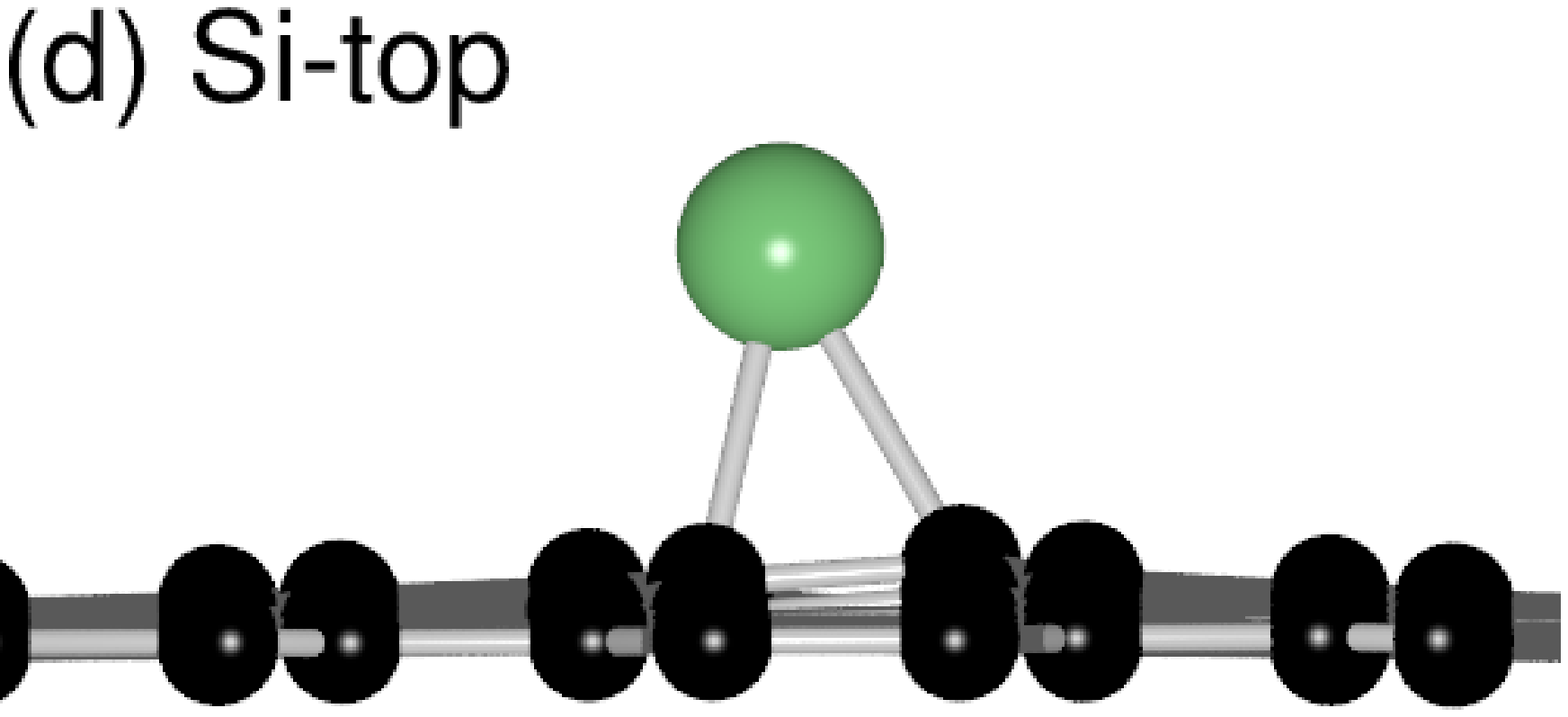}
\includegraphics[width=0.195\linewidth]{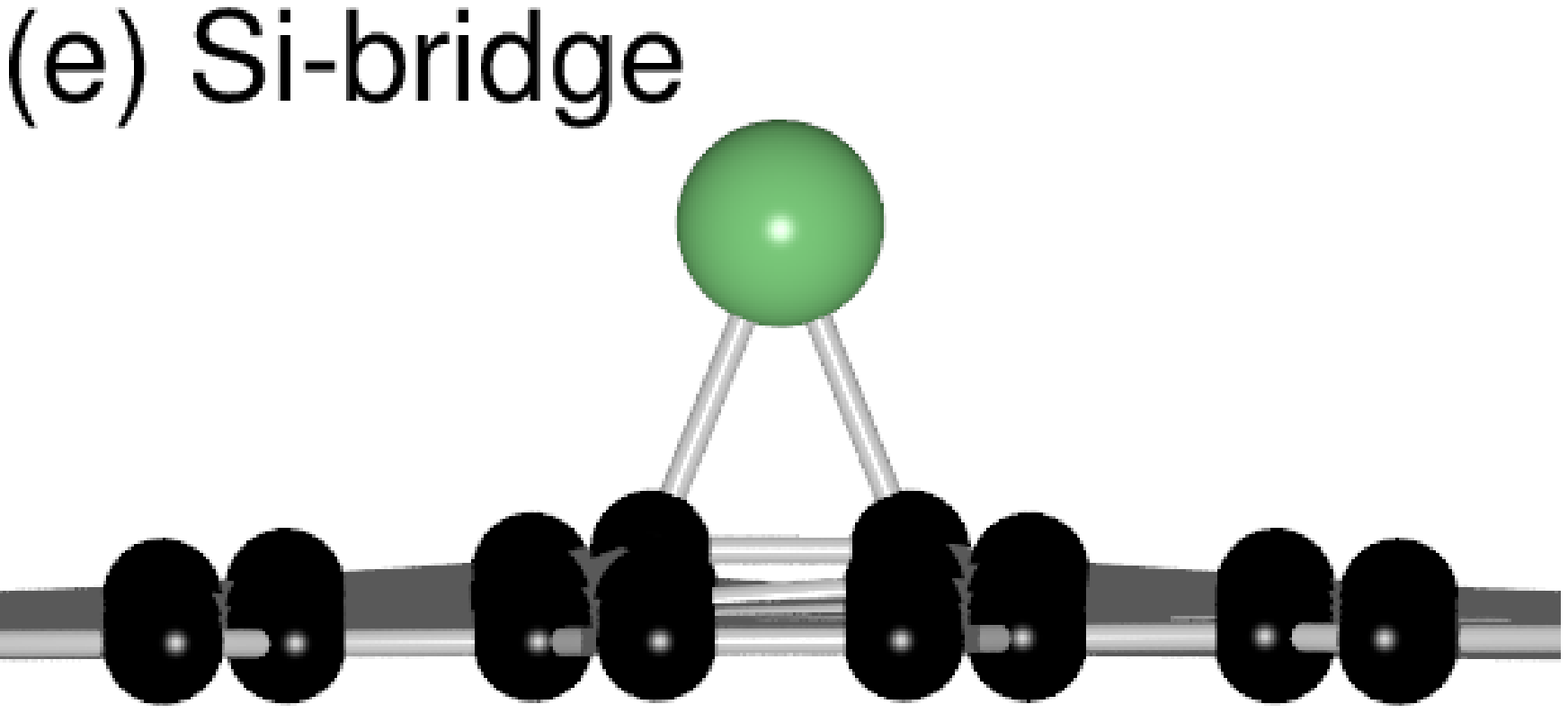}
\includegraphics[width=0.195\linewidth]{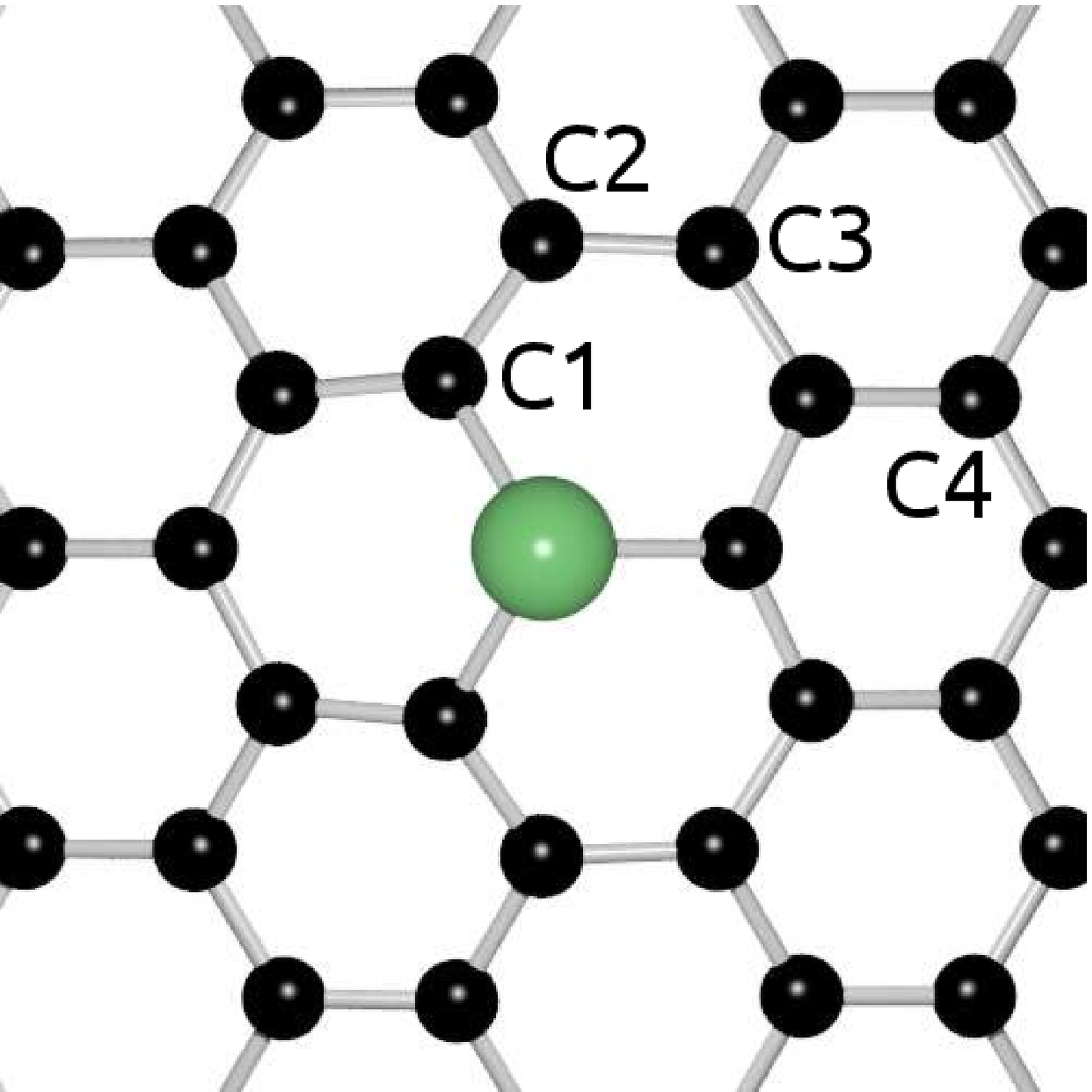}
\includegraphics[width=0.195\linewidth]{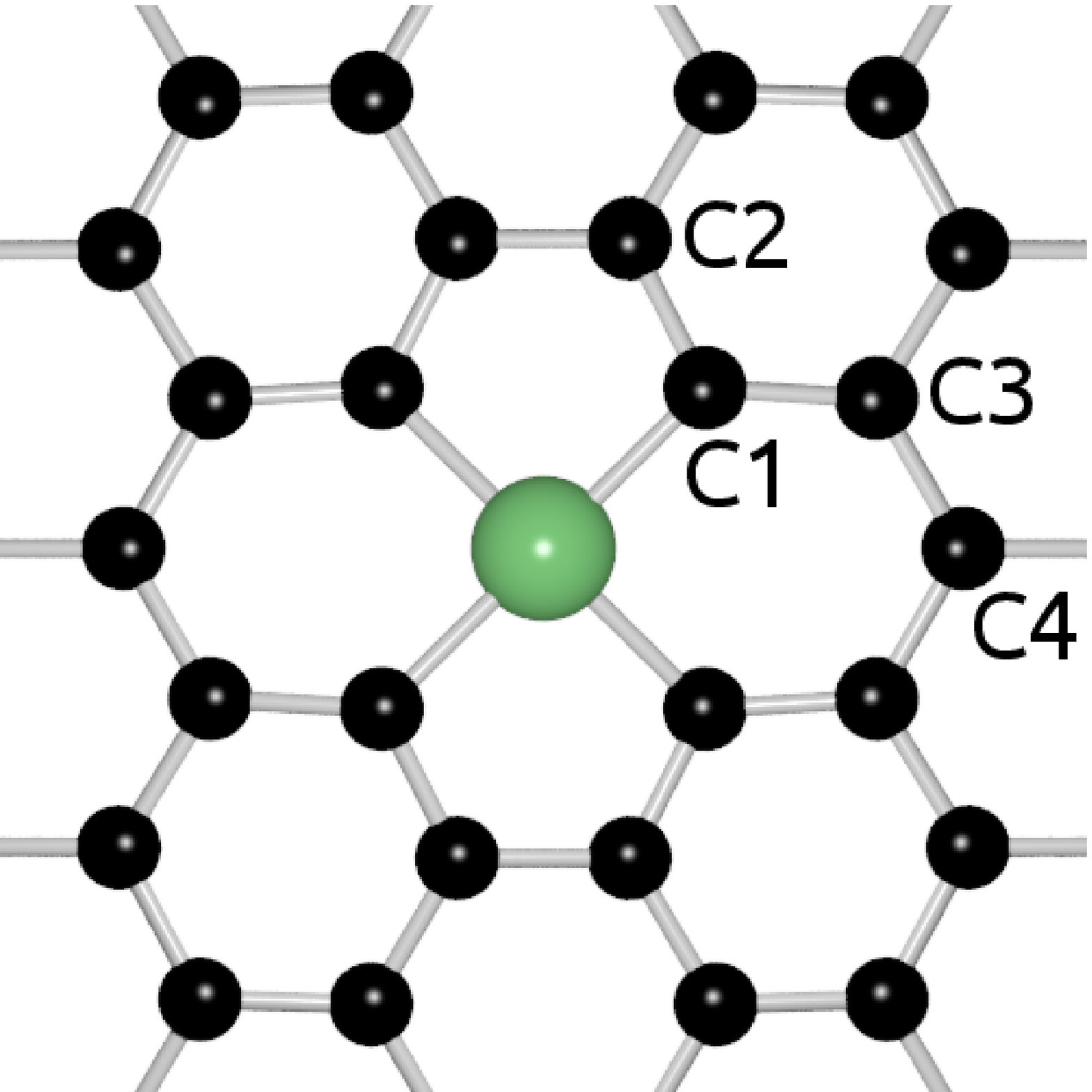}
\includegraphics[width=0.195\linewidth]{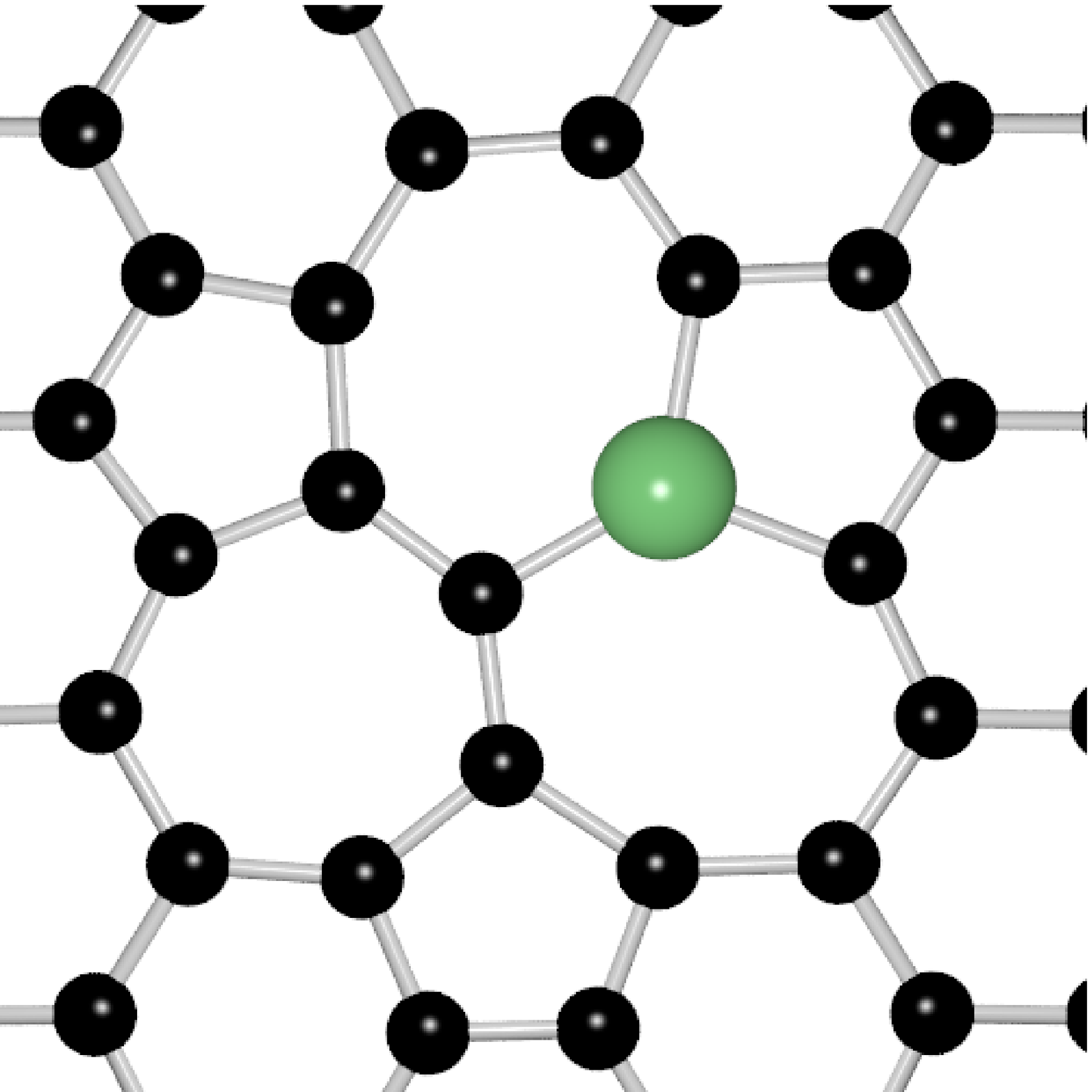}
\includegraphics[width=0.195\linewidth]{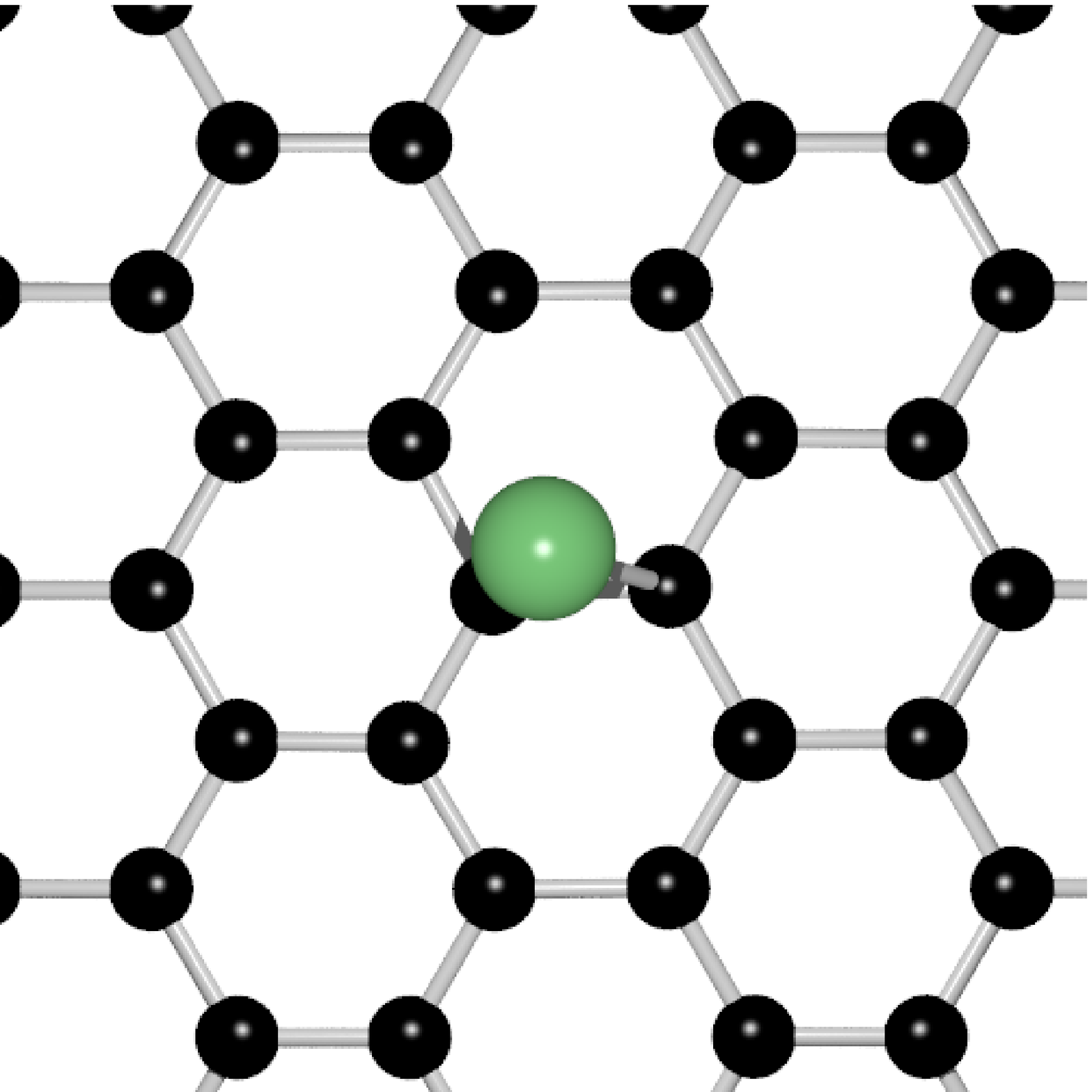}
\includegraphics[width=0.195\linewidth]{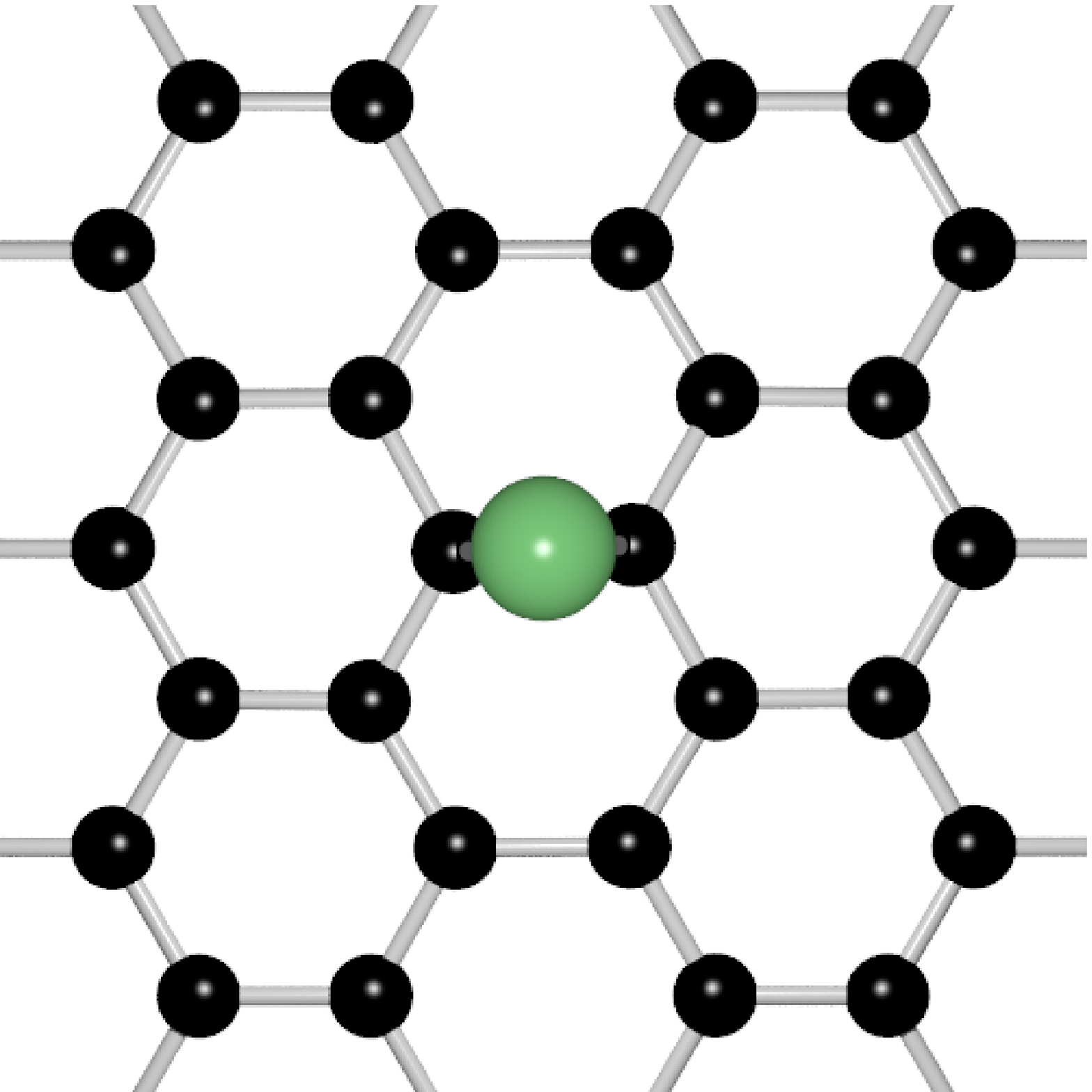}
\includegraphics[width=0.195\linewidth]{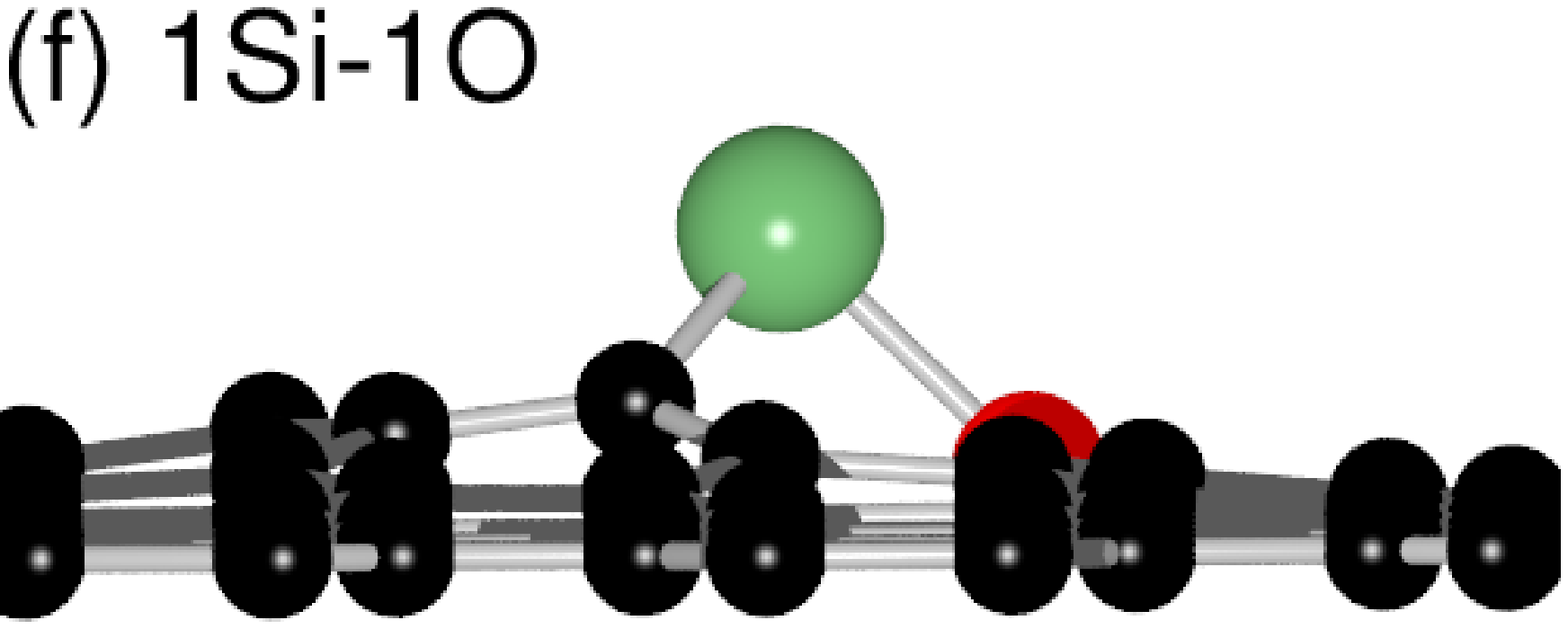}
\includegraphics[width=0.195\linewidth]{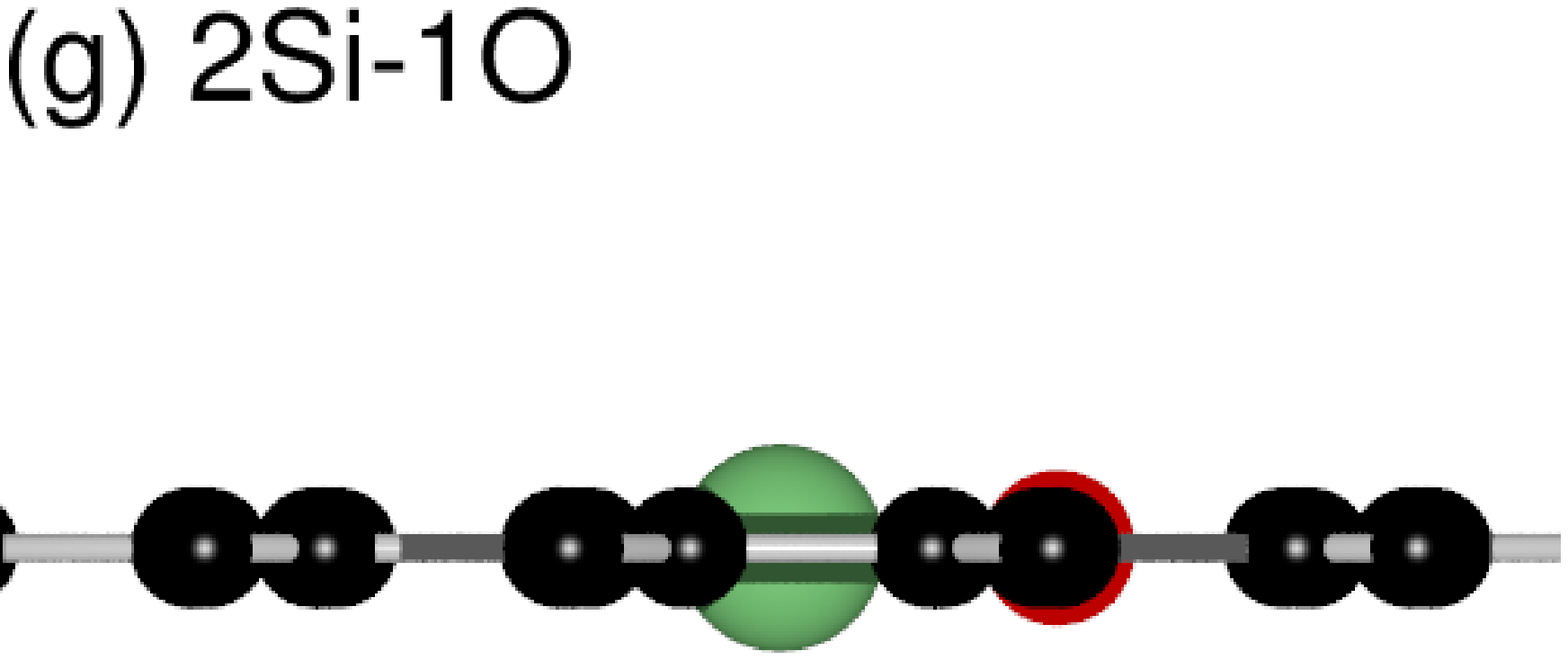}
\includegraphics[width=0.195\linewidth]{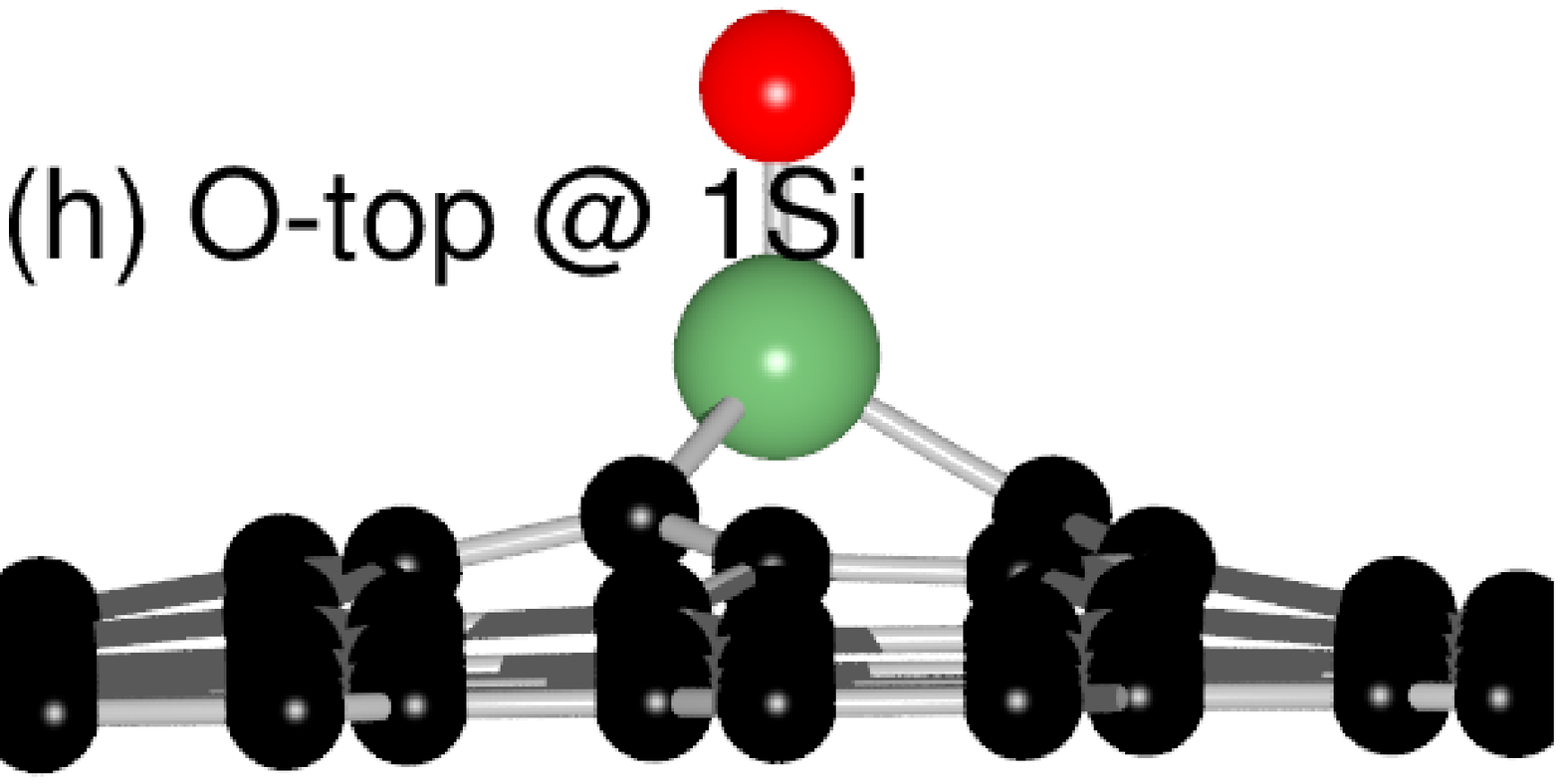}
\includegraphics[width=0.195\linewidth]{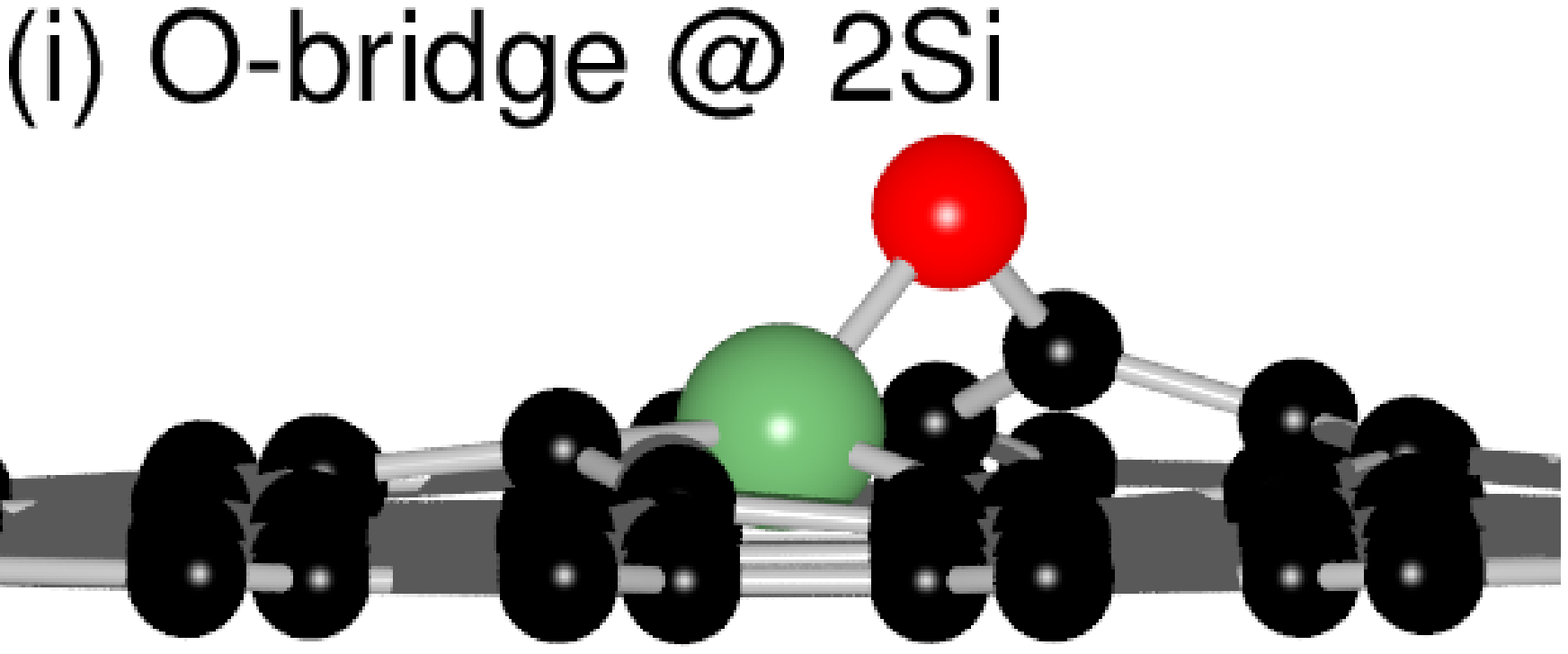}
\includegraphics[width=0.195\linewidth]{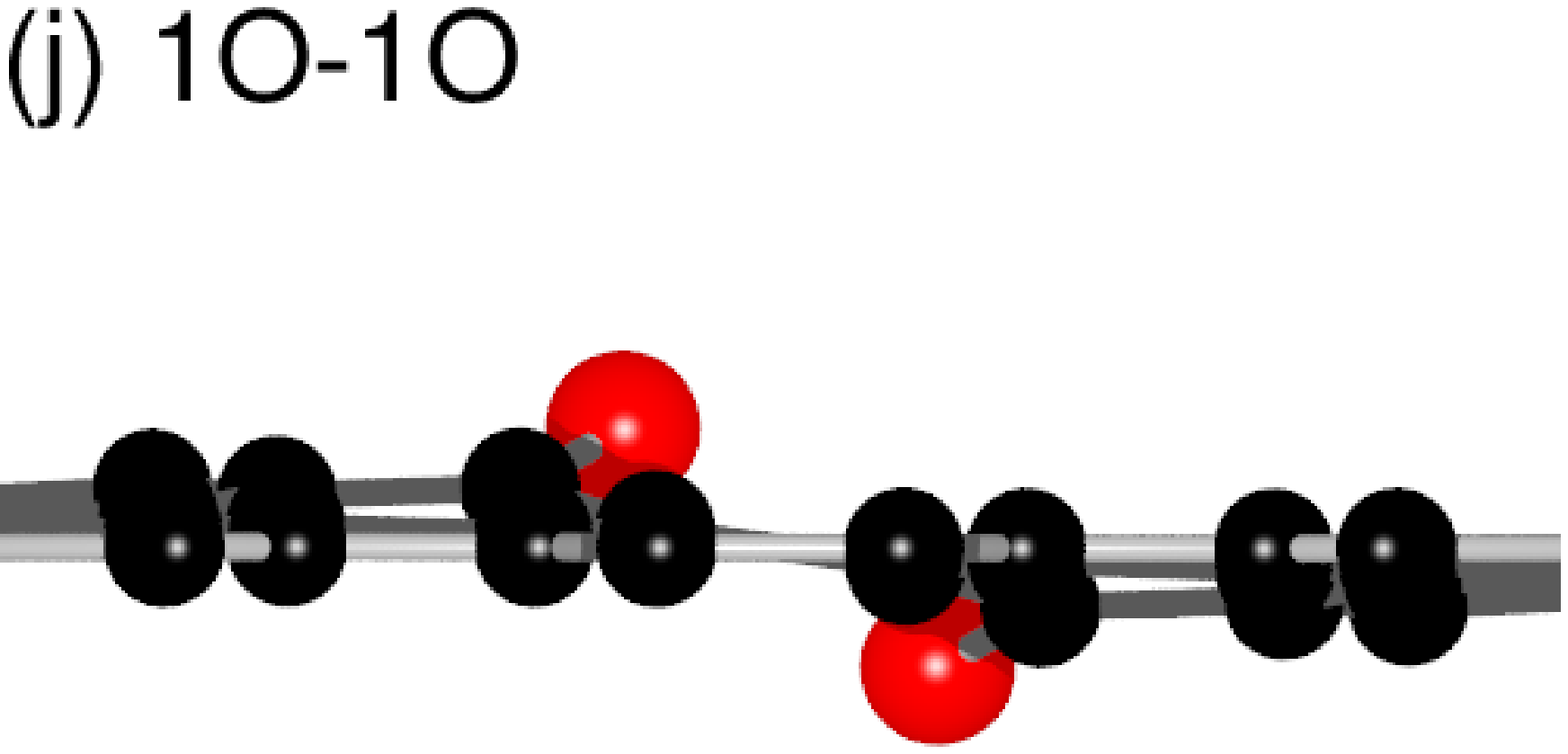}
\includegraphics[width=0.195\linewidth]{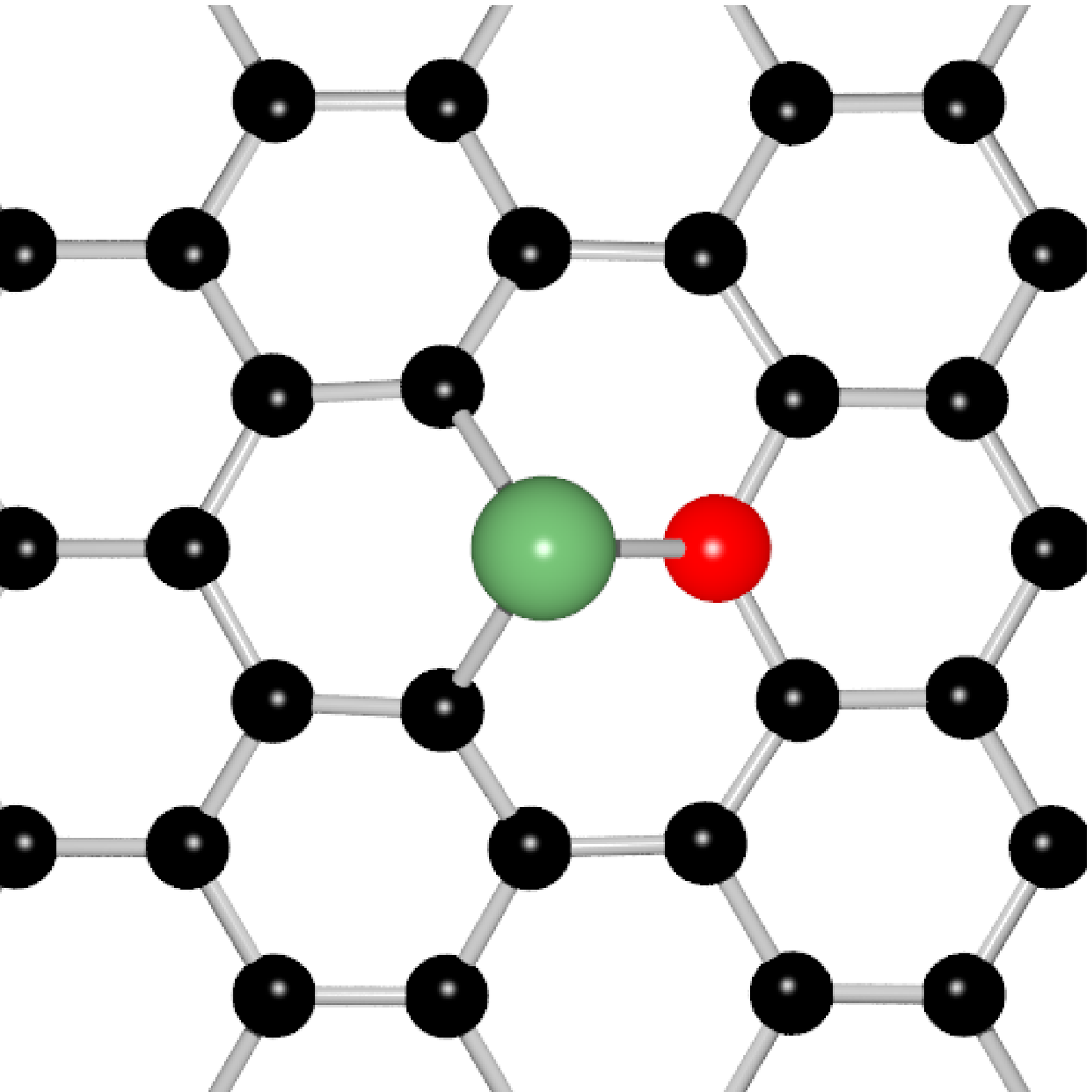}
\includegraphics[width=0.195\linewidth]{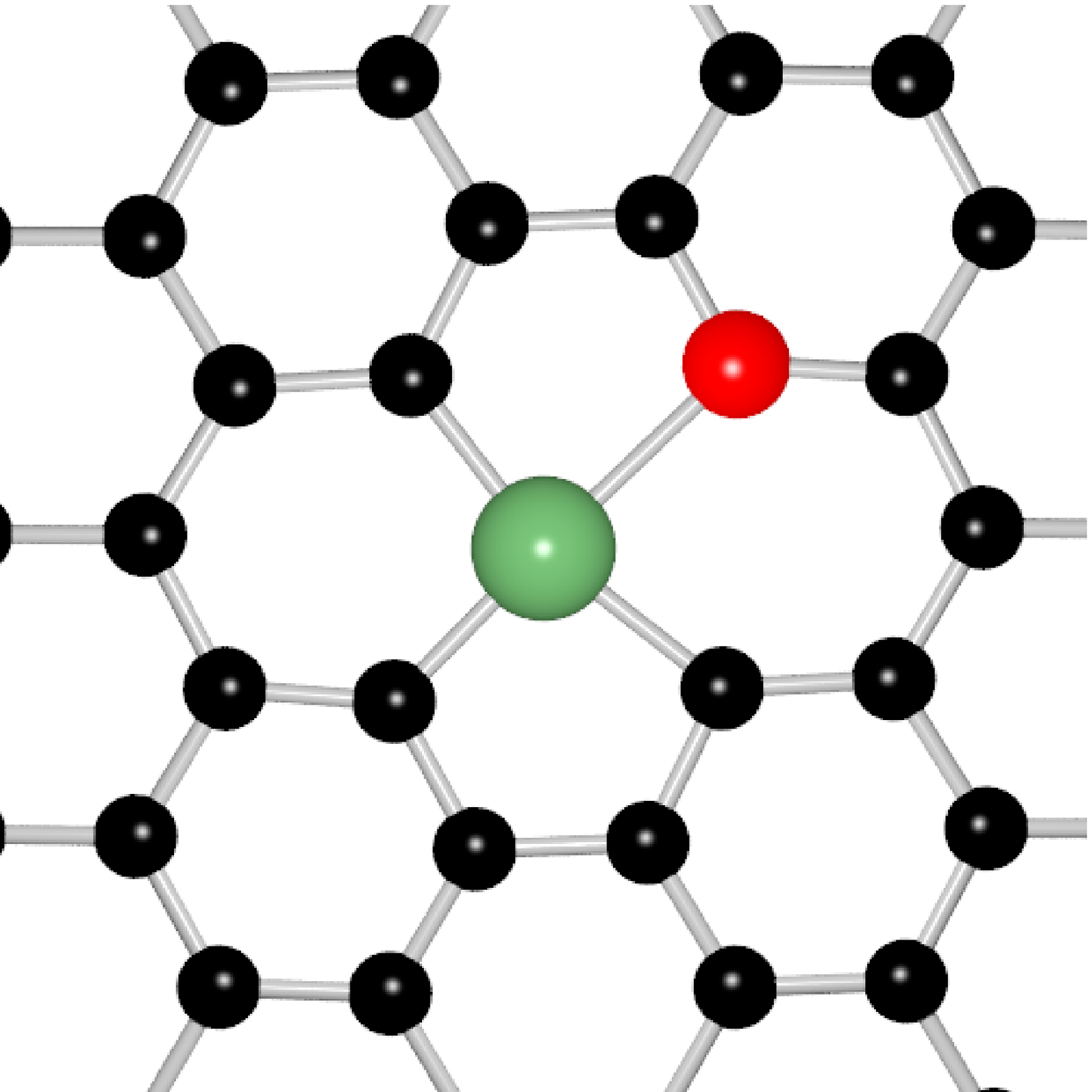}
\includegraphics[width=0.195\linewidth]{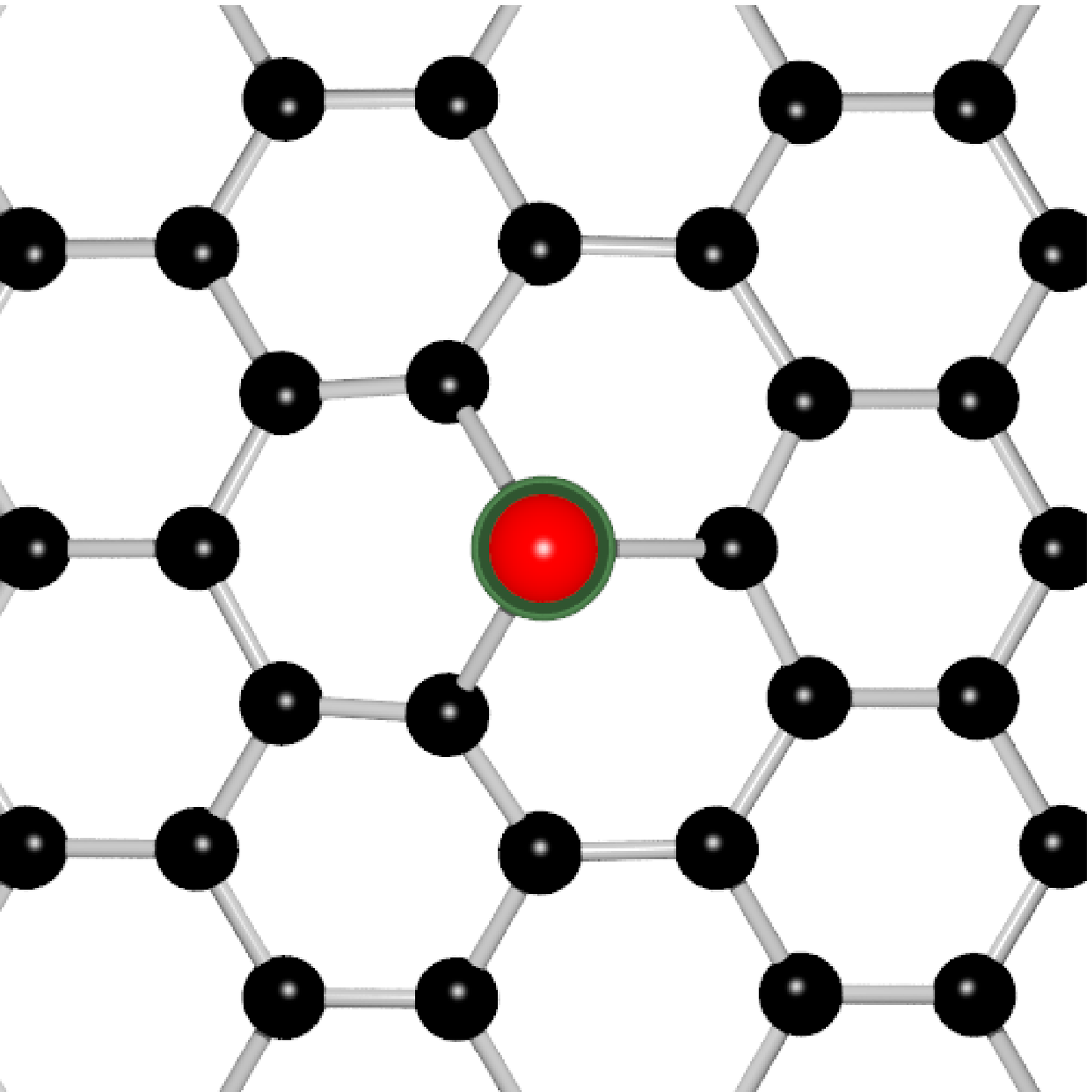}
\includegraphics[width=0.195\linewidth]{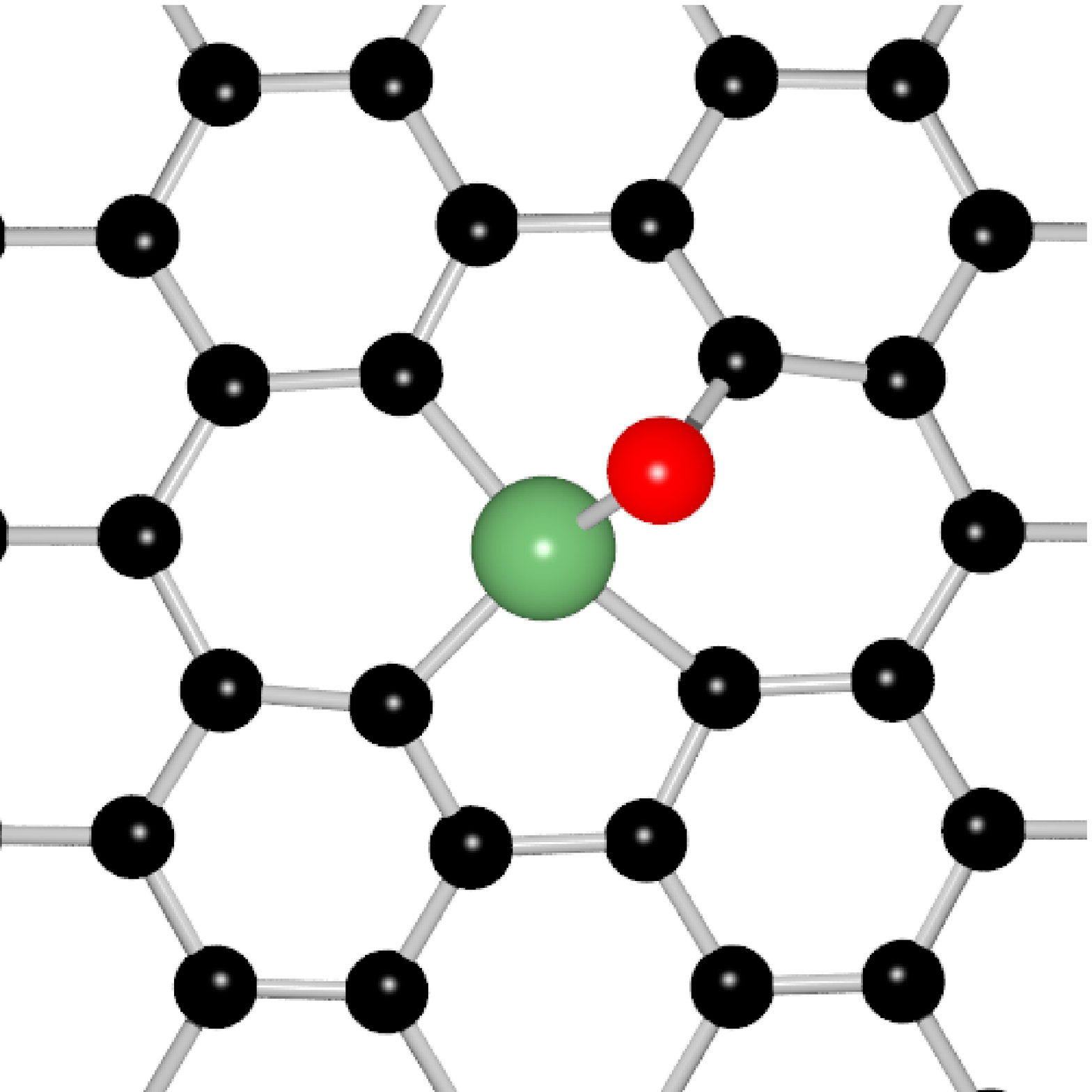}
\includegraphics[width=0.195\linewidth]{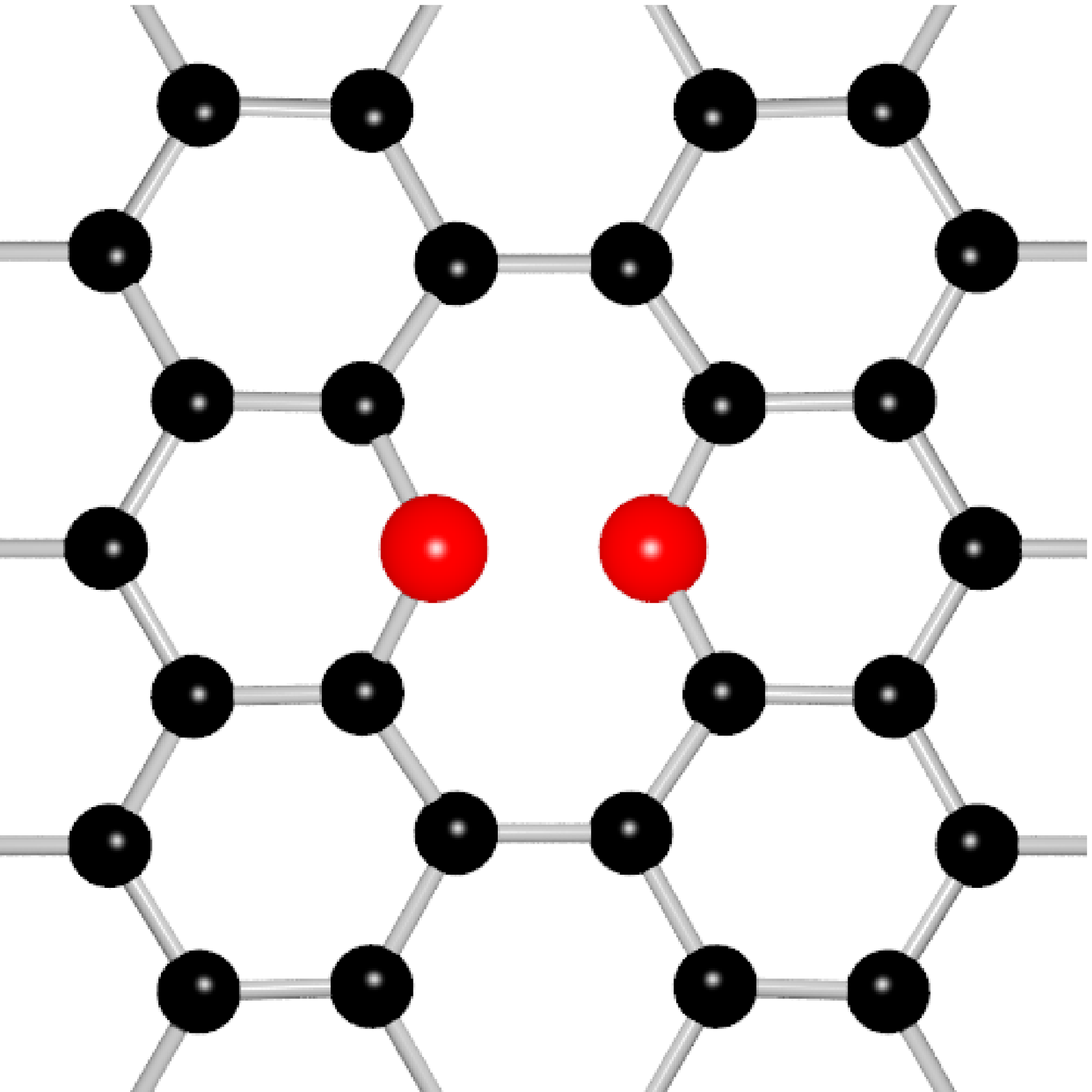}
\includegraphics[width=0.195\linewidth]{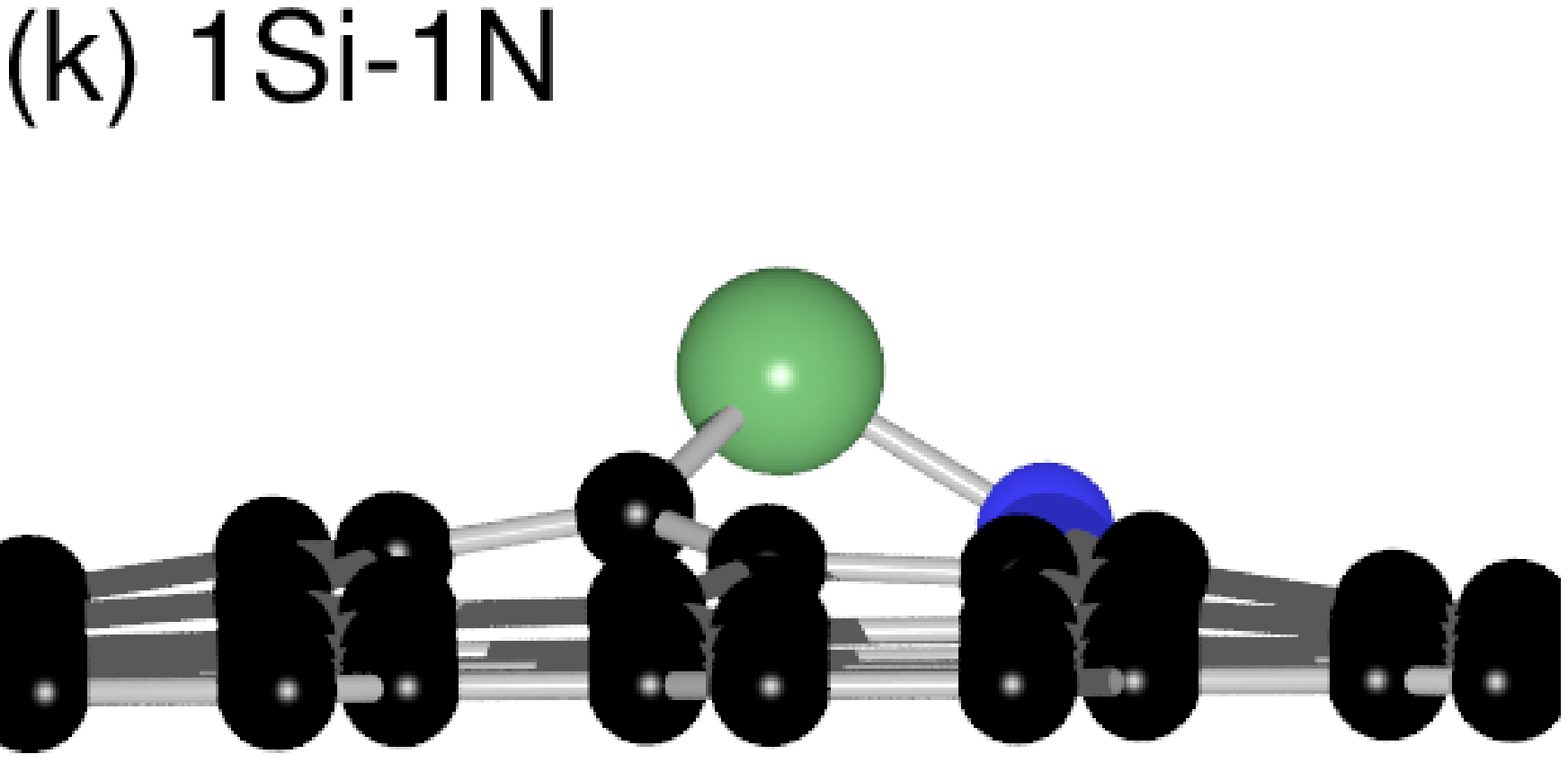}
\includegraphics[width=0.195\linewidth]{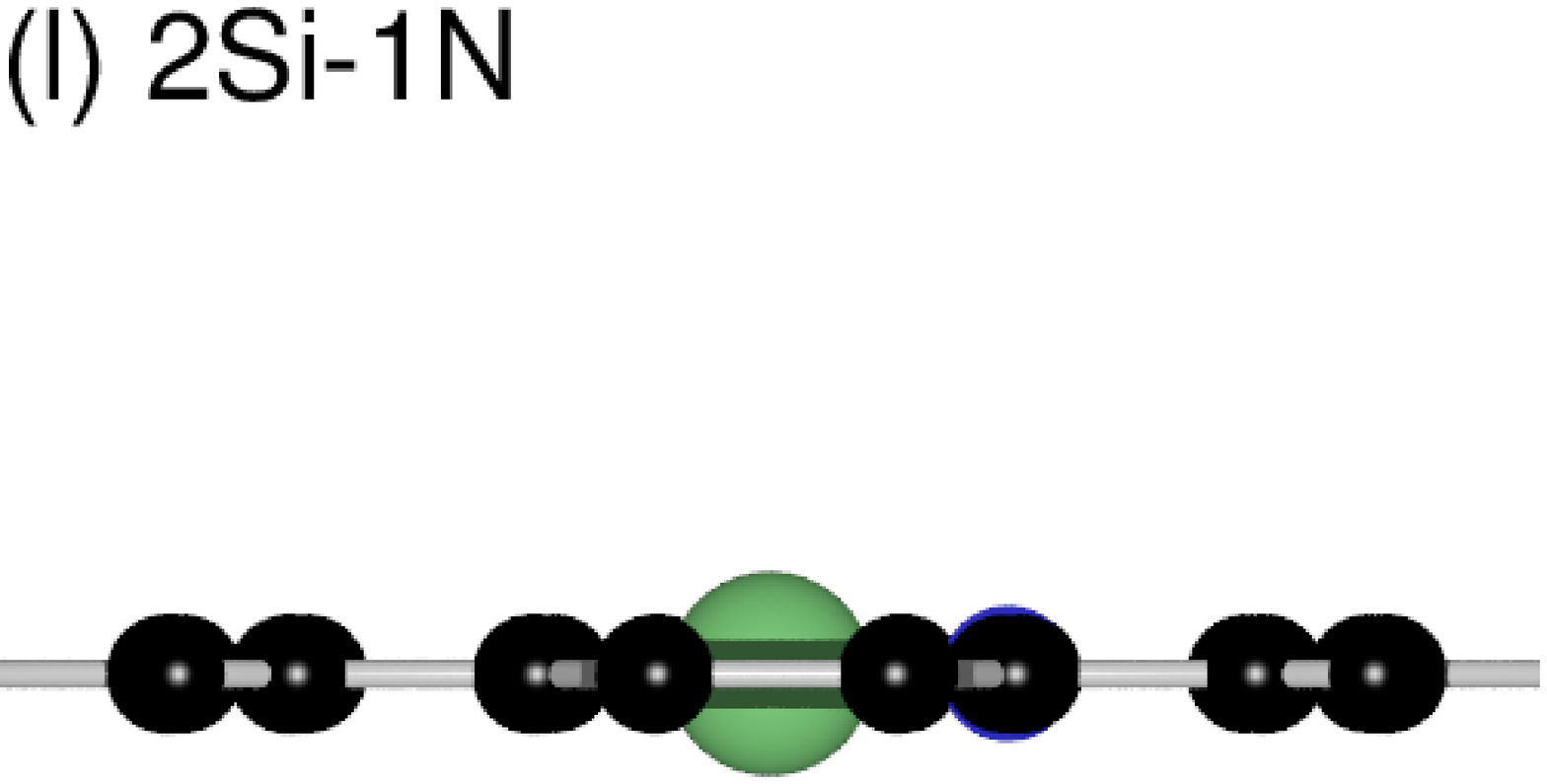}
\includegraphics[width=0.195\linewidth]{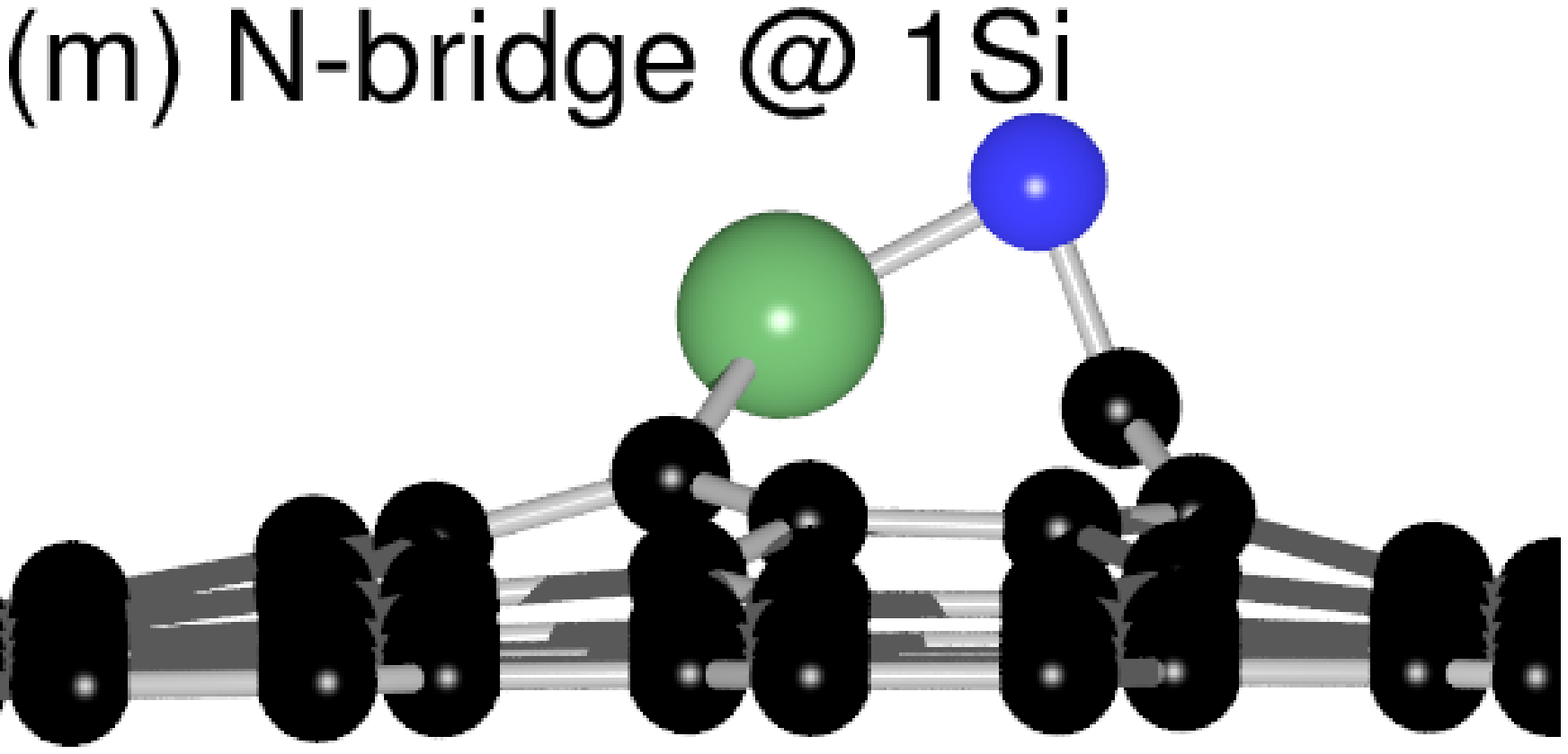}
\includegraphics[width=0.195\linewidth]{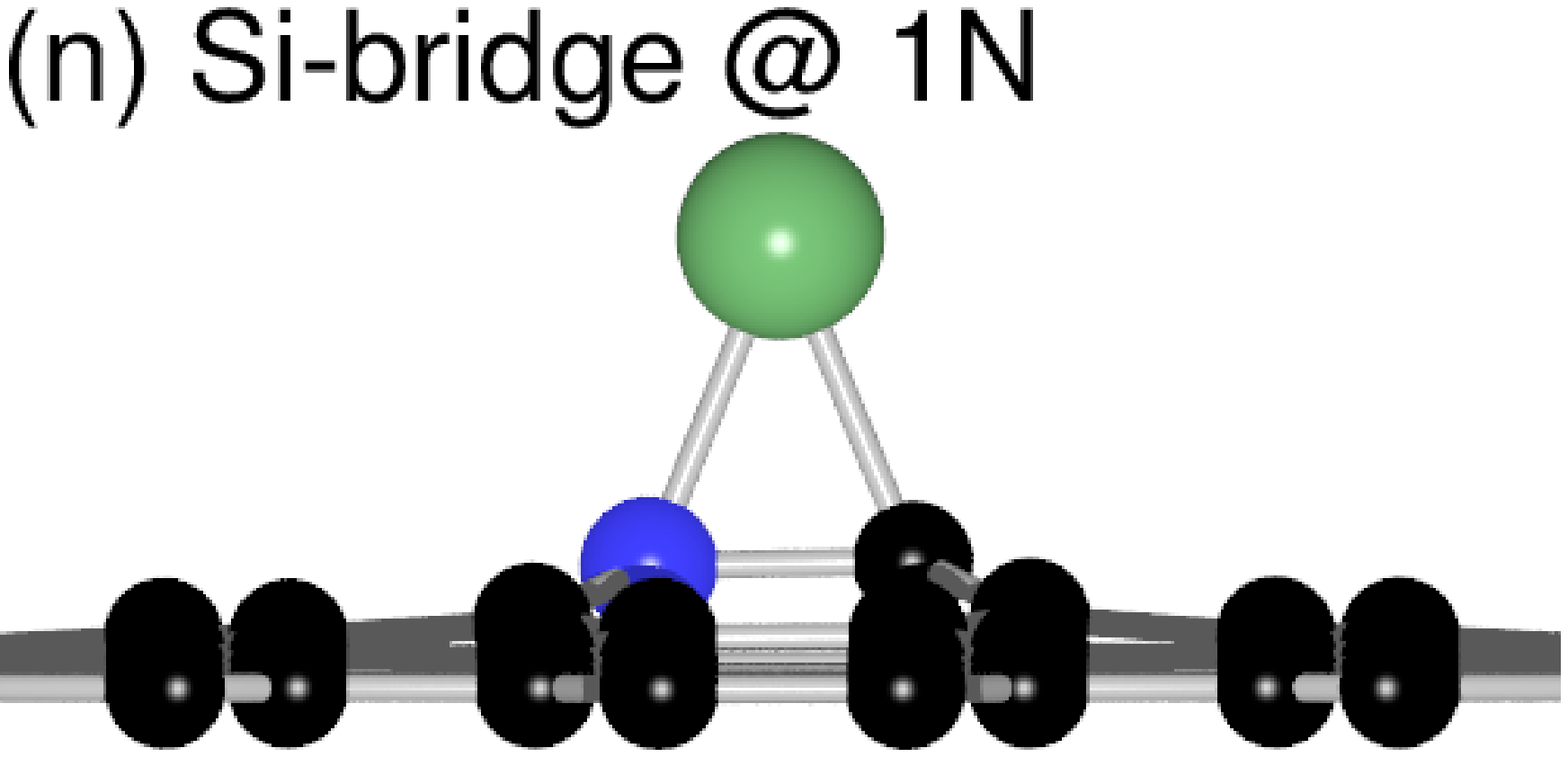}
\includegraphics[width=0.195\linewidth]{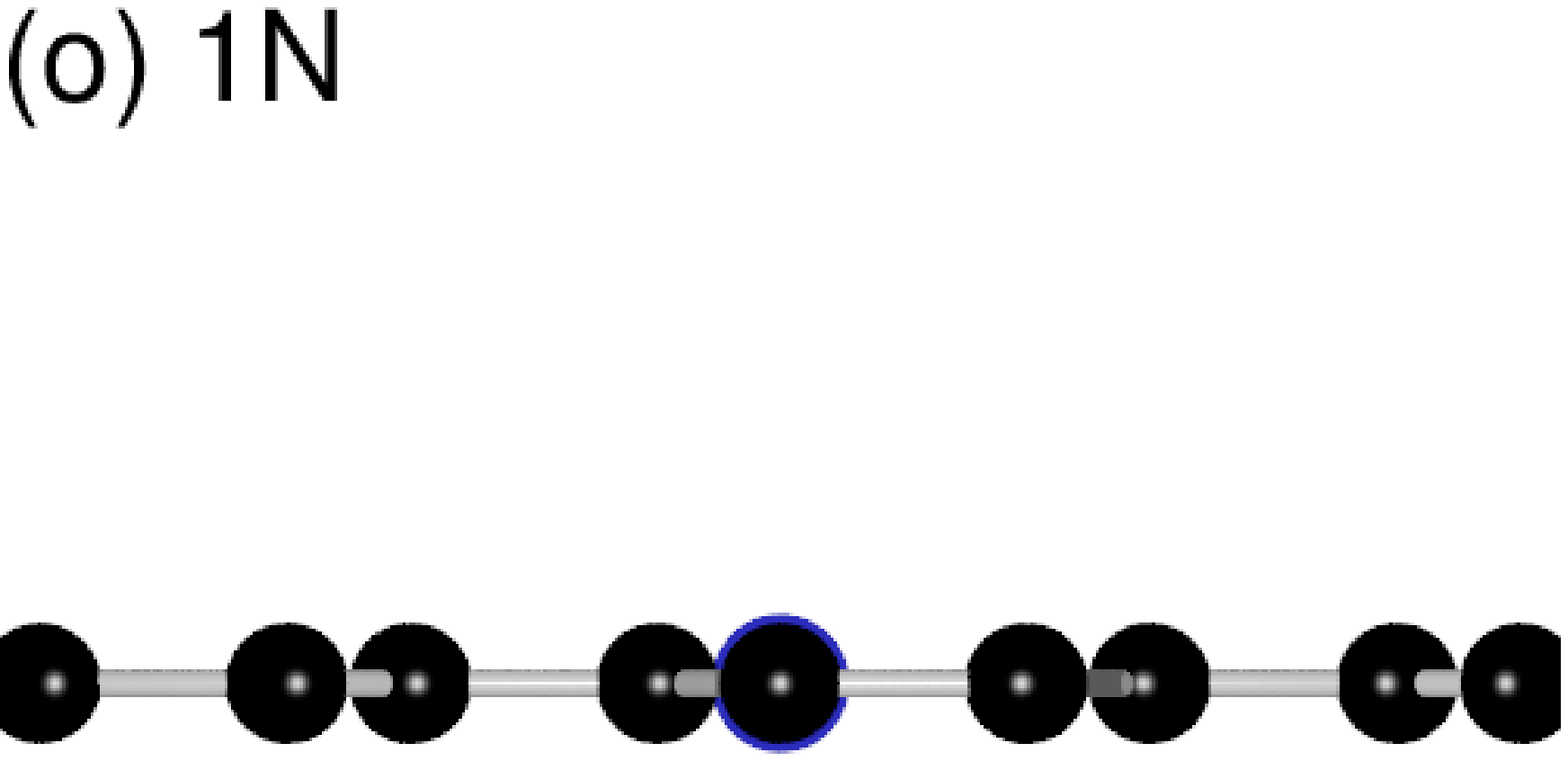}
\includegraphics[width=0.195\linewidth]{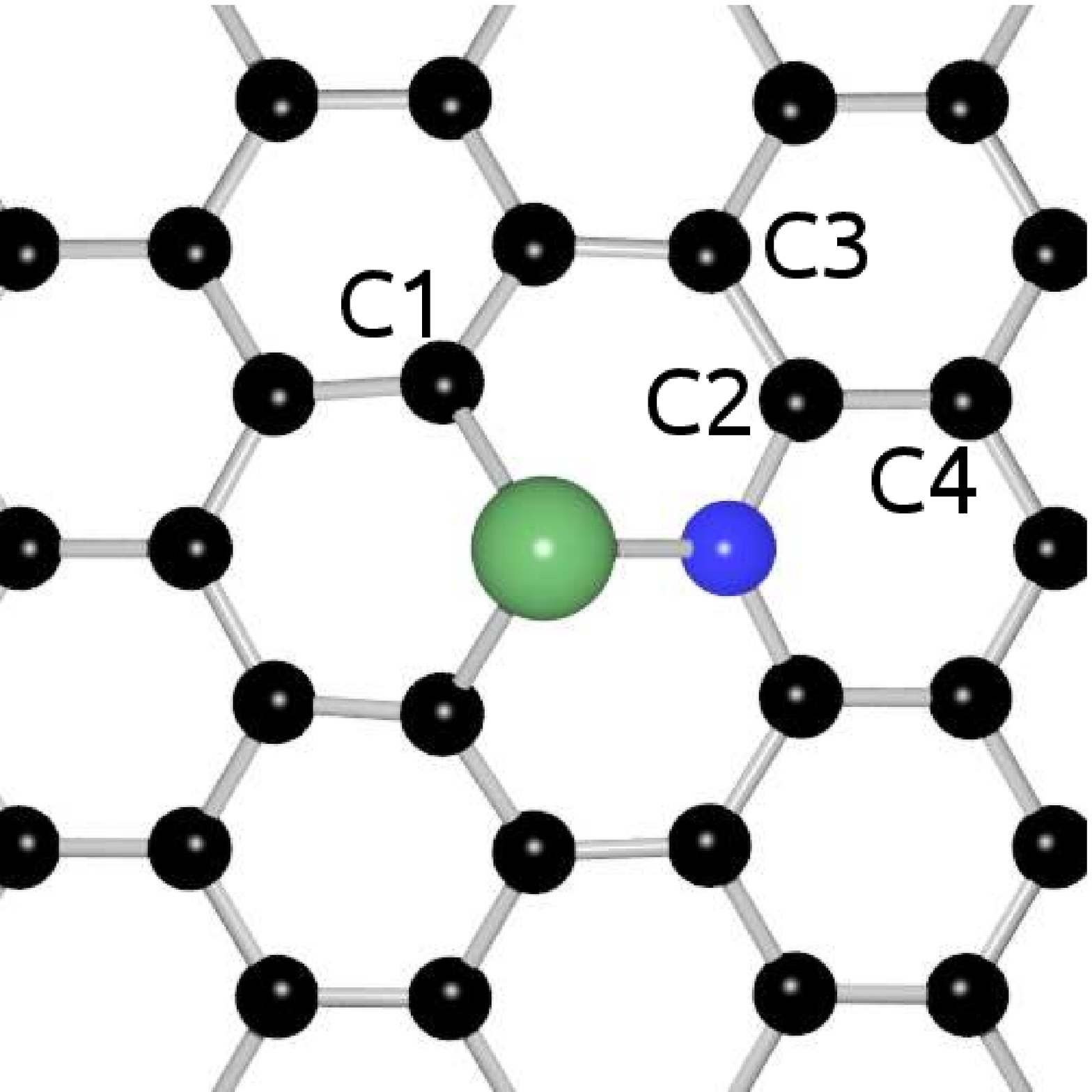}
\includegraphics[width=0.195\linewidth]{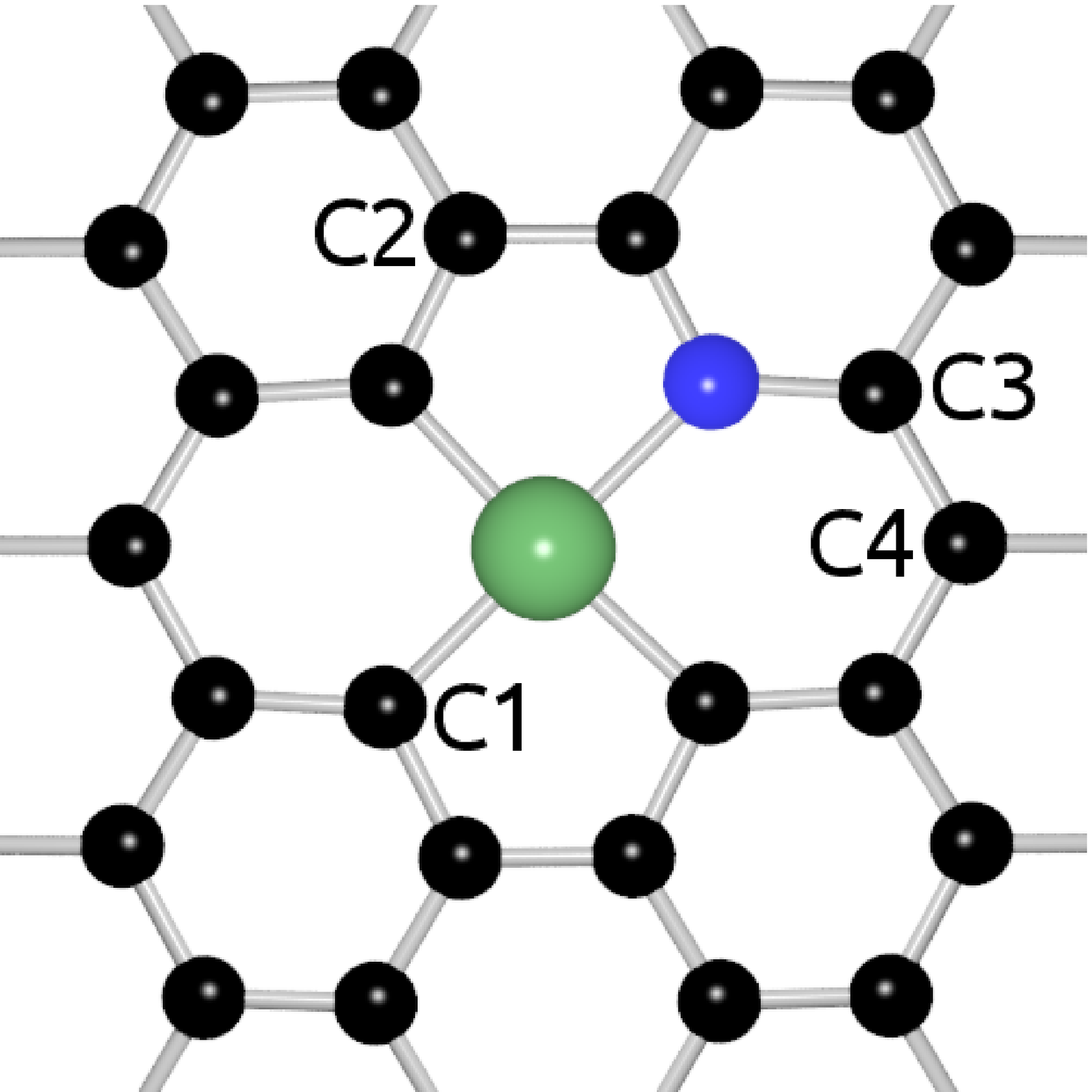}
\includegraphics[width=0.195\linewidth]{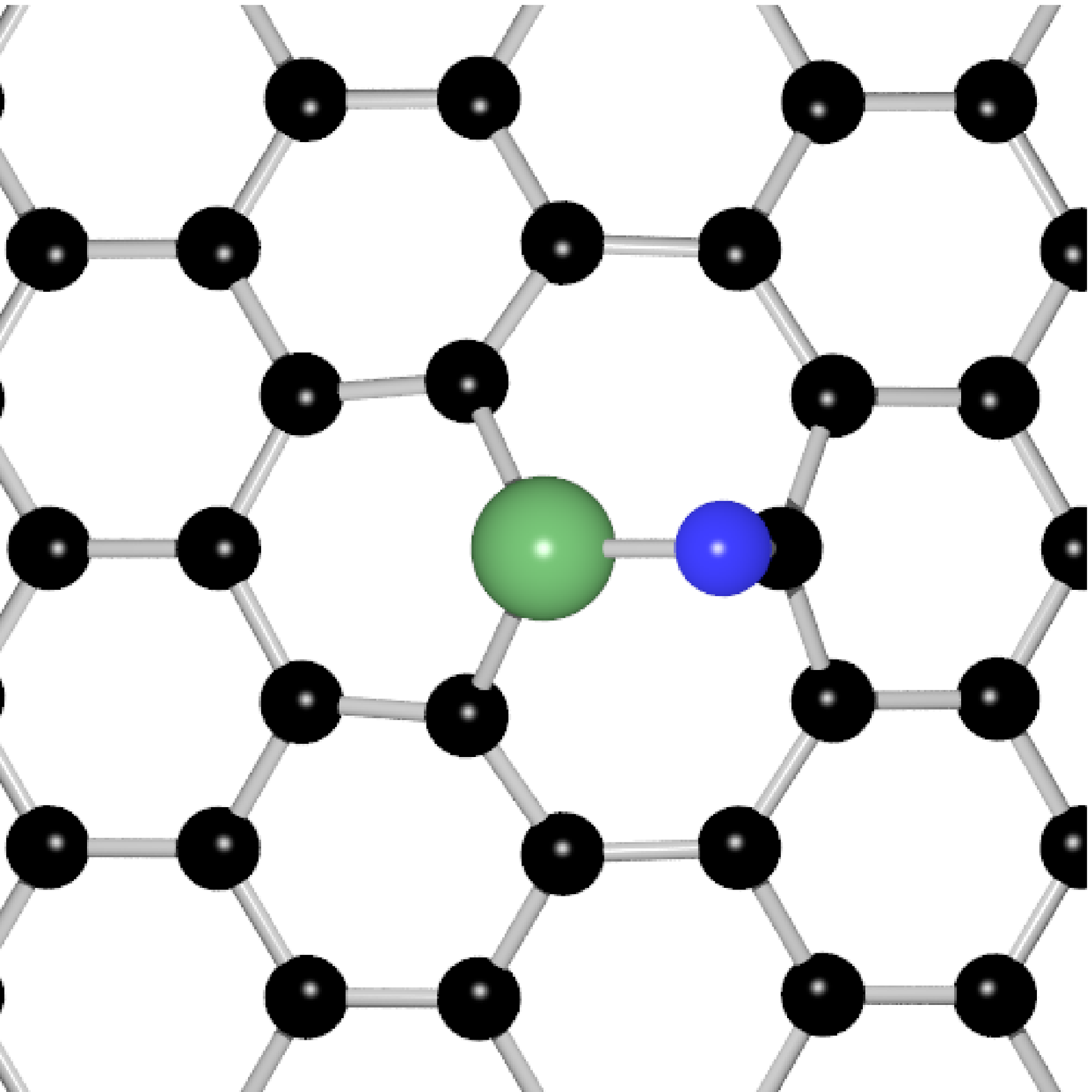}
\includegraphics[width=0.195\linewidth]{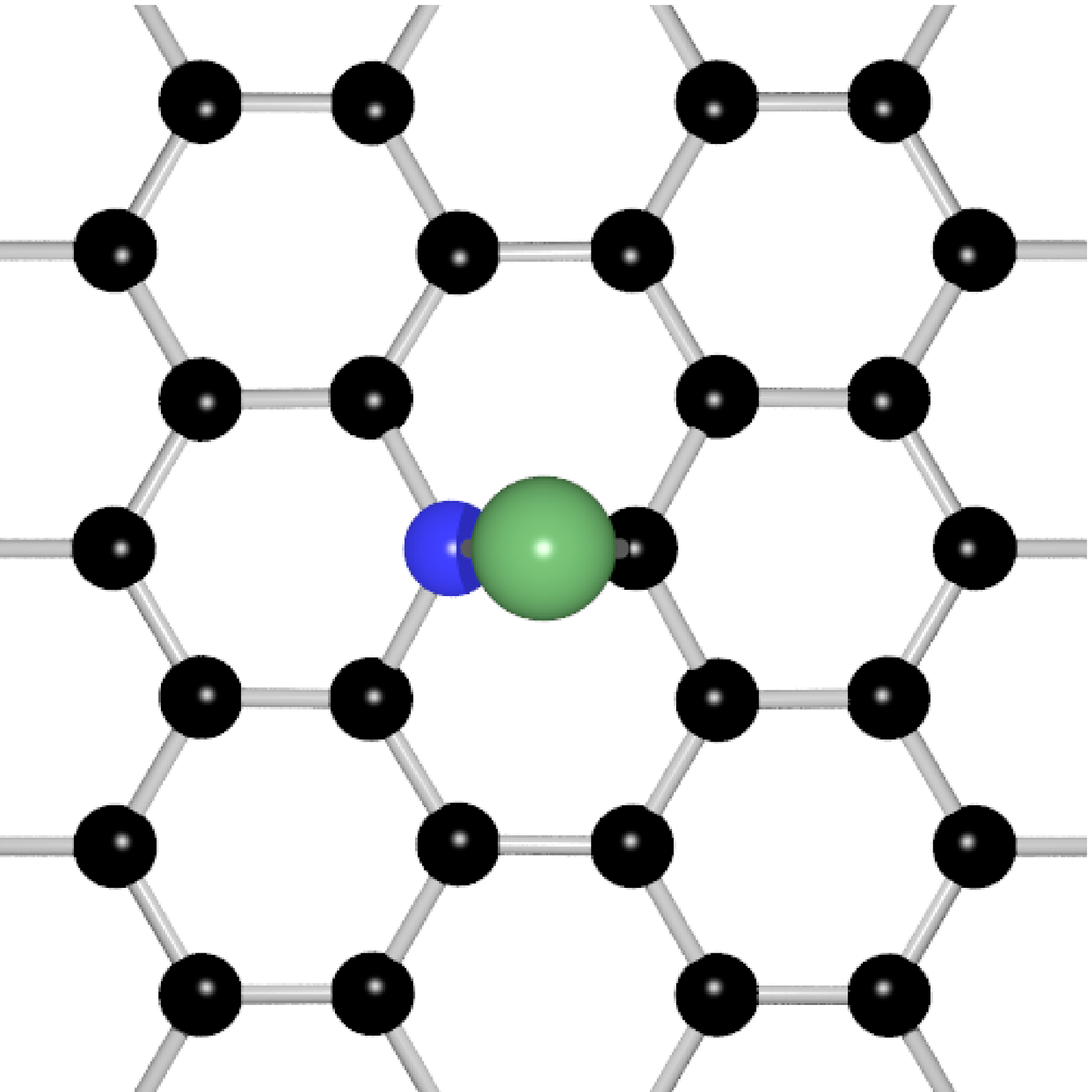}
\includegraphics[width=0.195\linewidth]{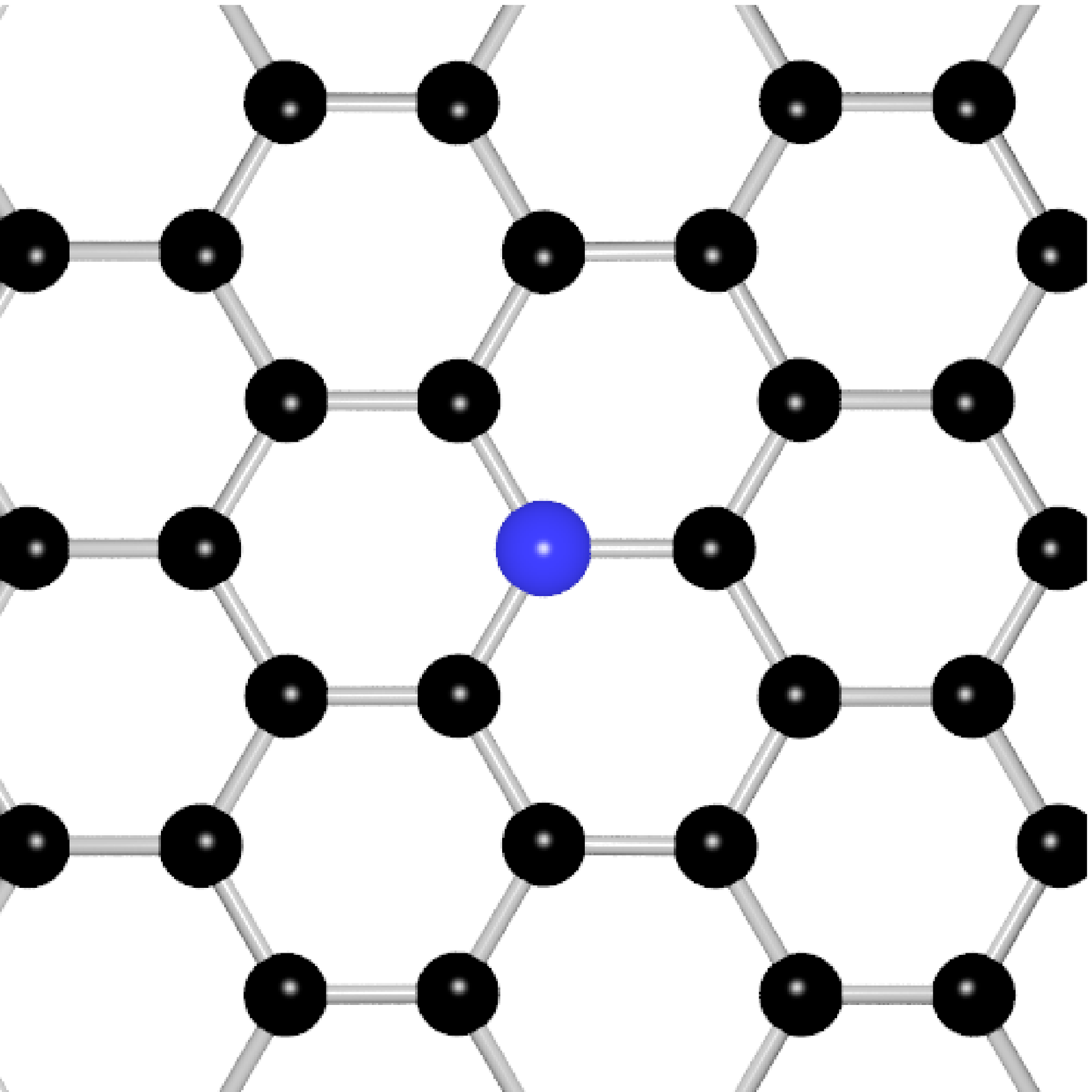}
\includegraphics[width=0.195\linewidth]{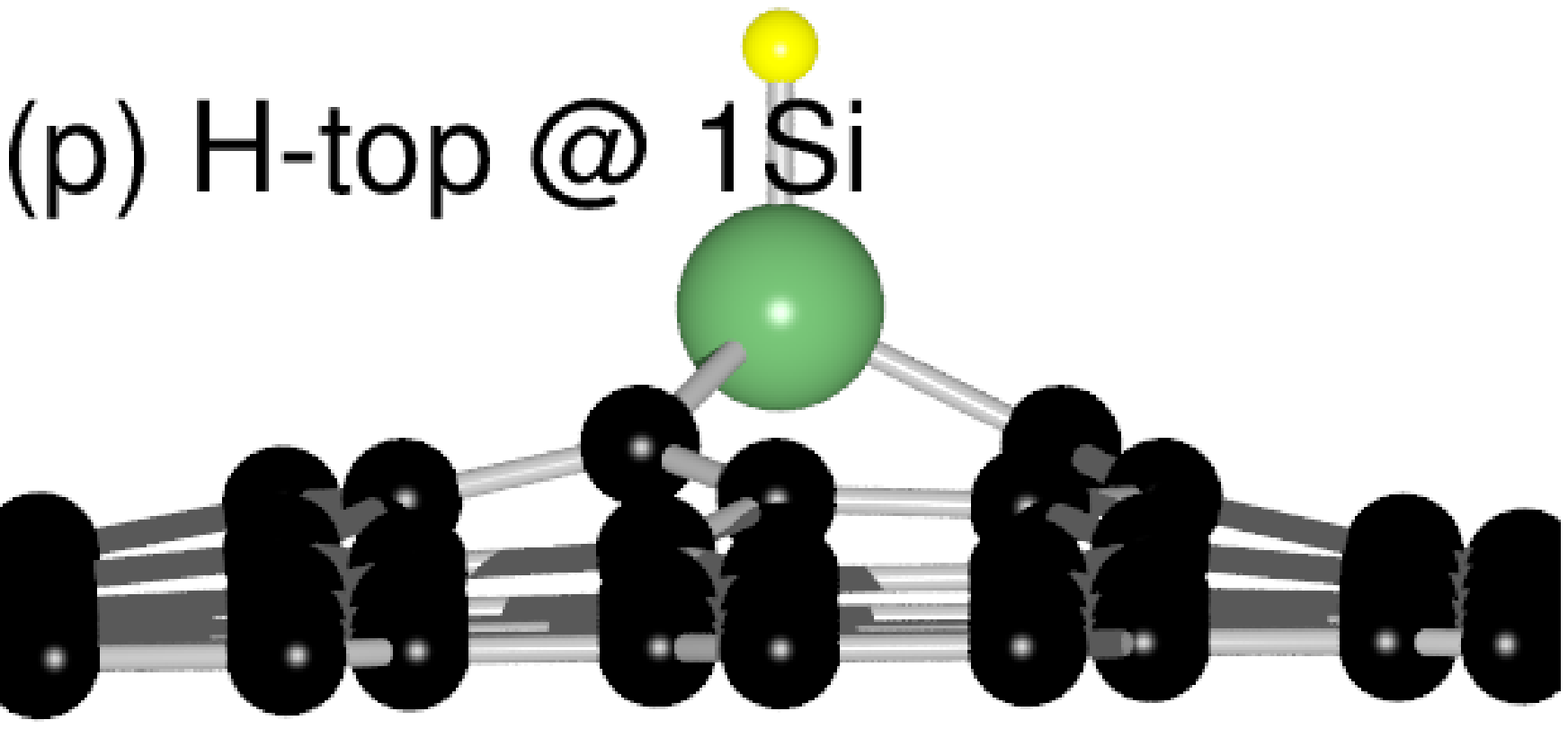}
\includegraphics[width=0.195\linewidth]{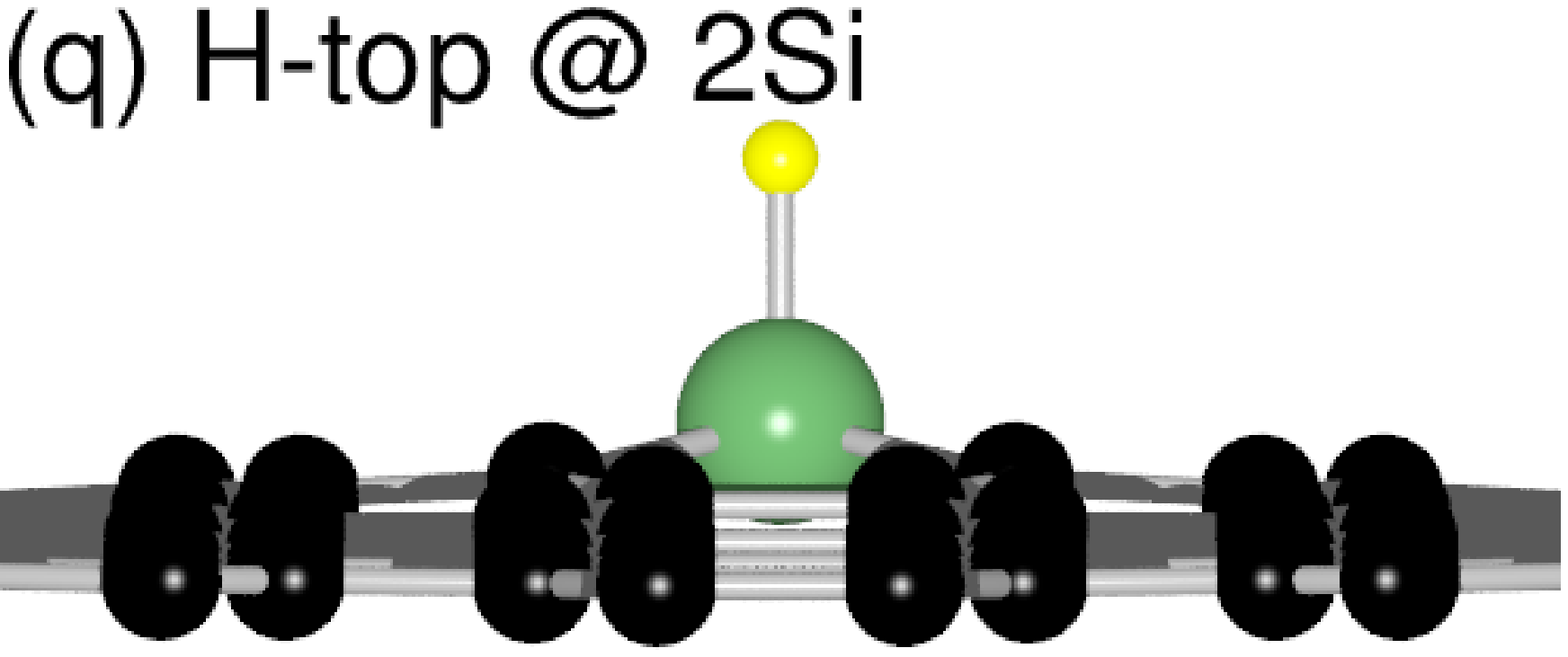}
\includegraphics[width=0.195\linewidth]{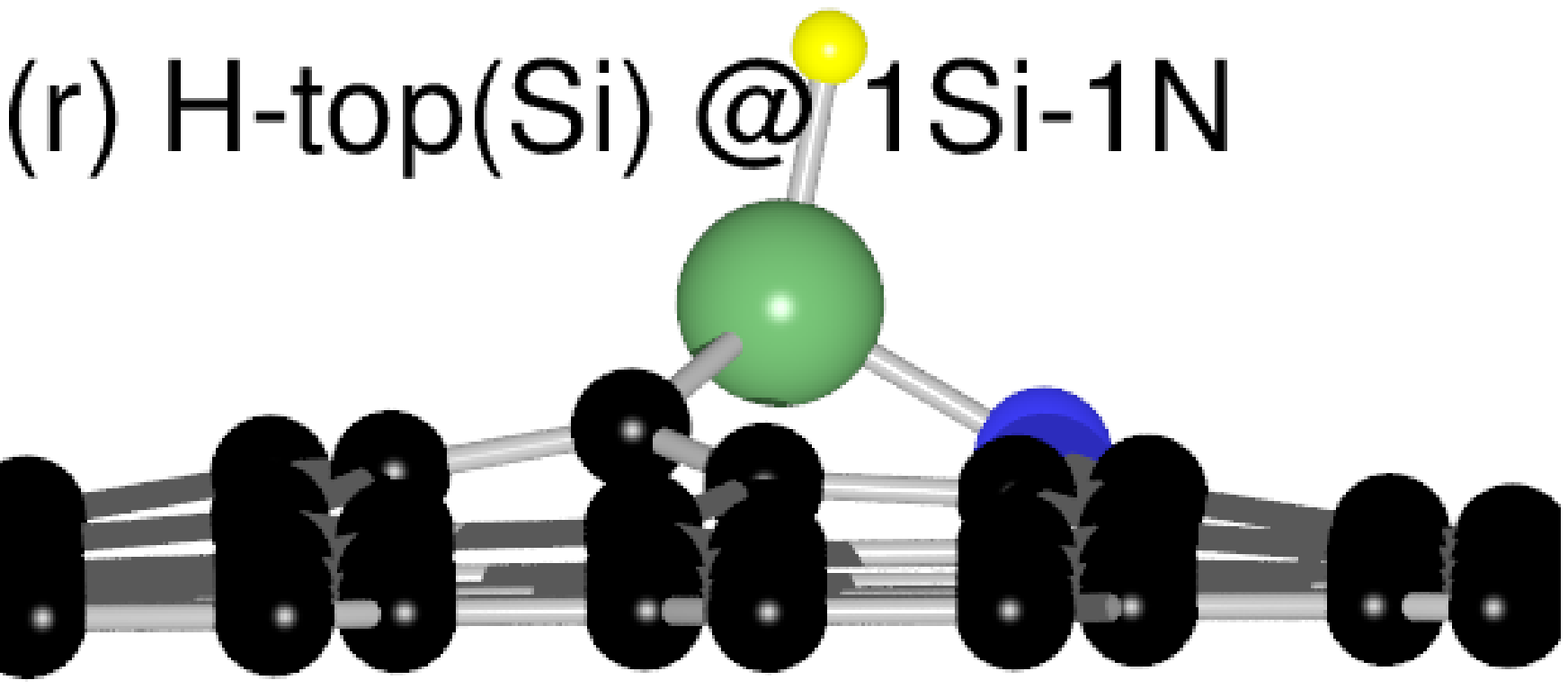}
\includegraphics[width=0.195\linewidth]{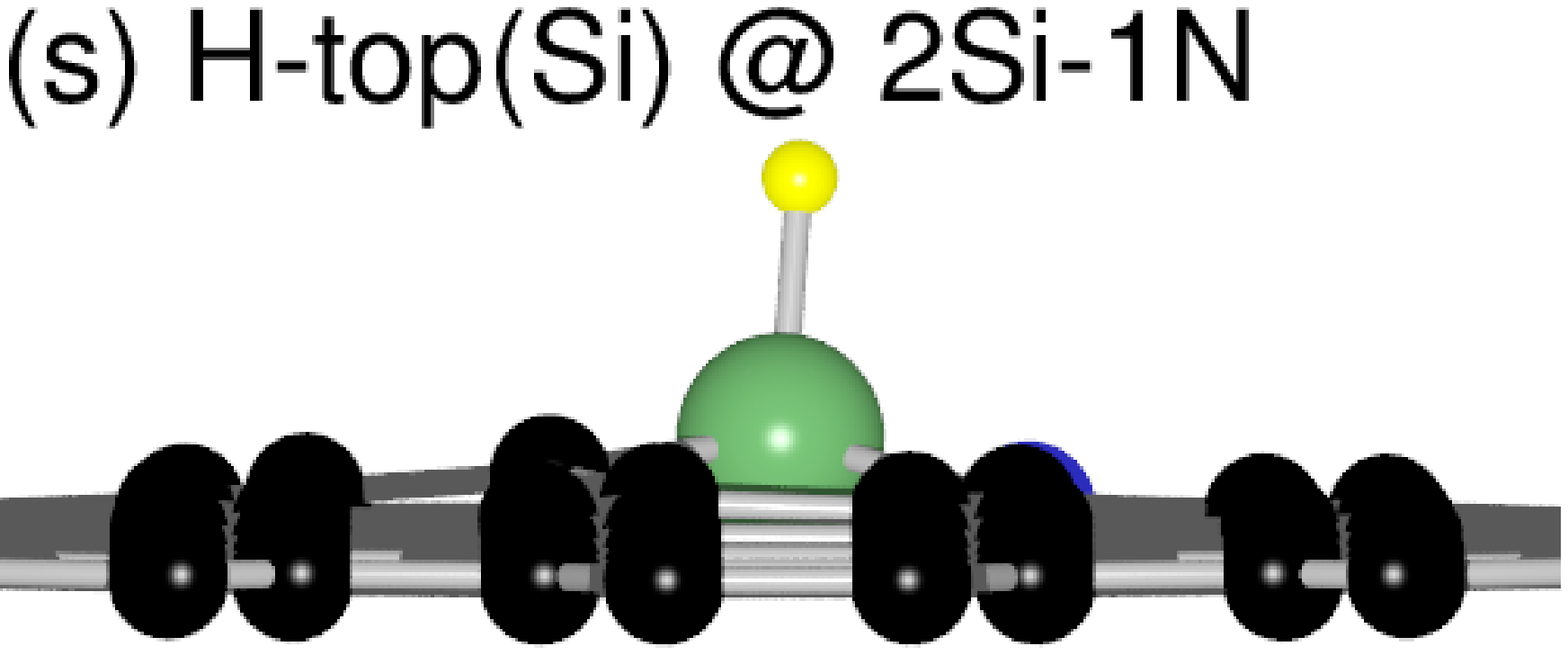}
\includegraphics[width=0.195\linewidth]{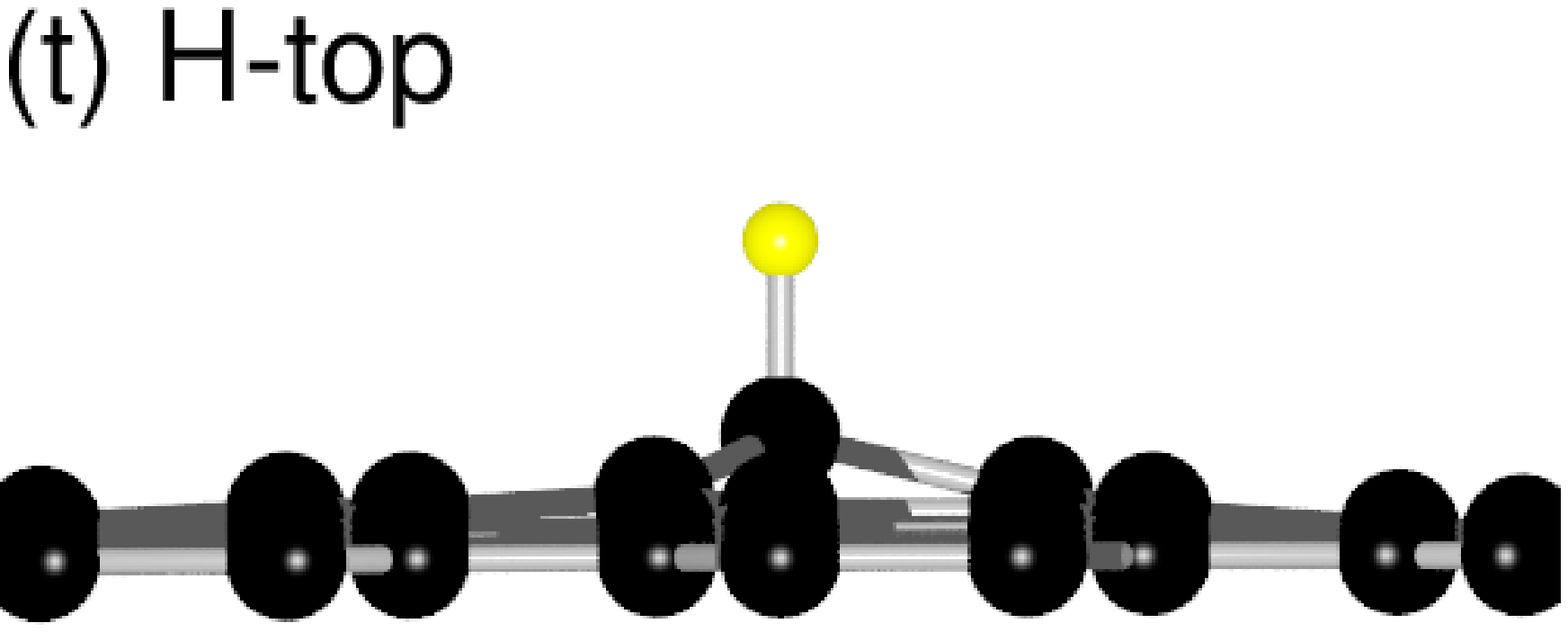}
\includegraphics[width=0.195\linewidth]{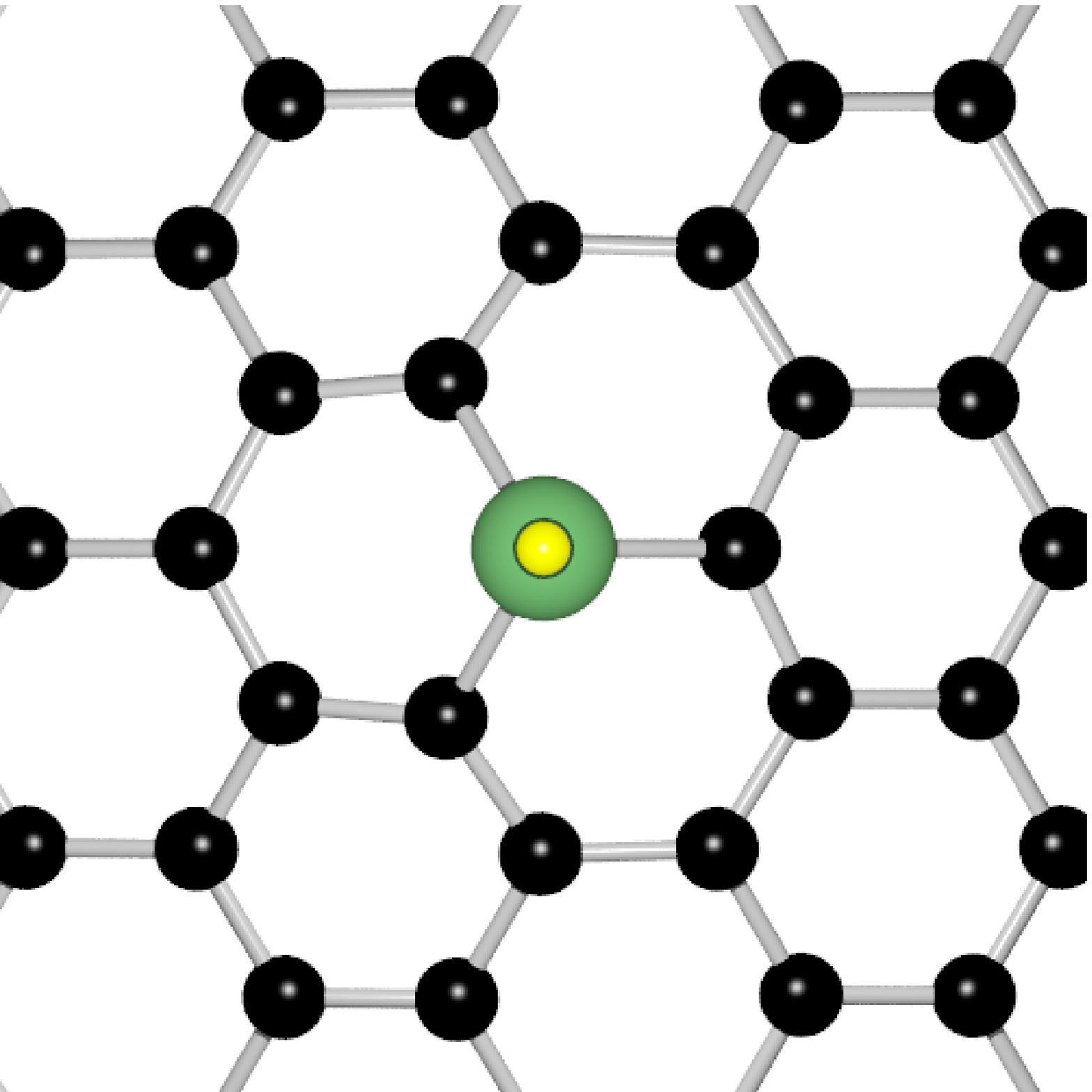}
\includegraphics[width=0.195\linewidth]{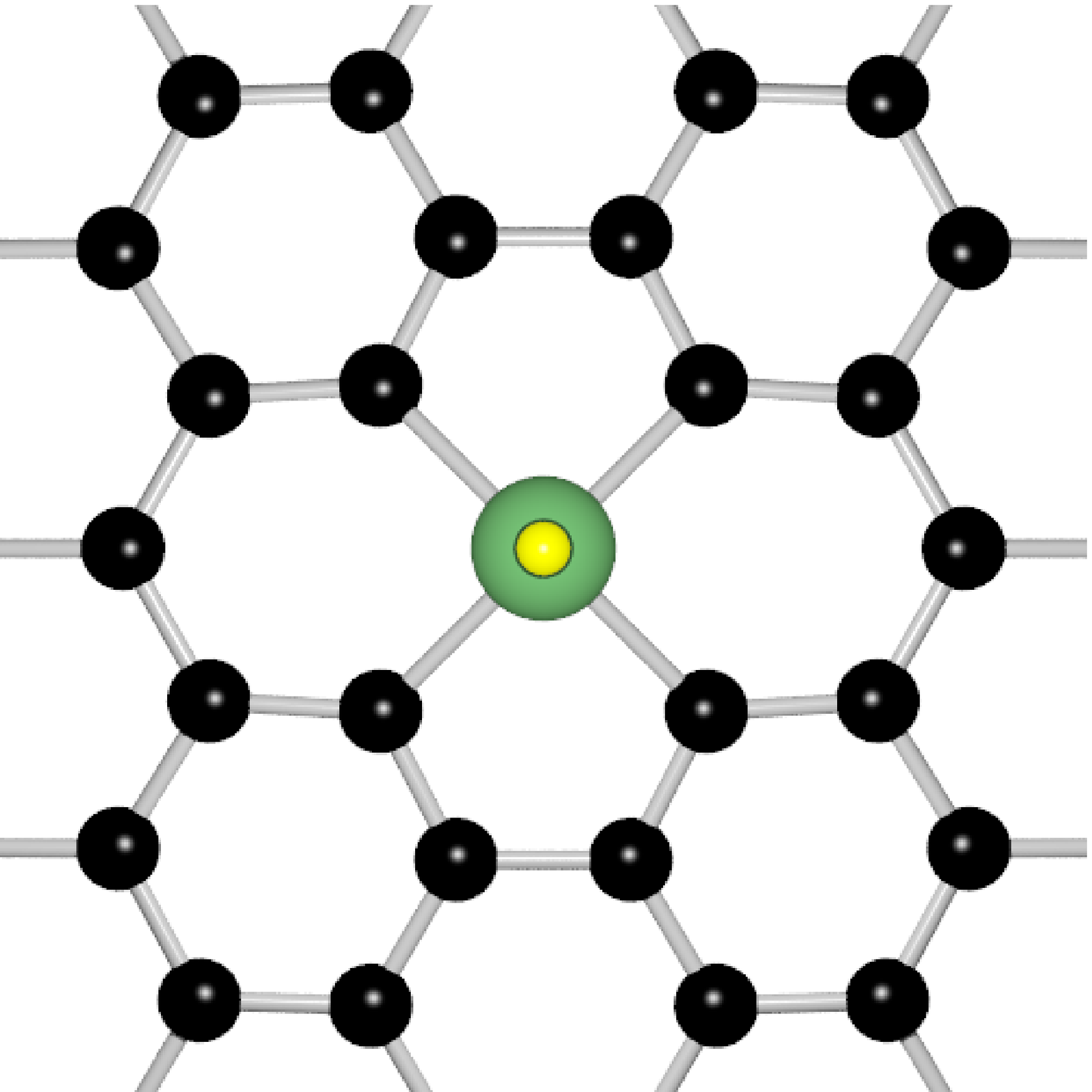}
\includegraphics[width=0.195\linewidth]{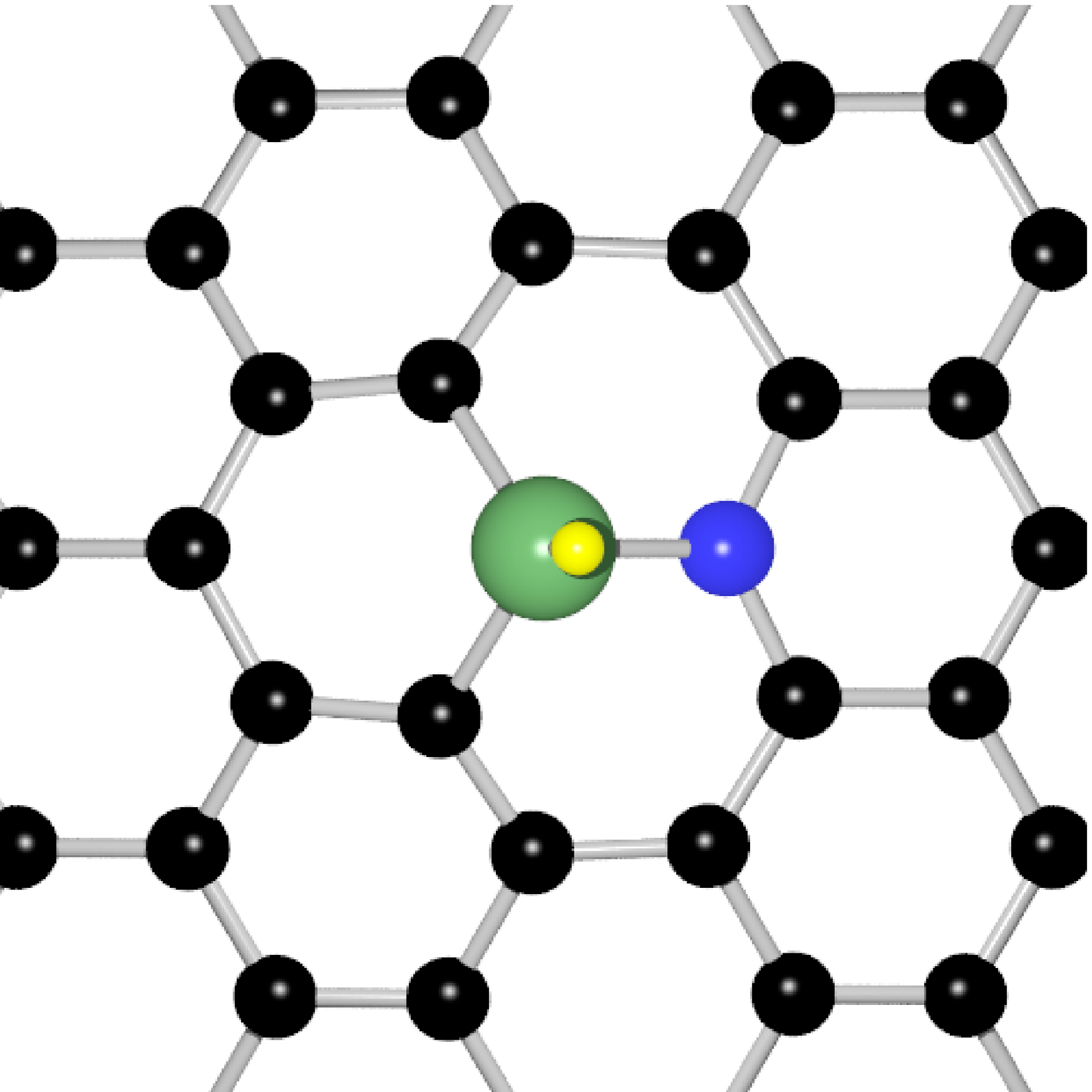}
\includegraphics[width=0.195\linewidth]{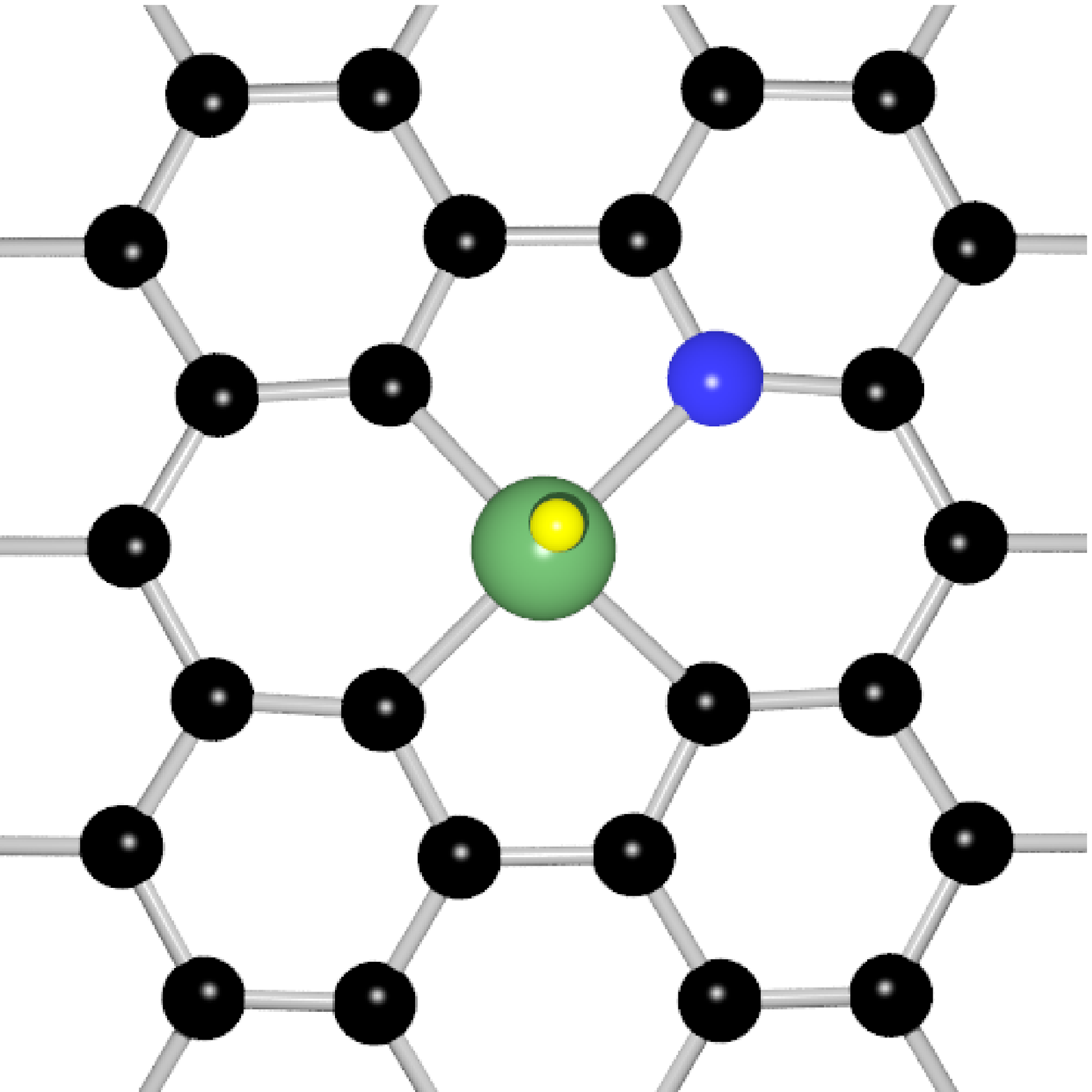}
\includegraphics[width=0.195\linewidth]{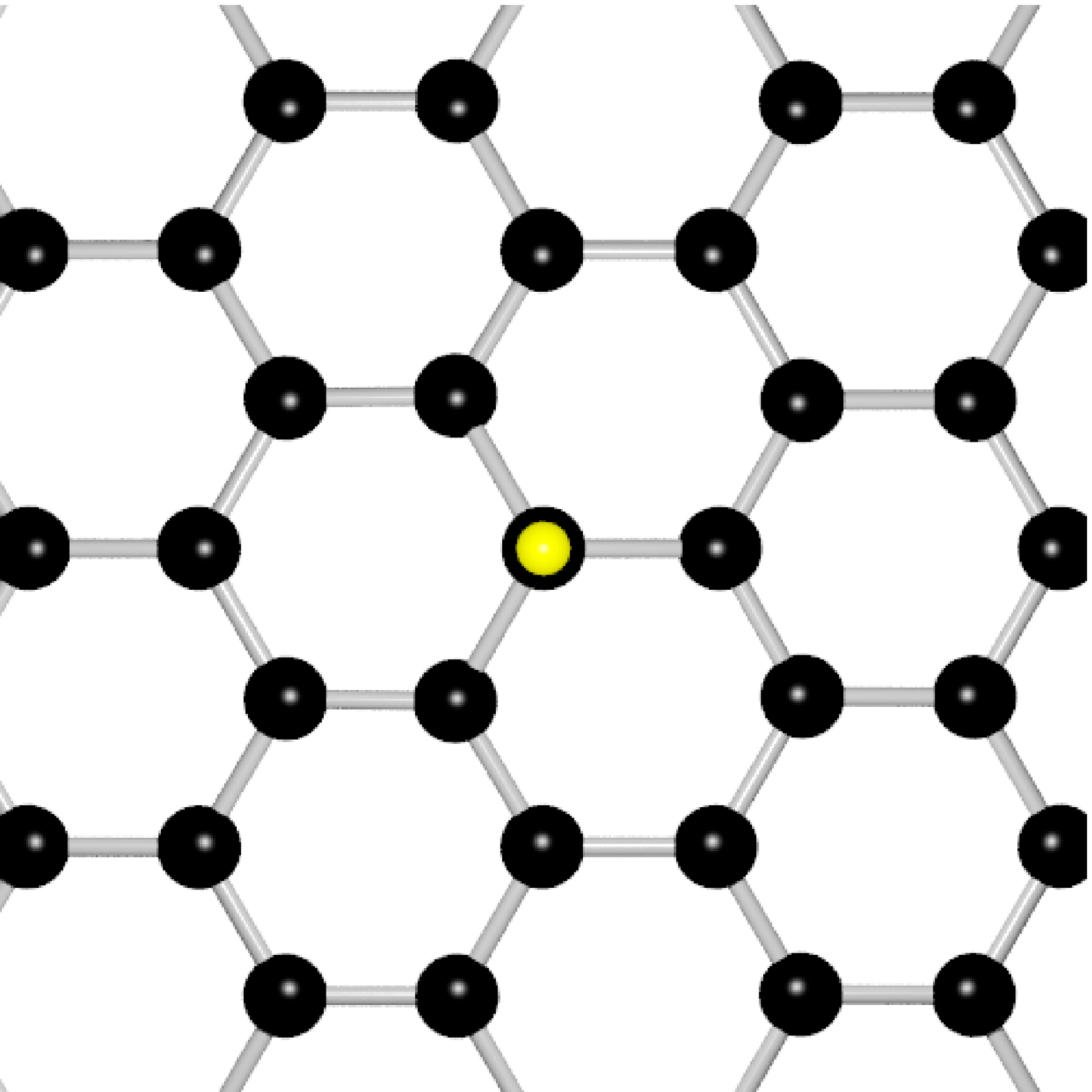}
\caption{(Color online). Atomic structures, top and side views,
for defects in graphene containing silicon (large green
spheres), oxygen (red spheres), nitrogen (blue spheres), and hydrogen
(small yellow spheres) atoms. The geometries are relaxed in the $11 \times
11$ supercells, and only the defect neighbourhood is shown. (a)-(e)
Silicon impurities. (f)-(j) Silicon-oxygen impurities (k)-(o)
Silicon-nitrogen impurities. (p)-(t) Hydrogen adatoms on silicon and
silicon-nitrogen impurities. \label{fig:geometries}} \end{figure*}

The studied defect cases and shorthand notations for labelling them are presented in Fig.~\ref{fig:geometries}. 
When a
single Si, N or O atom fills a monovacancy (1Vac) in the graphene
lattice, the substitutions are denoted as 1Si, 1N and 1O, and when it
fills a divacancy (2Vac), the substitutions are denoted as 2Si, 2N and
2O, respectively. 
A composite defect consisting of two defects, for instance two substitutions, located at
neighboring sites is denoted as 1Si-1O. For adatoms on pure graphene, we specify the absorption site like Si-top, and 
further when the adatom is on top of a defect, we combine the notation to form O-top@1Si for a oxygen adatom
on top of a silicon in a monovacancy.
Below, we first show the results for the defects containing only silicon and oxygen, see Figs.~\ref{fig:geometries}.(a)-(j).
After that, nitrogen is added, see Figs.~\ref{fig:geometries}.(k)-(o). Finally, the hydrogen adatom cases shown in Figs.~\ref{fig:geometries}.(p)-(t) are studied.

\subsection{Silicon and oxygen impurities}

Silicon and oxygen impurities can appear in graphene for example due
to contamination in growth processes where SiO$_2$ or SiC are present,
or due to purposeful post-synthesis treatment. To determine which
impurities are the most likely to occur, we relaxed various silicon and
oxygen point-defect atomic geometries and evaluated their formation energies. 

We first analyze the results for the silicon defects. The 1Si defect,
where a single carbon atom is substituted by a heavier silicon atom,
prefers the tetrahedral sp$^3$ type hybridization, resulting in an
out-of-plane silicon position, as seen in Fig.~\ref{fig:geometries}(a).
Consequently, the surrounding carbon atoms move accordingly, and the
resulting curvature takes several bond lengths to relax, making it
effectively a slightly extended defect.
On the other hand, the silicon atom in a 2Si
defect, as shown in Fig.~\ref{fig:geometries}(b), forms sp$^2$d hybrid
orbitals that bond with the four neighboring carbon atoms, and the
geometry is only slightly perturbed in the out-of-plane direction. The
defect geometries and orbital hybridizations of the 1Si and 2Si defects
have been established both by computational and experimental methods in
Refs. [\onlinecite{Zhou_2012, Ramasse_2012}]. 

\begin{table*}
\caption{Defect spin moments and formation energies.
The spin moments are for relaxed $11 \times 11$ supercells, and the formation energies are also given for the $8 \times 8$ supercell. Energies are in units of eV. The spin moment values inside brackets are atom-projected spin moments using the Hirshfeld analysis. \label{table:defect_infosheet}}
\begin{tabular}{c c c c r r c}
\hline
Impurity atoms & Defect &  Spin moment ($\mu_B$) & $E_\text{f}(8 \times 8)$ & $E_\text{f}(11 \times 11)$ & $E_\text{f}$(Ref.) \\
\hline
 -           & 1Vac                     & $1.44$ [$0.69$(C)] & $7.68$ & $7.65$ & $7.5$ \cite{Banhart_2011} \\
             & 2Vac                     & -                  & $7.56$ & $7.36$ & $8$ \cite{Banhart_2011} \\ 
\hline
Si           & Isolated Si              &$2.0$& $4.66$ & $4.66$ & \\
             & 1Si             & -   & $3.79$ & $3.77$ & \\
             & 2Si              & -   & $4.60$ & $4.57$ & $4.41$ \cite{Zhou_2012_2} \\
             & 1Si @ 555-777           & -   & $6.89$ & $6.96$ & \\
             & Si-top/bridge            & -   & $4.33$ & $4.33$ & \\
             & 1Si-1Si                  & -   & $5.99$ & $5.97$ & \\      
             & 1Si-2Si                  & -   & $6.09$ & $6.06$ & \\  
\hline
O            & Isolated O               &$2.0$& $2.92$  & $2.92$  & \\
             & 1O                       & -   & $2.43$  & $2.44$  & \\
             & 2O (1O-1Vac)             & -   & $3.03$  & $2.96$  & \\
             & O-bridge                 & -   & $0.43$  & $0.43$  & \\
             & 1O-1O                    & -   & $-1.10$ & $-1.12$ & \\
\hline
Si + O       & 1Si-1O                   & -   & $2.78$ & $2.77$ & \\
             & 2Si-1O                   & -   & $2.26$ & $2.22$ & \\
             & 1Si-2O                   & -   & $5.03$ & $4.98$ & \\
             & 2Si-2O                   & -   & $4.66$ & $4.52$ & \\
             & O-top @ 1Si              & -   & $1.43$ & $1.40$ & \\
             & O-bridge @ 2Si           & -   & $1.69$ & $1.66$ & \\
             & Si-bridge @ 1O           &     & $4.06$ & $4.04$ & \\
\hline
N            & Isolated N               &$3.0$& $5.28$ & $5.28$ & \\
             & 1N               & -   & $0.93$ & $0.91$ & $0.9$ \cite{Zhou_2012_2} \\
             & 2N (1N-1Vac)             & -   & $5.52$ & $5.45$ & \\
             & N-bridge                 & $0.72$ [$0.53$(N)] & $4.38$ & $4.37$ & $ $ \\
             & $\text{1N} \cdots \text{1N}$ (3rd-NN)     & -   & $1.91$ & $1.91$ & \\             
\hline
Si + N       & 1Si-1N                  & $1.00$ [$0.44$(Si), $0.07$(N)] & $3.34$ & $3.33$ & $ $ \\
             & 2Si-1N                   & -   & $3.25$ & $3.22$ & $2.59$ \cite{Zhou_2012_2}  \\
             & 1Si-2N                   & -   & $8.26$ & $8.21$ &  \\
             & 2Si-2N                   &     & $7.15$ & $7.14$ &  \\
             & N-bridge @ 1Si          & -   & $5.79$ & $5.75$ & $ $ \\
             & N-bridge @ 2Si          & $0.67$ [$0.02$(Si), $0.26$(N)] & $5.87$ & $5.83$ & $ $ \\
             & Si-bridge @ 1N           & $1.00$ [$0.74$(Si), $0.02$(N)] & $5.16$ & $5.16$ & $ $ \\
\hline
H            & Isolated H               &$1.0$ & $2.27$ & $2.27$ & \\
             & H-top                   & $1.00$ [$0.04$(H)] & $1.45$ & $1.45$ & $ $ \\
\hline
Si + H       & H-top @ 1Si              & -   & $3.23$ & $3.21$ & $ $ \\
             & H-top @ 2Si              & -   & $4.61$ & $4.56$ & $ $ \\
\hline
N + H        & H-top @ 1N               & -   & $2.68$ & $2.68$ & $ $ \\
\hline
Si + N + H   & H-top(Si) @ 1Si-1N       & -   & $2.03$ & $2.02$ & $ $ \\
             & H-top(N)  @ 1Si-1N       & -   & $4.54$ & $4.52$ & $ $ \\
             & H-top(Si) @ 2Si-1N       & -   & $2.91$ & $2.88$ & $ $ \\
             & H-top(N)  @ 2Si-1N       & -   & $4.12$ & $4.08$ & $ $ \\
\hline
\end{tabular}
\end{table*}
The defect formation energies are listed in Table~\ref{table:defect_infosheet}. The 1Si and 2Si defect formation energies
are $3.77$ eV and $4.57$ eV, respectively, based on the $11 \times 11$
supercell calculation. Increasing the supercell size typically lowers
the formation energy, 
except in some cases where the computational supercell
is too small, such that there is not enough space for bulk graphene
between the defects,
or due to other finite size effects \cite{Kotakoski06mvac}.
The chosen chemical
potentials $\mu_i$ affect the absolute values of the formation energies,
but when comparing the formation energies of defects with the same
number of atoms $n_i$, the $n_i \mu_i$ terms cancel in Eq.~\eqref{eq:def:formation_energy}. Therefore, regardless whether the
silicon atom migrates from the substrate or arrives as part of a
molecule exposed to the surface, as long as the source of the silicon
remains the same, 2Si has a higher formation energy than 1Si by
$E_{\text{f}}[\text{2Si}]-E_{\text{f}}[\text{1Si}] =
E[\text{2Si}]-E[\text{1Si}] + \mu_C \approx 0.8$ eV. The leftover carbon
atoms are assumed to form bulk graphene, but if they obtained an energy
higher than $\mu_C$, 1Si would become even more favorable compared to
2Si. On the other hand, if the defects were to form on the spot, the
energy per ejected carbon atom is lower for 2Si. In any case, the sp$^3$
hybridization in 1Si with three-fold coordination is energetically
preferred to the sp$^2$d hybridization in 2Si with four-fold
coordination, even if the 1Si defect causes a local curvature of the
graphene sheet. 

A divacancy can be relaxed further by a bond rotation to obtain a
555-777 defect with even lower energy \cite{Banhart_2011}. We
studied the case where a silicon atom substitutes one of the carbon
atoms at the 555-777 defect. It turned out that silicon atom prefers to
substitute not the middle site but a site neighboring it, as shown in
Fig.~\ref{fig:geometries}(c). However, the formation energy of such a
defect is much higher than that of 1Si and 2Si. 

A silicon adatom on pristine graphene takes a position 
on top of a carbon atom
(Si-top), but it is slightly shifted towards the hexagon of the graphene
backbone, see Fig.~\ref{fig:geometries}(d). Its formation energy is
$4.33$ eV, being between the 1Si and 2Si formation energies. It is only
barely lower than the formation energy of an isolated silicon atom that
is $4.66$ eV. Moreover, the total energy of the silicon adatom in a
bridge position on top of a carbon-carbon bond (Si-bridge), shown in
Fig.~\ref{fig:geometries}(e), is only a few meV higher, further
indicating that silicon adatoms are rather mobile on graphene. This, in
turn, implies that silicon adatoms are easily trapped by monovacancies
and divacancies, because the formation energies of 1Si or 2Si are much
lower than the sum of the formation energies of an isolated silicon atom
or Si-top/bridge, and 1Vac or 2Vac. The energy gain is roughly $7$-$8$
eV. Furthermore, a system consisting of a pair of neighbouring 1Si
defects (1Si-1Si), or a pair of neighbouring 1Si and 2Si defects
(1Si-2Si), has a total energy that is $1.6$ eV, or $2.3$ eV, lower
than the energy of two fully separated defects, respectively. Such composite defects
minimize the total distortion, such as the curvature of the graphene
lattice, and therefore reduce the total energy. However, silicon
substitution defects are not expected to be mobile (at least under 
ambient
conditions) after they have been formed.

Oxygen could also play an important role in
forming impurities in graphene. It is ubiquitous, and it is commonly present
during the graphene growth process or measurements especially on a
SiO$_2$ substrate. Oxygen in graphene prefers the bridged adatom
position (O-bridge, not shown) as opposed to filling a monovacancy or a
divacancy, 1O or 2O (not shown). Namely, the formation energy of an
O-bridge is about $2-3$ eV smaller than those of 1O and 2O, and also
about $0.8$ eV smaller than the case where the oxygen adatom is located
exactly on top of a carbon atom. Contrary to silicon, an oxygen atom is
too small to completely fill a divacancy, and a 2O defect is better
described as 1O-1Vac. Curiously, two neighbouring oxygen substitutions
(1O-1O), shown in Fig.~\ref{fig:geometries}(j), has a remarkably low
formation energy of $-1.12$ eV. However, there can be a high energy
barrier to reach such a configuration, as hinted by the high formation
energy of 2O (1O-1Vac) that is $2.96$ eV.

It is not a surprise that, when considering composite defects containing
both silicon and oxygen, silicon prefers to form 1Si and 2Si defects,
and oxygen prefers the adatom position. However, the lowest formation
energy of $1.40$ eV is obtained for an oxygen adatom located on top of a
1Si defect (O-top @ 1Si), shown in Fig.~\ref{fig:geometries}(h), and not
in a bridged position. The oxygen adatom bonds with the silicon atom in
1Si much more preferably than with graphene, and surprisingly, also more
preferably than forming an O$_2$ molecule. An oxygen bridge located on a
2Si defect (O-bridge @ 2Si), shown in Fig.~\ref{fig:geometries}(i), also
has an unexpectedly low formation energy of $1.69$ eV. These results
indicate that oxygen is captured at the silicon impurities.

When embedding oxygen in the graphene lattice, the composite defects
such as 1Si-1O and 2Si-1O, shown in Fig.~\ref{fig:geometries} (f) and (g),
have comparably small formation energies as well, $2.78$ eV and $2.26$
eV, respectively. Thus it costs less energy to form oxygen substitutions
next to 1Si or 2Si than in pristine graphene. There are many possible
configurations that could be called 1Si-2O or 2Si-2O, but they all have
systematically at least about $2$ eV higher energies than 1Si-1O and
2Si-1O. The same goes for reversing the roles of silicon and oxygen,
namely a defect consisting of a silicon adatom located near a 1O defect
has a much larger formation energy. Otherwise, the silicon and oxygen
pair well to form composite point-defects.

\subsection{Nitrogen doping and finite spin moments}

\begin{figure}[tb]
	\includegraphics[width=\linewidth]{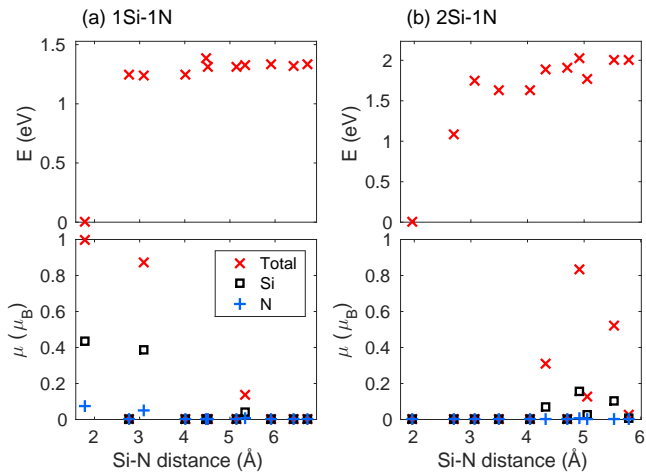}
	\caption{(Color online). The total energies $E$ and spin moments $\mu$ for the (a) 1Si-1N and (b) 2Si-1N defects in relaxed $11 \times 11$ supercells, where the N atom is separated to lattice sites further away from the Si atom. The zero energy values correspond to the nearest-neighbour cases of the 1Si-1N defect in (a) and the 2Si-1N defect in (b). The Si and N atom-projected spin moments are evaluated using the Hirshfeld analysis.
\label{fig:total_energies_of_Si_N_separation}}
\end{figure}

A nitrogen impurity in graphene dopes its surroundings, which can be
used to tune the electronic properties of graphene.
A nitrogen substitution defect (1N), shown in Fig.
\ref{fig:geometries}(o), is energetically by far the most stable defect
configuration that contains nitrogen only.
The differences in formation energy to a bridged nitrogen
adatom (N-bridge, not shown) and to a divacancy substitution (2N or
1N-1Vac, not shown) are $3.4$ eV and $4.6$ eV, respectively. Since a
nitrogen atom has one more electron than a carbon atom, and a
substitutional nitrogen atom forms sp$^2$ hybridized bonds in graphene,
some of the electron density is donated to the nearby atoms, effectively
doping the neighbourhood \cite{Lambin_2012}. Thus, we are mostly
interested in the 1N-doped silicon impurities, to find how the added
nitrogen impurity changes the properties of the silicon point-defects.

The straightforward case, with neighboring silicon and nitrogen
substitutions (1Si-1N), shown in Fig.~\ref{fig:geometries}(k), has a
comparably low formation energy of $3.33$ eV. Interestingly, even if
neither 1Si or 1N is magnetic individually, the 1Si-1N defect has a
finite total spin moment of $1.00 \mu_B$. Table
\ref{table:defect_infosheet} also shows the atom-projected Hirshfeld
spin moments, indicating that almost half of the total spin density
resides near the silicon atom. This is explained by that the odd
electron originating from the nitrogen atom cannot pair and forms a
spin-polarized impurity state centered around the silicon atom. The
nitrogen doping does not distort the atomic structure especially much
compared to the 1Si case, suggesting that the silicon still hybridizes
as sp$^3$. In fact, as the nitrogen atom is able to form sp$^2$ bonds,
the 1Si-1N defect is comparable to the 1Si defect with the distinction
of the nitrogen doping. Thus, in a sense, the 1N defect dopes the 1Si
defect and creates a magnetic moment. However, the difference in the
total energies of the spin-polarized and spin-unpolarized solutions is
rather small, $72$ meV in the $11 \times 11$ supercell, meaning that one
has to reach low temperatures not to mix these states.

The nitrogen doped 2Si defect, namely 2Si-1N shown in Fig.
\ref{fig:geometries}(l), does not have a finite spin moment. However,
its formation energy is extremely low, $3.22$ eV, which is $0.11$ eV
lower than that of the 1Si-1N defect. Furthermore, the 2Si-1N defect is
relaxed perfectly in-plane, unlike the 2Si defect. Both the 1Si-1N and
2Si-1N defects have been observed in experiments \cite{Zhou_2012,
Zhou_2012_2}, reflecting the fact that their formation energies are low
in comparison to any competing defects. In addition, Zhou et
al.\cite{Zhou_2012_2} evaluated the energies required to remove either
the silicon or the nitrogen atom from the 2Si-1N configuration,
obtaining the high values of $7.45$ eV and $7.01$ eV, respectively,
which further highlights the stability of these defects.

Besides the 1Si-1N defect, some of the nitrogen-doped defects are
magnetic, such as the bridged nitrogen adatoms on graphene (N-bridge,
not shown) and on 2Si (N-bridge @ 2Si, not shown), and even a silicon
bridge on 1N (Si-bridge @ 1N), shown in Fig.~\ref{fig:geometries}(n).
Curiously, a nitrogen bridge on 1Si (N-bridge @ 1Si), shown in Fig.
\ref{fig:geometries}(m), is not magnetic even if a nitrogen bridge on
graphene is. However, all their formation energies are higher than those
of 1Si-1N and 2Si-1N, but still lower than having the adatoms moved on
graphene. In this sense, nitrogen and silicon adatoms are attracted to
1Si, 2Si and 1N impurities, and then they are possibly relaxed to 1Si-1N
and 2Si-1N defects.

We further tested both the 1Si-1N and 2Si-1N defects  by separating the
silicon and nitrogen atoms from the nearest-neighbor sites. In practice,
the nitrogen atom was swapped with a carbon atom farther away in the lattice,
after which the geometry was fully relaxed. The resulting total energies
are shown in Fig.~\ref{fig:total_energies_of_Si_N_separation} (top) as a
function of the distance between the silicon and nitrogen atoms. For
both 1Si-1N and 2Si-1N, separating the silicon and nitrogen atoms costs
more than $1$ eV in energy. Intuitively, the 1Si and 1N defects being in the
nearest-neighbor sites causes the least amount of distortion in the
honeycomb lattice and there is only minimal associated energy cost. This
also validates the nearest-neighbor configurations 1Si-1N and 2Si-1N as
the most stable, and therefore the most important to study. In the large
Si-N distance limit, the energies in Fig.
\ref{fig:total_energies_of_Si_N_separation} approach the formation
energies of separated 1Si and 1N, and 2Si and 1N defects, namely the
values $E_{\text{f}}[\text{1Si-1N}]+1.35$ eV and
$E_{\text{f}}[\text{2Si-1N}]+2.26$ eV, respectively. In this sense, 1Si
and 2Si defects separated by more than about $3 \text{\AA}$ from a 1N
defect can already be considered almost fully separated in terms of the
energetics.

The total and atom-projected Hirshfeld spin moments for the defect
configurations with the separated silicon and nitrogen defects are shown
in Fig.~\ref{fig:total_energies_of_Si_N_separation} (bottom). The
systems with  1Si and 1N defects are magnetic practically only if the
silicon and nitrogen atoms reside as the nearest- or third-nearest
neighbors. On the other hand, the 2Si and 1N defect configurations
obtain finite spin moments only if the silicon and nitrogen atoms are
farther than about $4 \text{\AA}$ away from each other. However, as the
separation distance grows, it is expected that the system becomes
spin-unpolarized, as in the case of individual 2Si and 1N defects.

Since the 1Si-1N defects are magnetic, it is meaningful to ask how the
defect spin moments interact and align in a system with many such
defects. To study this, we put two 1Si-1N defects in a $11 \times 11$
supercell such that the defects formed a honeycomb superlattice of their
own. Based on calculations using the default \textit{light} basis for
each atom, due to the large number of possible configurations, it turned
out that the 1Si-1N defects prefer antiferromagnetic ordering,
regardless of which sublattices the impurity atoms reside on, and whether the
defect out-of-plane positions are in the same direction or not. The
antiferromagnetic ordering is consistently tens of meV lower in energy
compared to the ferromagnetic or spin-unpolarized solutions. The energy
differences are rather small, and the spin moments can be assumed to
interact only weakly especially in low defect concentrations.

\subsection{Hydrogen adatoms on silicon impurities}

Interestingly, hydrogen adatoms located on top of silicon and
silicon-nitrogen impurities have remarkably low formation energies, see
Table \ref{table:defect_infosheet}. Hydrogen adatoms on graphene
(H-top), shown in Fig.~\ref{fig:geometries}(t), have an energy penalty of $1.45$ eV per
hydrogen atom, when using a H$_2$ molecule as a reference for the
chemical potential $\mu_H$ in Eq. \eqref{eq:def:formation_energy}.
However, the formation energies of silicon impurities actually decrease
when a hydrogen adatom is added on top the silicon impurity. The
decrease in energy is the most notable for the 1Si and 1Si-1N defects,
being $0.56$ eV and $1.31$ eV, respectively. The 1N-doping plays a
significant role by lowering the energy even further, even though the
hydrogen adatom is clearly located on top of the silicon atom, see Fig.
\ref{fig:geometries}(p) and (s). The low formation energies overall hint that
the silicon-nitrogen defects could be reactive not only to hydrogen but
also to various molecules as well. This is the prerequisite for using the
defects as sensors.

A hydrogen adatom on graphene has a total spin-moment of $1.0 \mu_B$. This
follows from the fact that the hydrogen atom bonds to a carbon atom,
passivating the carbon $p_z$ orbital and effectively resulting in an
empty site as seen from the point of view of the $\pi$ electron bands. By
Lieb's theorem\cite{Lieb_1989}, a finite spin moment will result from
the sublattice imbalance. Quite surprisingly, magnetization does not
occur when the hydrogen adatom is located on top of a 1Si defect, or a
1Si-1N defect. In fact, all the silicon and silicon-nitrogen impurities
with a hydrogen adatom considered here, see Table
\ref{table:defect_infosheet}, are nonmagnetic. Therefore hydrogen
adatoms, which readily trap at the silicon and silicon-nitrogen
impurities, passivate any defect-induced magnetism.

\subsection{Electronic band structures}

\begin{figure}[tb!]
	\includegraphics[width=0.45\linewidth]{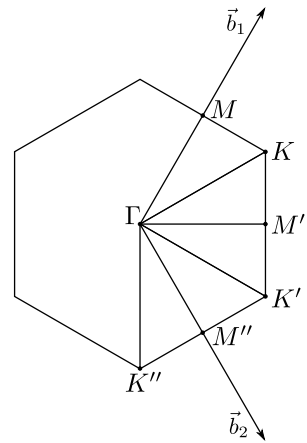}
	\caption{The first Brillouin zone of the computational supercells. The reciprocal primitive vectors are denoted as $\vec{b}_1$ and $\vec{b}_2$. When evaluating the band structure, we focus on the triangular paths between nearby high symmetry points, namely $\Gamma M K \Gamma$, $\Gamma M^{\prime} K^{\prime} \Gamma$ and $\Gamma M^{\prime \prime} K^{\prime \prime} \Gamma$. \label{fig:brillouin}}
\end{figure}

\begin{figure*}[tb!]
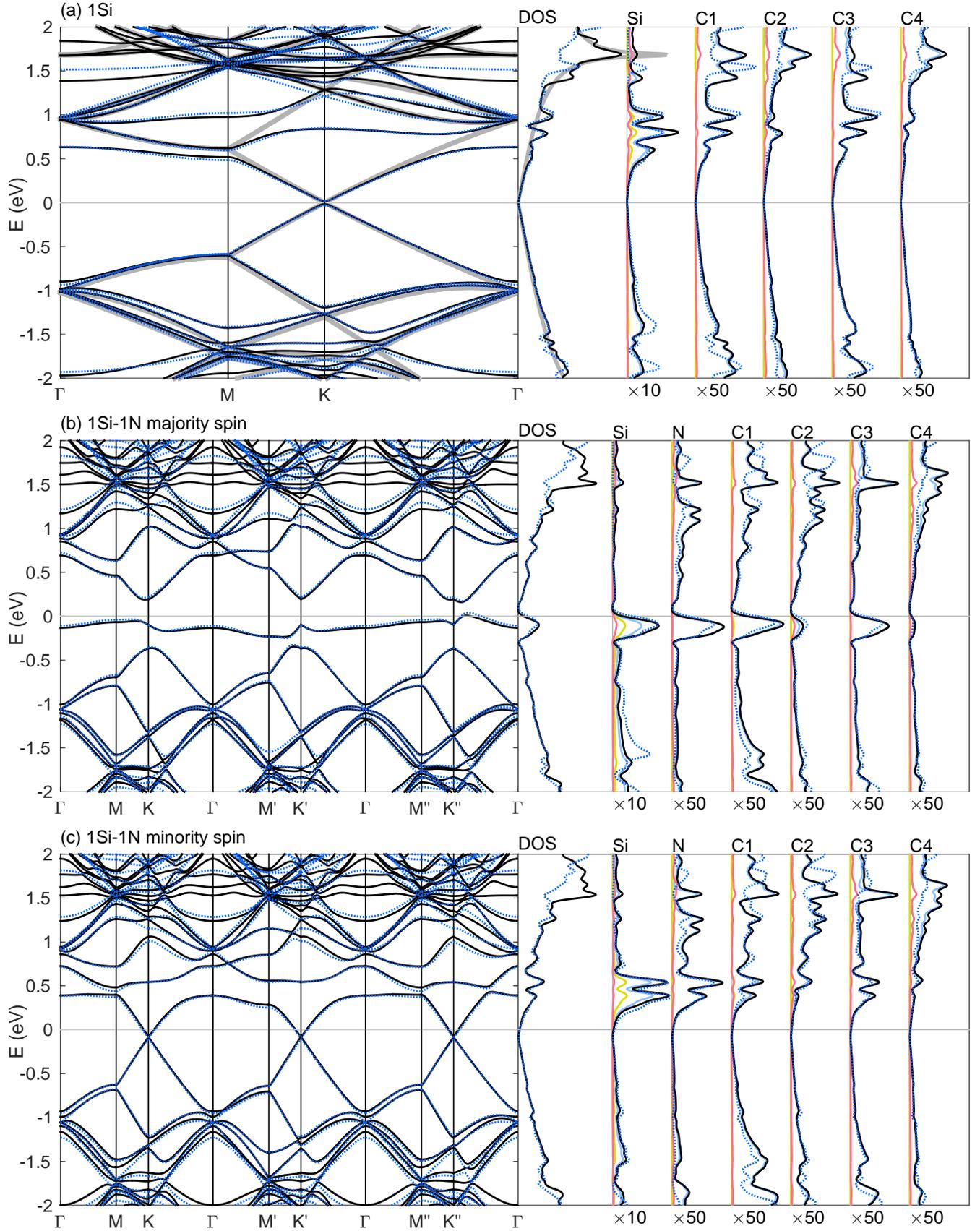

	\includegraphics[width=0.99\linewidth]{4a_1Si_bands_and_pDOS}
	\includegraphics[width=0.99\linewidth]{4b_1Si_1N_spin_up_bands_and_pDOS}
	\includegraphics[width=0.99\linewidth]{4c_1Si_1N_spin_dn_bands_and_pDOS}
	\caption{(Color online). Band structures, DOS and atom-projected pDOS of $8 \times 8$ supercells with (a) a 1Si defect, and with a spin-polarized 1Si-1N defect showing the (b) majority and (c) minority spin components separately. The DFT results are the solid lines, and the fitted tight-binding model results are the dotted lines. The pristine graphene band structure and DOS (thick grey lines) are shown in (a). The pDOS curves are the total atom-projected pDOS (black), the s-type pDOS (yellow), the p-type pDOS (light blue), and the d-type pDOS (red). The DOS and pDOS have been broadened using Gaussian broadening of $30$ meV. Zero energy is at the Fermi energy, the charge neutrality point. The atom labels match those in Figs. \ref{fig:geometries}(a) and (k). \label{fig:band_structures_1Si}}
\end{figure*}

\begin{figure*}[tb!]
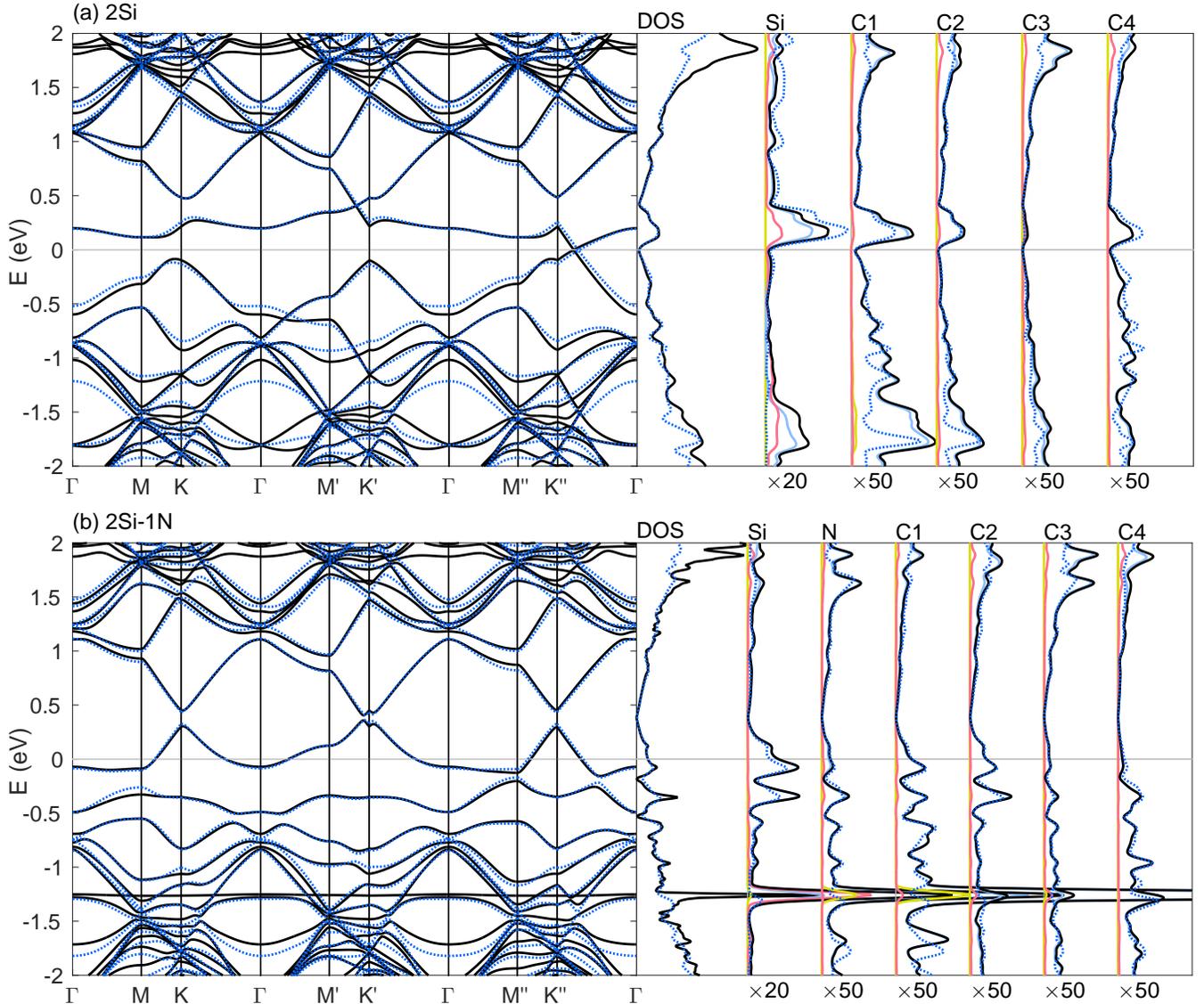

	\includegraphics[width=0.99\linewidth]{5a_2Si_bands_and_pDOS}
	\includegraphics[width=0.99\linewidth]{5b_2Si_1N_bands_and_pDOS}
	\caption{(Color online). Band structures, DOS and atom-projected pDOS for (a) the 2Si and (b) the 2Si-1N defects. The DFT results are the solid lines, and the fitted tight-binding model results are the dotted lines. The pDOS plots show the total atom-projected pDOS (black), the s-type pDOS (yellow), the p-type pDOS (light blue), and the d-type pDOS (red). The $8 \times 8$ supercell has been used, and DOS and pDOS have Gaussian broadening of $30$ meV. Zero energy is the Fermi energy, the charge neutrality point. The atom labels match those in Fig.~\ref{fig:geometries}(b) and (l).\label{fig:band_structures_2Si}}
\end{figure*}

In the following, we studied the DFT band structures and DOS of systems
with $8 \times 8$ unit cells of graphene that contain a single 1Si, 2Si,
1Si-1N, or 2Si-1N defect. This corresponds to an infinite, periodic
array of the impurities. The band energies are evaluated in the
reciprocal space along the line segments between the nearby high
symmetry points in the first Brillouin zone (BZ), as illustrated in Fig.
\ref{fig:brillouin}. From the outset, we restricted the study only
to one of the halves of the first BZ, since time reversal symmetry is
not broken, and therefore inversion around the $\Gamma$ point, namely
$\mathbf{k} \mapsto -\mathbf{k}$, does not alter state energies. This
also holds for the majority and minority spin channels in the
spin-polarized case.

The band structure, DOS and atom-projected partial DOS (pDOS) of the
system with a 1Si defect are shown in Fig.
\ref{fig:band_structures_1Si}(a). The band structure closely matches the
pristine graphene band structure (thick grey line). In fact, one sees in
the presented energy range
the pristine graphene occupied and unoccupied
$\pi$ electron bands that have been folded multiple times in order to
match the $8 \times 8$ supercell as a periodic unit. A supercell with a defect
cannot be unfolded back into smaller periodic unit cells, as
there are energy gaps between the bands at the high symmetry points.
Curiously, with supercells of sizes $p \times p$ where $p$ is divisible
by three, the $K$ and $K'$ points are actually mapped onto the $\Gamma$
point, and there would be no energy gap at $\Gamma$ \cite{Lambin_2012,
Martinazzo_2010}. However, in our case with $p=8$, there is a small
energy gap at the $K$ point in the 1Si band structure, but otherwise the
Dirac cone seems to be unperturbed.

Consequently, the 1Si DOS is similar to the graphene DOS, indicated by
the thick grey line in Fig.~\ref{fig:band_structures_1Si}(a), except
that there are additional states at energies around $0.8$ eV. The
opening of energy gaps in the band structure results in double peaks in
the DOS if the peak broadening is sufficiently small. But most
importantly, the peaks in the DOS are in general explained by the
flatter energy bands, for instance resulting from localized impurity
states that are only weakly $\mathbf{k}$-dependent. However,
substituting a carbon atom by a valence isoelectronic silicon atom does
not introduce a new impurity band, but rather the $\pi$ electron bands
become perturbed. Peaks are then formed as there are stronger resonances
at certain energies that are also expected to depend on the dimensions
of the periodic array of defects. The pDOS around the 1Si defect, in
Fig.~\ref{fig:band_structures_1Si}(a), shows that the peaks in the
density at around $0.8$ eV are mostly localized at the silicon atom and
the nearest and third-nearest carbon atoms C1 and C3, see Fig.
\ref{fig:geometries}(a) for the atom labels. We can also see that carbon
atoms exhibit only p-type orbital occupations as expected, but the
silicon atom has some contribution from s- and even d-type orbitals.

The 1Si-1N band structure and DOS are shown in Fig.
\ref{fig:geometries}(b) and (c) for the majority and minority spin
components from a calculation with collinear spin. The paths in the
reciprocal space along which the band energies are shown now include the
three triangular paths $\Gamma M K \Gamma$, $\Gamma M^{\prime}
K^{\prime} \Gamma$, and $\Gamma M^{\prime \prime} K^{\prime \prime}
\Gamma$ due to breaking of the graphene lattice symmetries. There is a
spin moment of $0.98 \mu_B$ that corresponds to an almost fully occupied
band in the majority spin channel that is unoccupied in the minority
spin channel. The origin of the spin moment is the nitrogen atom that
introduces the odd electron that is attracted to the silicon atom. As a
consequence, there is an almost flat impurity band just below the Fermi
energy, and a clear bump in the DOS and atom-projected pDOS. The density
is mostly localized at the silicon atom with some weight also at nearby
atoms of the adjacent sublattice. In addition to the p-type orbital
occupations, the impurity state also exhibits some s- and d-type orbital
occupation at the silicon atom, and small d-type occupations even at the
second-nearest neighbor carbon atom C2.

For both spin components, the band energies along the three paths $\Gamma M K \Gamma$, $\Gamma M^{\prime} K^{\prime} \Gamma$ and $\Gamma M^{\prime \prime} K^{\prime \prime} \Gamma$ have only minor differences, suggesting that there is no strong directional dependence. Therefore, the electronic states at the nitrogen atom are expected to be similar to the states at the carbon atoms that are nearest-neighbours of the silicon atom. In fact, the band structure of the minority spin component, shown in Fig.~\ref{fig:geometries}(c), closely resembles the band structure of the 1Si defect. The Dirac cone at the $K$-point is intact, being only slightly shifted to the occupied side due to the nitrogen doping. However, with the larger system of $11 \times 11$ graphene unit cells, the spin moment is already $1.00 \mu_B$, and therefore the Dirac cone in the minority channel is expected at the Fermi energy. We can call such a material half-semimetal, since the majority spin channel has only an impurity band just below the Fermi energy, and the band structure of the minority spin channel is that of a semimetal. Furthermore, the main qualitative difference between the pDOS of periodic arrays of 1Si and 1Si-1N defects is that the silicon pDOS shows two peaks in the 1Si-1N minority channel, whereas the silicon pDOS at the 1Si defect has three peaks. Moreover, there is no clear occupation of d-type orbitals in the minority channel of the 1Si-1N defect system.

The band structure, DOS and pDOS of a system with a 2Si defect are shown in Fig.~\ref{fig:band_structures_2Si}(a). The breaking of the sublattice and the three-fold rotational symmetries have a profound effect on the band structure. Namely, there is a strong directional dependence on the wave vectors $\mathbf{k}$, and even the Dirac cones are slightly shifted from the $K$, $K'$ and $K''$ points. Furthermore, there is a clear bump in the DOS roughly at $0.2$ eV. This shows in the pDOS as localized density from p- and d-type orbitals at the silicon atom and as density from p-type orbitals at the nearby carbon atoms. Interestingly, there are states in the occupied side that do not have much density at the silicon atom, but they do have density at the nearby carbon atoms. Therefore there are also impurity states surrounding the silicon site.

The band structure of a system with a 2Si-1N defect, shown in \ref{fig:band_structures_2Si}(b), is necessarily doped due to the odd electron from the nitrogen atom. There is a band at the Fermi energy that is half filled, and as a result the conical dispersions are shifted roughly by $0.4$ eV to the unoccupied side. In fact, the 2Si-1N band structure looks more like doped 1Si band structure than 2Si band structure. Compared to 2Si band structure, where the bands have strong directional dependency, the valence bands in the 2Si-1N band structure are almost flat resulting in many distinct peaks in the DOS and pDOS in the defect neighborhood. These states are again not only localized impurity states at the silicon but also at the nearby carbon atoms, especially the nearest-neighbour carbon atom C1. Curiously, there is an almost perfectly flat band at $-1.26$ eV, implying a state localized at the defect, and that the related orbitals have a symmetry other than that of the $\pi$ electrons.

\section{Tight-binding results}

\subsection{Tight-binding models}

We constructed tight-binding models for the silicon and silicon-nitrogen point-defects on graphene to further study their electronic and transport properties. This way we could simulate larger systems with random defect configurations, which is beyond typical first principles approaches.

The tight-binding model with an orthogonal state per each atom site, and simple hoppings between the sites, describes the graphene $\pi$ and $\pi^*$ electron bands remarkably well \cite{Wallace_1947, Castro_Neto_2009}. The tight-binding Hamiltonian is written as
\begin{equation}
 \hat{H} = \sum_i \varepsilon_i \: \hat{c}^{\dagger}_i \hat{c}_i + \Bigg[ \sum_{i<j} t_{ij} \: \hat{c}^{\dagger}_i \hat{c}_j  + \text{H.c.} \Bigg] ,
\end{equation}
where $\hat{c}^\dagger_i$ ($\hat{c}_i$) creates (annihilates) a fermion in a state $i$, and H.c. stands for Hermitian conjugate. Here, the coefficients $­\varepsilon_i$ describe on-site potentials, and $t_{ij}$ are hoppings between sites. We denote $t_1, t_2, t_3$ as the nearest-, second-nearest-, and third-nearest neighbor hoppings, respectively.

The tight-binding models for systems containing defects were found by modifying the hoppings and on-site potentials in the vicinity of the defects, and fitting the outcome to the first principles band structures. Furthermore, we checked that the resulting tight-binding local density of states (LDOS) matched the first principles atom-projected partial DOS around the Fermi energy, at least qualitatively. We used the $8 \times 8$ supercells that have a single defect each. The band structure fitting was performed in the first Brillouin zone in a uniform grid of $50 \times 50 \times 1$ k-points. We took 12 bands into the fitting procedure, the six highest valence bands and the six lowest conduction bands, determined at the $\Gamma$ point, except for the 2Si-1N defect, where only four valence and four conduction bands were taken in order to avoid the flat band at $-1.26$ eV.

\subsection{Tight-binding parametrization}

First the carbon-carbon (C-C) hoppings $t_1, t_2, t_3$ were obtained by fitting to the pristine graphene band structure. We allowed $t_1$ and $t_2$ to be free fitting parameters to scale the unit of energy and to freely break the electron-hole symmetry by tuning $t_2$. We included the third-nearest neighbor hoppings $t_3$ by assuming that $t_3 = t_1 (0.18/2.7)$, where the factor is motivated by previous graphene tight-binding models \cite{Hancock_2010}. If $t_3$ was not constrained, altering $t_1$ and $t_3$ simultaneously would produce a continuous set of models with good fits. The resulting C-C hoppings are $(t_1, t_2, t_3) = (-2.855, -0.185, -0.190)$ eV, which were used for all carbon-carbon hoppings in our tight-binding calculations.

We expect that the properties of both the silicon and silicon-nitrogen defects can be reproduced around the Fermi energy using simple tight-binding models. Namely, silicon and carbon atoms are valence isoelectronic, and nitrogen is also able to form sp$^2$ bonds. The relevant $\pi$ and $\pi^*$ electron bands are then perturbed by the defects in the effective tight-binding description. Moreover, we chose a minimal set of tight-binding parameters to enhance the physical significance of each parameter. We took only the nearest-neighbor hoppings between the Si and C atoms, and between the Si and N atoms, to be nonzero. The hoppings between C and N atoms were taken identical to those of the C-C hoppings, which is similar to the nitrogen substitution model in Ref. \onlinecite{Lambin_2012}. The C on-site potentials were set to zero, whereas the Si and N on-site potentials could take non-zero values. 

\begin{table}
\caption{Tight-binding parameters for the 1Si, 1Si-1N, 2Si and 2Si-1N defects. The C-N hoppings are taken as the same as the C-C hoppings, and the C on-site potentials are assumed as zero. The values are in units of eV.  \label{table:tight_binding_infosheet}}
\begin{tabular}{ l c  r r  r r }
\hline
Defect & Spin polarization & \multicolumn{2}{c}{Hopping $t_1$:}                  & \multicolumn{2}{c}{Potential $\varepsilon$:} \\
       &               & \multicolumn{1}{c}{C-Si} & \multicolumn{1}{c}{Si-N} & \multicolumn{1}{c}{Si} & \multicolumn{1}{c}{N}  \\
\hline
1Si      & no       & $-1.123$ &           & $0.118$  &          \\
1Si-1N   & majority & $-0.776$ & $-0.919$  & $-0.539$ & $-4.874$ \\
         & minority & $-0.921$ & $-1.483$  & $0.165$  & $-5.182$ \\
         & no       & $-0.819$ & $-1.419$  & $-0.351$ & $-5.393$ \\
2Si      & no       & $-0.849$ &           & $2.087$  &          \\
2Si-1N   & no       & $-1.627$ & $-0.894$  & $2.722$  & $-2.833$ \\
\hline
\end{tabular}
\end{table}

With the parameter constraints and fitting method described above, the
resulting tight-binding parameters for the 1Si, 1Si-1N, 2Si, and 2Si-1N
defects are listed in Table \ref{table:tight_binding_infosheet}. We
provide spin-unpolarized and spin-polarized parametrization of the
1Si-1N defect with distinct models for the majority and minority spin
components. Comparing to first principles, the tight-binding
models produced remarkably accurate band structures, DOS, and LDOS, as
shown by the dotted lines in Figs. \ref{fig:band_structures_1Si} and
\ref{fig:band_structures_2Si}. The tight-binding Fermi energies are at
$0.537$ eV, $0.593$ eV, $0.392$ eV, and $0.274$ eV for the systems with
a 1Si, 1Si-1N (spin-polarized), 2Si, and 2Si-1N defect, respectively.
Below, we use these values as $E=0$ point, even if the defect concentration
changes.

The most visible signature of the 1Si defect is the opening of the
energy gaps between the lowest and second lowest conduction bands at the
$\Gamma$ point, as well as between the second and third lowest
conduction bands at the $M$ and $K$ points. This can be explained by the
weaker Si-C hopping compared to the C-C hopping value. An on-site
potential at the silicon atom also breaks the translational symmetry,
resulting in small energy gaps at many band crossings. In the case of
the 2Si defect, the C-Si hopping is even weaker, and we set a repulsive
on-site potential at the silicon atom, since the sp$^2$d hybridization
takes more electrons to form bonds with the four neighboring carbon
atoms. For the defects containing nitrogen, we used a rather low on-site
potential at the nitrogen sites, because a nitrogen substitution dopes
its surroundings by sharing some of its electron density. For the 2Si-1N
defect, we did not include the state corresponding to the extremely flat
band in the model.

Curiously, a 1Si-1N defect is spin-polarized. We obtained the tight-binding Hamiltonians for both the majority, $\hat{H}_1$, and the minority, $\hat{H}_2$, spin components individually. To explicitly include the electron spin in the Hamiltonian, we can write
\begin{equation}
\hat{H} = \hat{H}_1 \otimes \ketbra{\sigma_1}{\sigma_1} + \hat{H}_2 \otimes \ketbra{\sigma_2}{\sigma_2} = \bar{H} \otimes \Id + \hat{\Delta} \otimes \hat{\sigma}_{\hat{z}_1},
\end{equation}
where the electron spin is first projected onto the defect majority and minority spin states, $\ket{\sigma_1}$ and $\ket{\sigma_2}$, and we have defined $\bar{H}=(\hat{H}_1 + \hat{H}_2)/2$, $\hat{\Delta}=(\hat{H}_1-\hat{H}_2)/2$, and $\hat{\sigma}_{\hat{z}_1} = \ketbra{\sigma_1}{\sigma_1} - \ketbra{\sigma_2}{\sigma_2}$. Due to the last term in the Hamiltonian, the electron spin can flip, if it is not aligned with the defect spin moment. The strength of the spin flipping is proportional to the difference of the majority and minority tight-binding parameters of $\hat{\Delta}$. In our case, the tight-binding models for the two spin components are not necessarily consistent by e.g. assuming a mean-field Hubbard model, see Ref.~[\onlinecite{Soriano_2011}], where only the on-site potentials can be spin-dependent. However, the obtained majority, minority and spinless tight-binding models for the 1Si-1N defect are reasonable in a sense that the parameters are qualitatively similar, and the values of the spinless parameters are somewhere between the values of the majority and minority parameters. Furthermore, it seems that at least the C-Si hopping needs to be slightly spin-dependent in order to obtain a qualitatively good fit of the band structures.

\subsection{Randomly positioned defects}

The tight-binding models constructed above can be used to study large
systems containing millions of atoms, much larger than what can be
treated by state-of-the-art first principles simulations. The large
system sizes are needed to eliminate finite-size effects. Below, we
study the electronic and transport properties of supercells with $2000
\times 1000$ atoms in total and with randomly placed defects at a defect
concentration of $0.5\%$. Moreover, we take ensemble averages over
several defect configurations to ensure that the results are converged
in the large system limit. It is, however, sufficient to average over
only roughly five defect configuration realizations, as the systems considered here
are large enough for much of the random fluctuations to average out.

\begin{figure}[tb]
\includegraphics[width=0.98\linewidth]{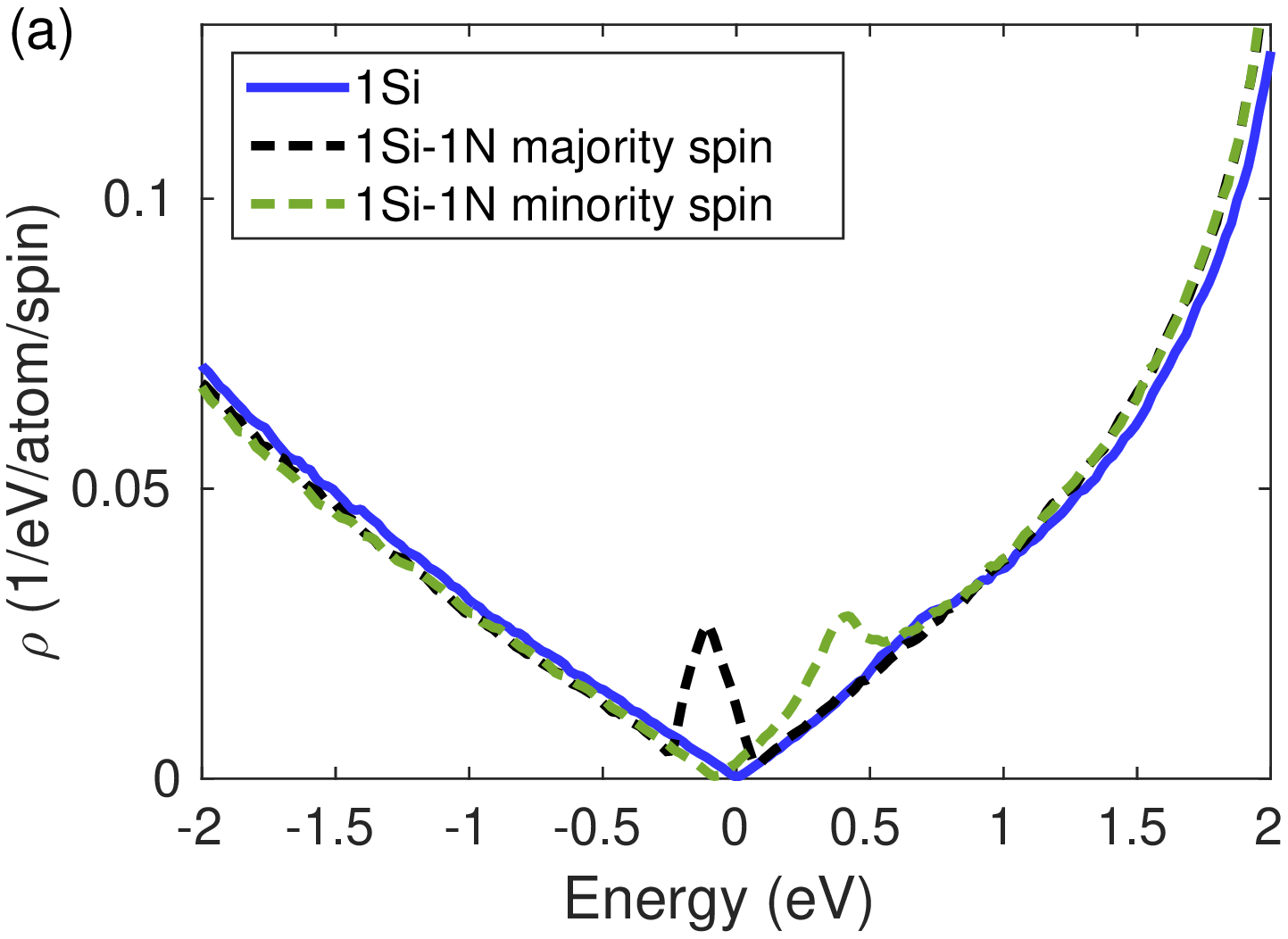}
\includegraphics[width=0.98\linewidth]{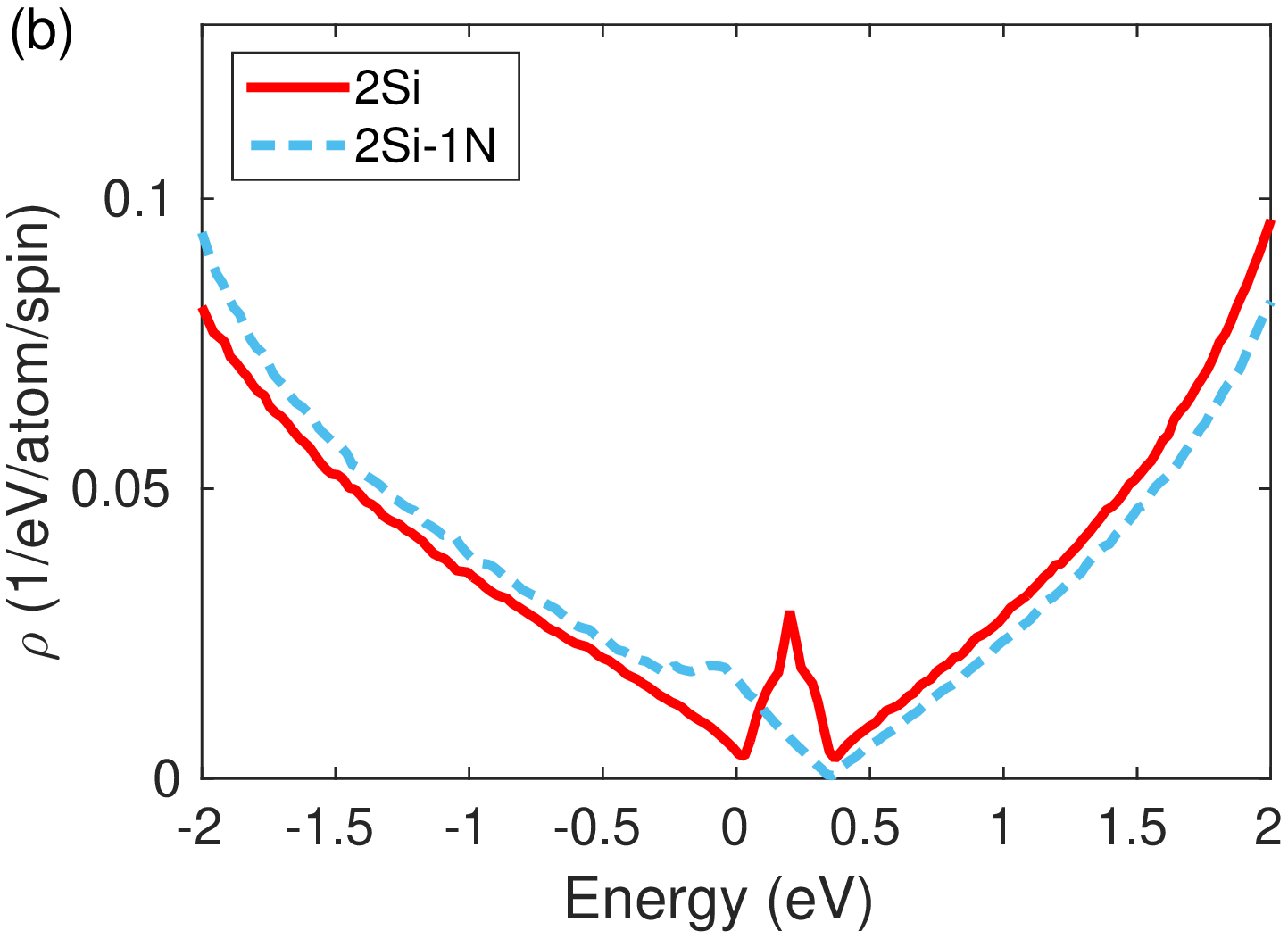}
\caption{(Color online).  Tight-binding density of states for a $2000\times 1000$ atom system with $0.5\%$ random defect concentration.\label{fig:DOS}}
\end{figure}

The DOS of such large systems with randomly placed 1Si, 1Si-1N, 2Si and
2Si-1N defects are shown in Fig.~\ref{fig:DOS}. There is a strong
similarity to the DOS of a periodic supercell with one defect presented in Figs.
\ref{fig:band_structures_1Si} and \ref{fig:band_structures_2Si}.
However, averaging over random defect configurations smoothens the
individual peaks of the periodic case, and one obtains the
characteristic features of the DOS. Still, of course, the band structure
of the periodic supercell and the DOS of a large system are closely
related: whenever there are relatively flat impurity bands, there are
impurity states at the corresponding energies, manifesting themselves as
humps in the DOS. 

It is surprising that the DOS of a system containing $0.5\%$ of 1Si
defects shows only a very weak deviation from the DOS of pristine
graphene, whereas both the majority and minority spin channels of the
1Si-1N defect exhibit clear peaks. Namely, the resonant impurity states
of the majority (minority) spin reside on the occupied (unoccupied) side, at negative (positive) energies. Similarly, the resonant impurity states of the 2Si defects
correspond to positive energies. The 2Si-1N defect DOS in
Fig.~\ref{fig:DOS}(b) is again only weakly perturbed from the pristine
graphene DOS, but it is clearly doped by the nitrogen substitutions. It
seems that the spiky DOS of the periodic defect case, as shown in
Fig.~\ref{fig:band_structures_2Si}(b), smoothens to a slight elevation
around $E=0$ eV.

\subsection{Electronic transport}

The tight-binding models also enabled us to perform transport
calculations with large system sizes. We used the real-space
Kubo-Greenwood method (RSKG) \cite{Mayou_1988, Mayou_1995, Roche_1997,
Triozon_2002} which scales linearly with respect to the total number of
atoms and allows for the simulation of truly two-dimensional graphene
with many randomly distributed defects. Our implementation is
significantly accelerated by using graphics processing units (GPUs)
\cite{Fan_2014}.

The RSKG method can be used to efficiently compute the intrinsic
conductivity $\sigma(E, t)$ as a function of the Fermi energy $E$ and
correlation time $t$. In most cases, the maximum of $\sigma(E, t)$ over
$t$ represents a good approximation of the semiclassical conductivity
$\sigma_{\text{sc}}(E)$. The semiclassical conductivity is the
conductivity of a system where quantum-mechanical interference phenomena
leading to weak and strong localization are suppressed. The elastic mean
free path $l_{\text{e}}(E)$ can be then calculated using the Einstein
relation for diffusive transport
\begin{equation}
\label{equation:sigma}
\sigma_{\text{sc}}(E)=\frac{1}{2}e^2 \rho(E)v(E)l_{\text{e}}(E),
\end{equation}
where $v(E)$ is the Fermi velocity and $\rho(E)$ is the total DOS.

In systems with some given defect type, both the semiclassical conductivity and the elastic mean free path depend on the defect concentration $n$. Here, $n$ was set to match the silicon concentration, and it is fixed to $0.5\%$. The defects were positioned and oriented randomly in the graphene lattice, as would be expected in real materials.

\begin{figure*}[ht!]
\includegraphics[width=0.3149\linewidth]{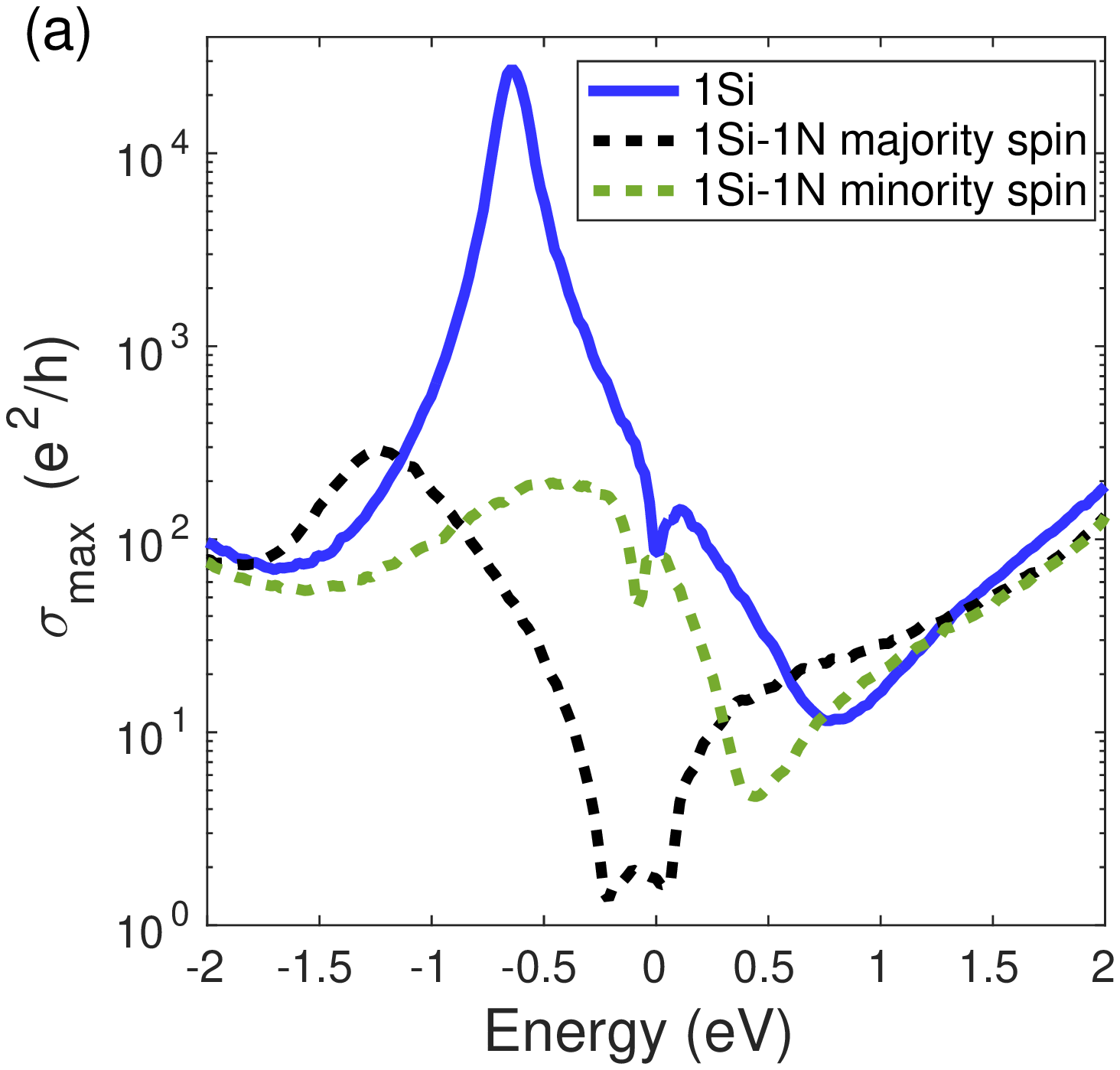}
\includegraphics[width=0.3149\linewidth]{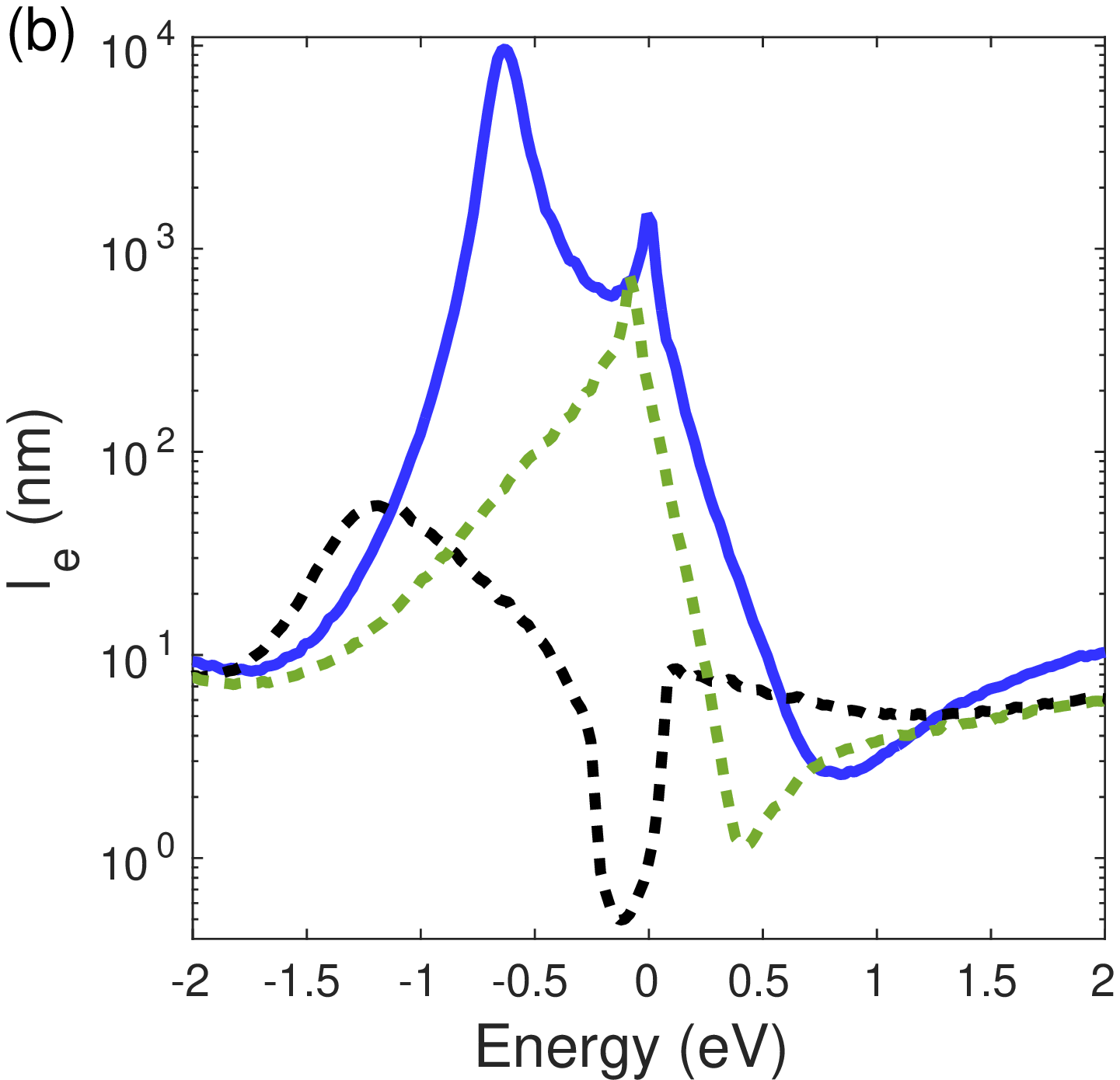}
\includegraphics[width=0.3149\linewidth]{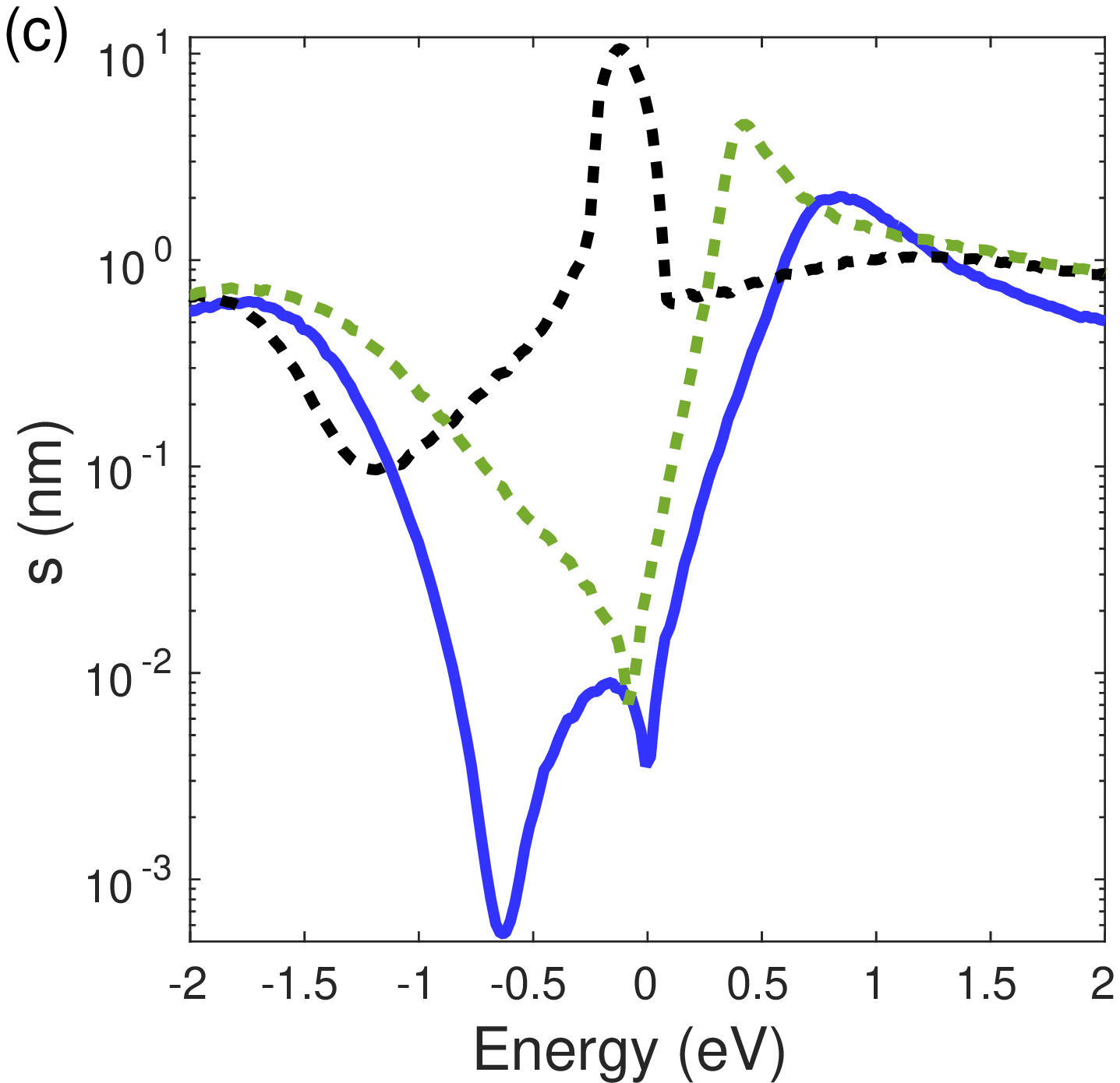}
\includegraphics[width=0.3149\linewidth]{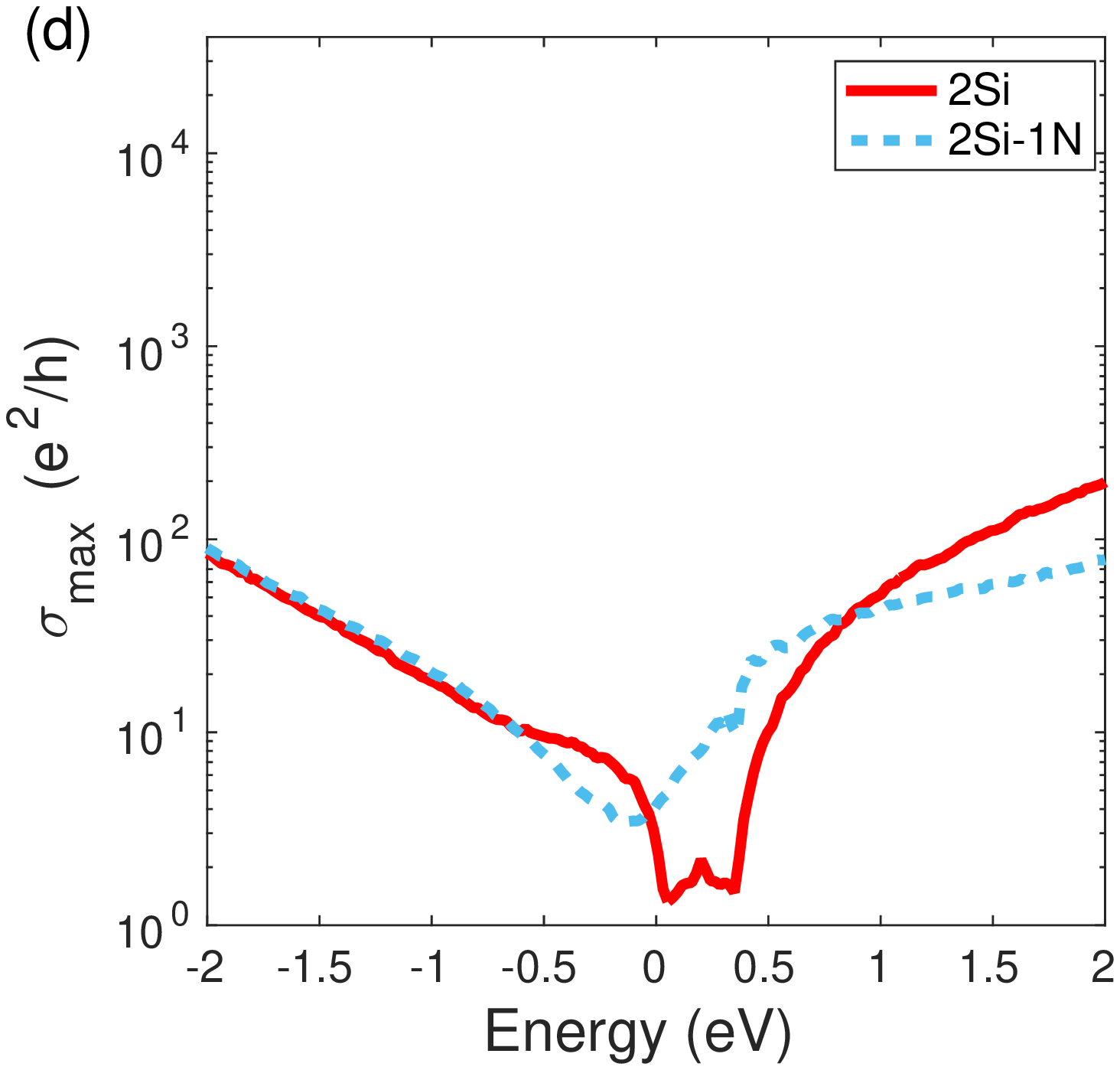}
\includegraphics[width=0.3149\linewidth]{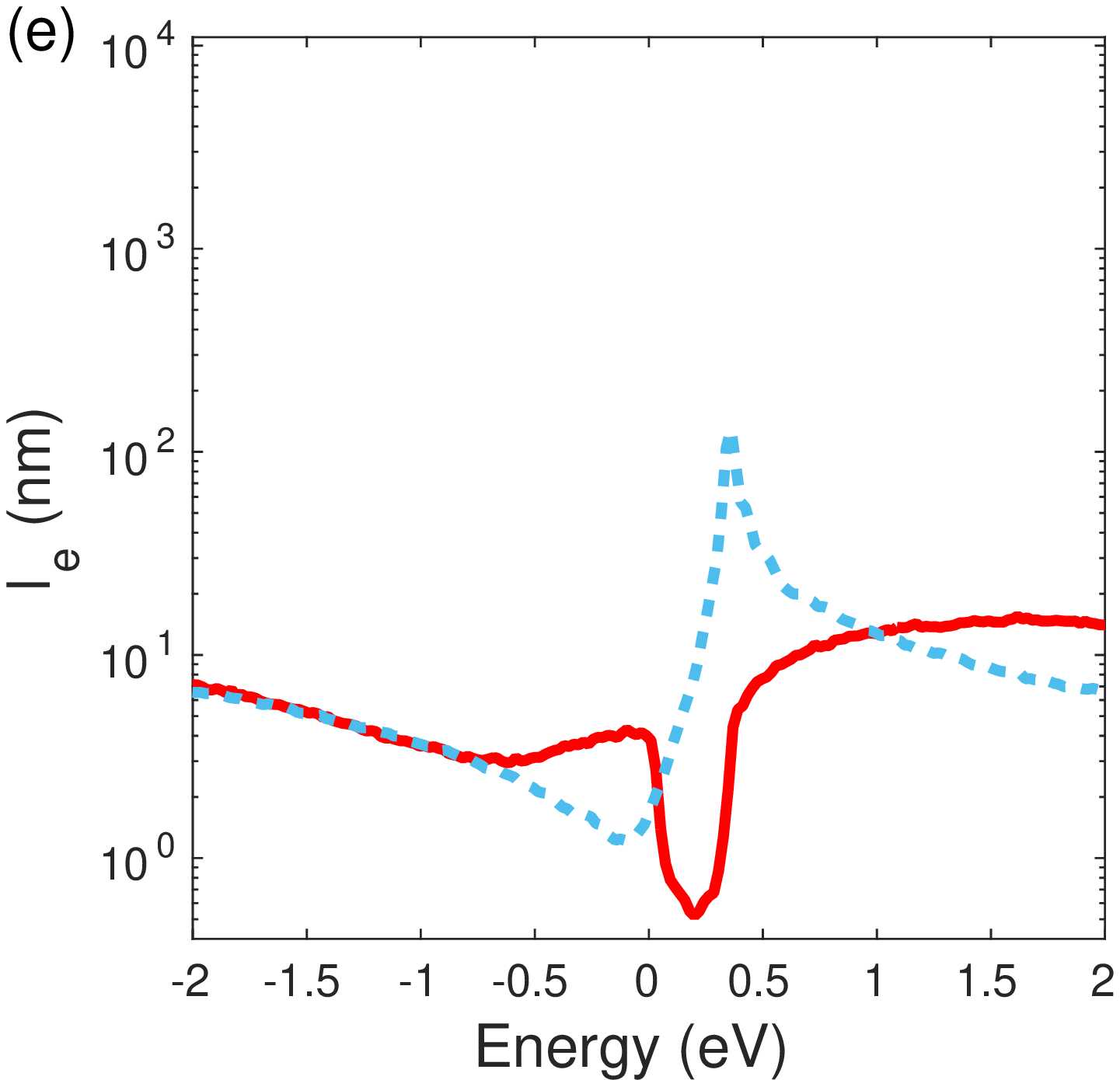}
\includegraphics[width=0.3149\linewidth]{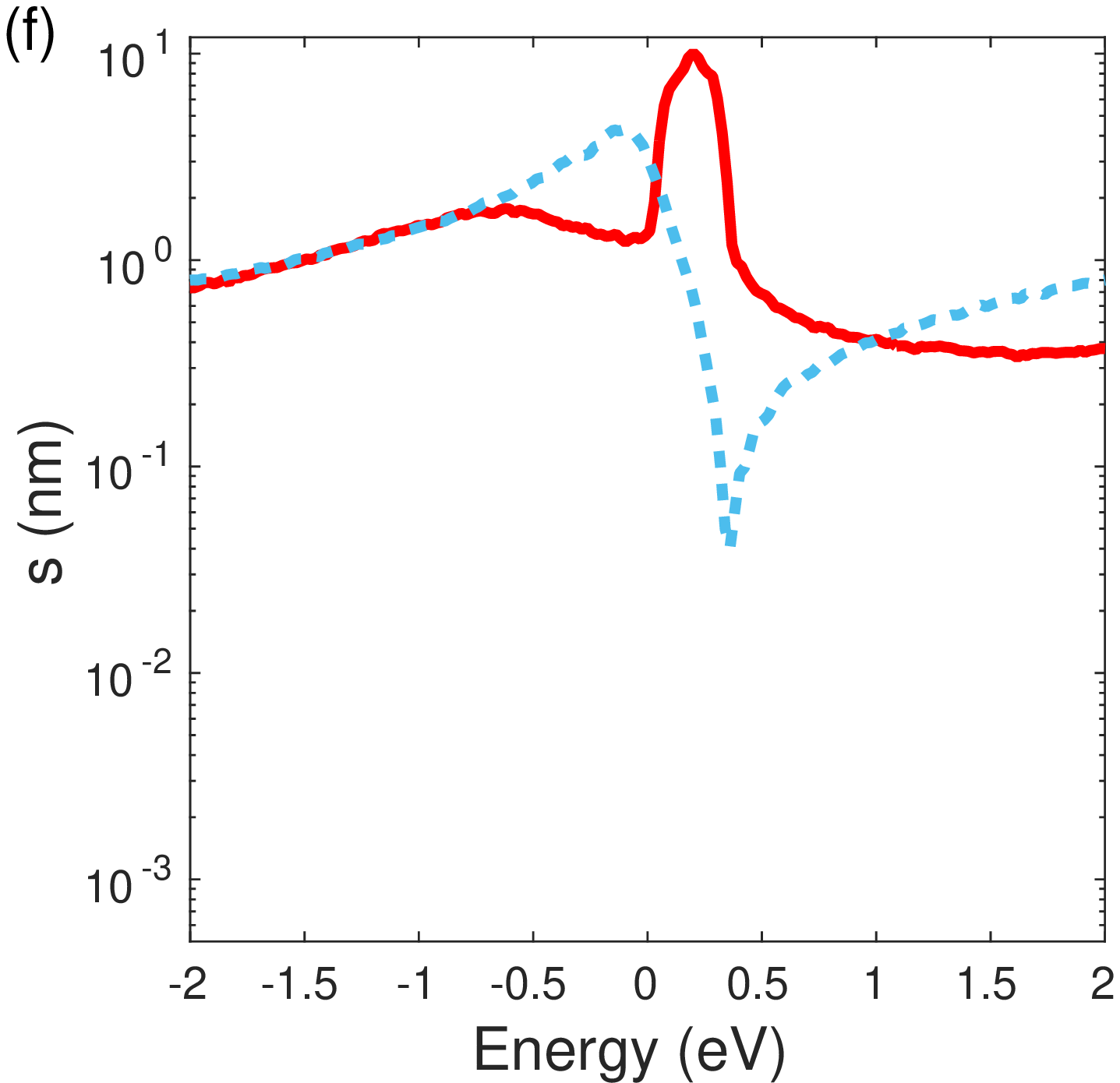}
\caption{(Color online). Electronic transport results for the 1Si, 1Si-1N with ferromagnetic ordering, 2Si and 2Si-1N defects at concentration of $0.5\%$. (a) and (d) Semiclassical conductivity. (b) and (e) Elastic mean free path. (c) and (f) Scattering cross section. \label{fig:trans}}
\end{figure*}

The defects act as resonant scatterers\cite{Ni_2010} that limit the charge carrier conduction around the resonant energies, resulting in local minima in the semiclassical conductivity. The semiclassical conductivities for systems with many randomly positioned 1Si and 1Si-1N defects are presented in Fig.~\ref{fig:trans}(a) and with many randomly positioned 2Si and 2Si-1N defects in Fig.~\ref{fig:trans}(d). The system with 1Si defects is again closest to pure graphene, showing a weak minimum only around $E=0.8$ eV. Furthermore, the peak at  $E=-0.64$ eV should be even more singular than Fig.~\ref{fig:trans}(a) shows, because in our simulation the intrinsic conductivity was still increasing at the largest correlation times considered. For the spin-polarized 1Si-1N defects, Fig.~\ref{fig:trans}(a) shows the semiclassical conductivity separately for the majority and minority tight-binding models. Using the same model to describe all defects corresponds in this case to the ferromagnetic ordering of the defect spin moments that would point along the direction of an external field. Both spin channels have smaller semiclassical conductivities at energies that match the peaks in their respective DOS. Then, similarly to the DOS, the conductivities of the spin channels are completely different around zero energy, suggesting that ferromagnetically ordered 1Si-1N defects would lead to distinct transport properties that depend on the spin states of the charge carriers. The 2Si and 2Si-1N defects exhibit clearly smaller semiclassical conductivities than the 1Si and 1Si-1N defects, and the minima in conductivity are again found at energies that match the peaks in the DOS.

From the semiclassical conductivity, we could calculate the elastic mean free path $l_e (E)$, shown in Fig.~\ref{fig:trans}(b) and (e). For the 1Si case, the smallest values of $l_e$ are around a few nanometers while the system with 1Si-1N defects exhibits values that are even smaller than 1 nm for a single spin channel. Thus the 1Si-1N defect is a strong scatterer of the charge carriers. Interestingly, the difference between the spin channels is very large around zero energy, being around three orders of magnitude. The systems with 1Si defects, on the other hand, exhibit elastic mean free paths that exceed one micrometer, which is a surprising result for such a high defect concentration of $0.5\%$. The elastic mean free paths in systems with 2Si and 2Si-1N defects are rather short in the energy window shown in Fig.~\ref{fig:trans}(e), as expected based on the low conductivities.

To quantify the transport properties with any small value of defect concentration \cite{Uppstu_2012}, we calculated the scattering cross section. After $l_{\text{e}}(E)$ is obtained for a given concentration $n$, the effective scattering cross section with dimension of [length] in two dimensions is evaluated as \cite{Uppstu_2012}
\begin{equation}
\label{equation:s}
s(E) = \frac{1}{l_{\text{e}}(E) n_{\text{d}}},
\end{equation}
where $n_{\text{d}}=\frac{4n}{3\sqrt{3}a^2}$ is the 2D number density of the defects, and $a$ is the carbon-carbon distance. For relatively small $n$, which is often the case in real experiments, $s(E)$ is largely independent of $n$.

The scattering cross sections $s(E)$ are shown in Fig.~\ref{fig:trans}(c) and (f). The large values of the scattering cross section for one spin channel in systems with 1Si-1N defects demonstrate further that the scattering by this defect type is highly spin-dependent. In general, the scattering cross section shows that the silicon and silicon-nitrogen defects scatter charge carriers in clearly distinct ways. The 1Si defects cause barely any scattering below $E=0$ eV and scatter moderately at around $E=0.84$ eV, whereas the 1Si-1N defects scatter the majority spin charge carriers very strongly at $E=0$ eV and the minority spin charge carriers moderately at $E=0.42$ eV. In the 2Si case, scattering is very strong between $E=0$ eV and $E=0.4$ eV where, in turn, the 2Si-1N defects cause only little scattering.

\begin{figure}[tb]
\includegraphics[width=0.99\linewidth]{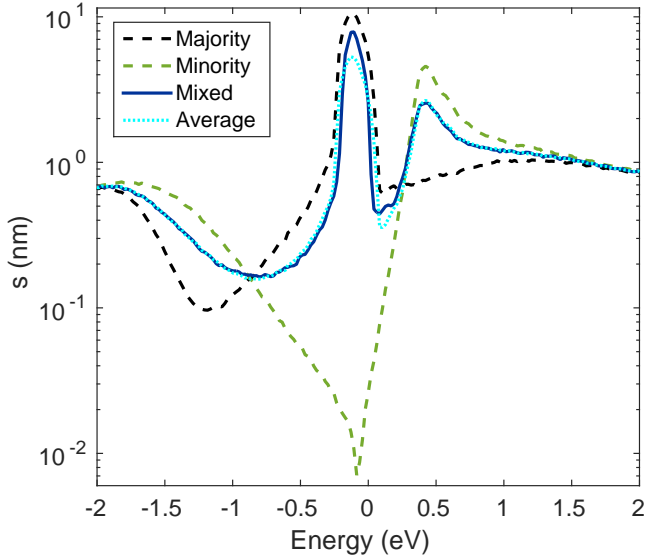}
\caption{(Color online). The scattering cross section for the spin-polarized 1Si-1N defect. The results labelled as \textit{majority} and \textit{minority} use only the corresponding tight-binding model at a defect concentration of $0.5\%$. The result labelled as \textit{mixed} has the defects randomly chosen to be described by either the majority or the minority tight-binding models. The \textit{average} is the mean of the \textit{majority} and \textit{minority} scattering cross sections, according to Mathiessen's rule.
\label{fig:mixed_transport}}
\end{figure}

So far we have shown the transport results for ferromagnetically ordered 1Si-1N defects, where all the impurity spin moments point in the same direction. In this case, the charge carrier spin basis can be chosen to align with the impurity spins, and the majority and minority spin components can be solved independently. On the other hand, when the external field is weak, the 1Si-1N defects prefer antiferromagnetic ordering.  However, the interaction energies between defect spin moments are rather small, roughly tens of meV, and the temperatures in experiments might not be low enough to manifest clear magnetic ordering, leading to random orientations of the impurity spin moments. The charge carrier spin basis cannot be simultaneously aligned with many random impurity spin moments, which can lead to processes where the spin of the charge carrier flips in a scattering event. These might be relevant for spin relaxation in graphene \cite{Kochan_2014}.

Our RSKG implementation does not allow impurity spins to point in random directions, but we were able to simulate the case where the defects are positioned randomly in the graphene lattice, and each impurity spin is randomly taken to point either up or down. This corresponds to assigning the majority or the minority model for each defect randomly. In a sense, this can be called antiferromagnetic ordering of the defects, even though antiferromagnetism is not enforced locally. The scattering cross section in this case with mixed 1Si-1N majority and minority defects, labelled as ``mixed'', is shown in Fig.~\ref{fig:mixed_transport}. Scattering occurs now at both resonances from the majority and minority DOS, as one would naturally expect. Furthermore, the results show that one can average the scattering cross sections of the two spin channels to obtain a total scattering cross section that is in good agreement with the ``mixed'' case. This is a special case of Mathiessen's rule, with equal defect densities of two defect types.

\begin{figure}[tb!]
\includegraphics[width=\linewidth]{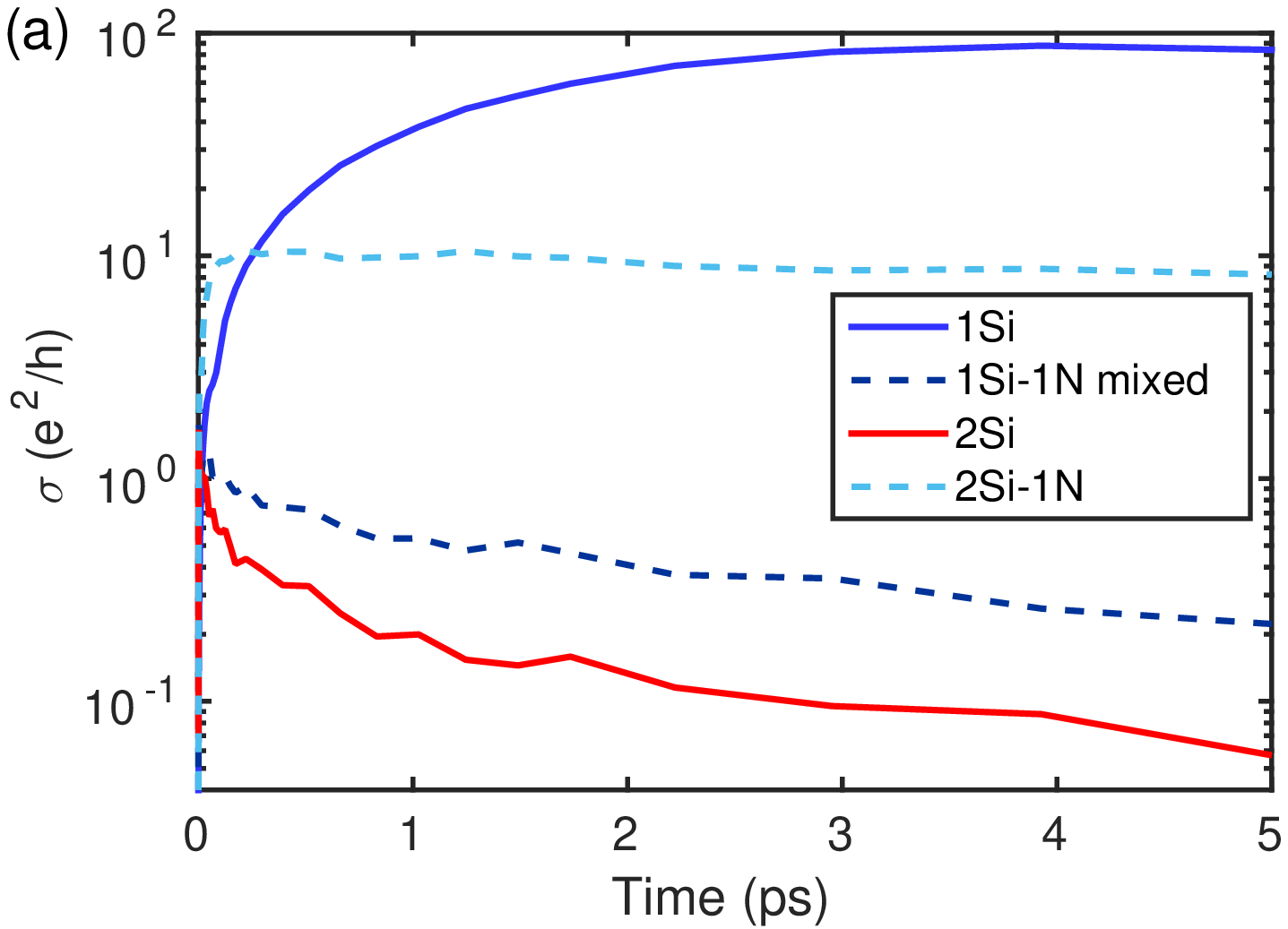}
\includegraphics[width=\linewidth]{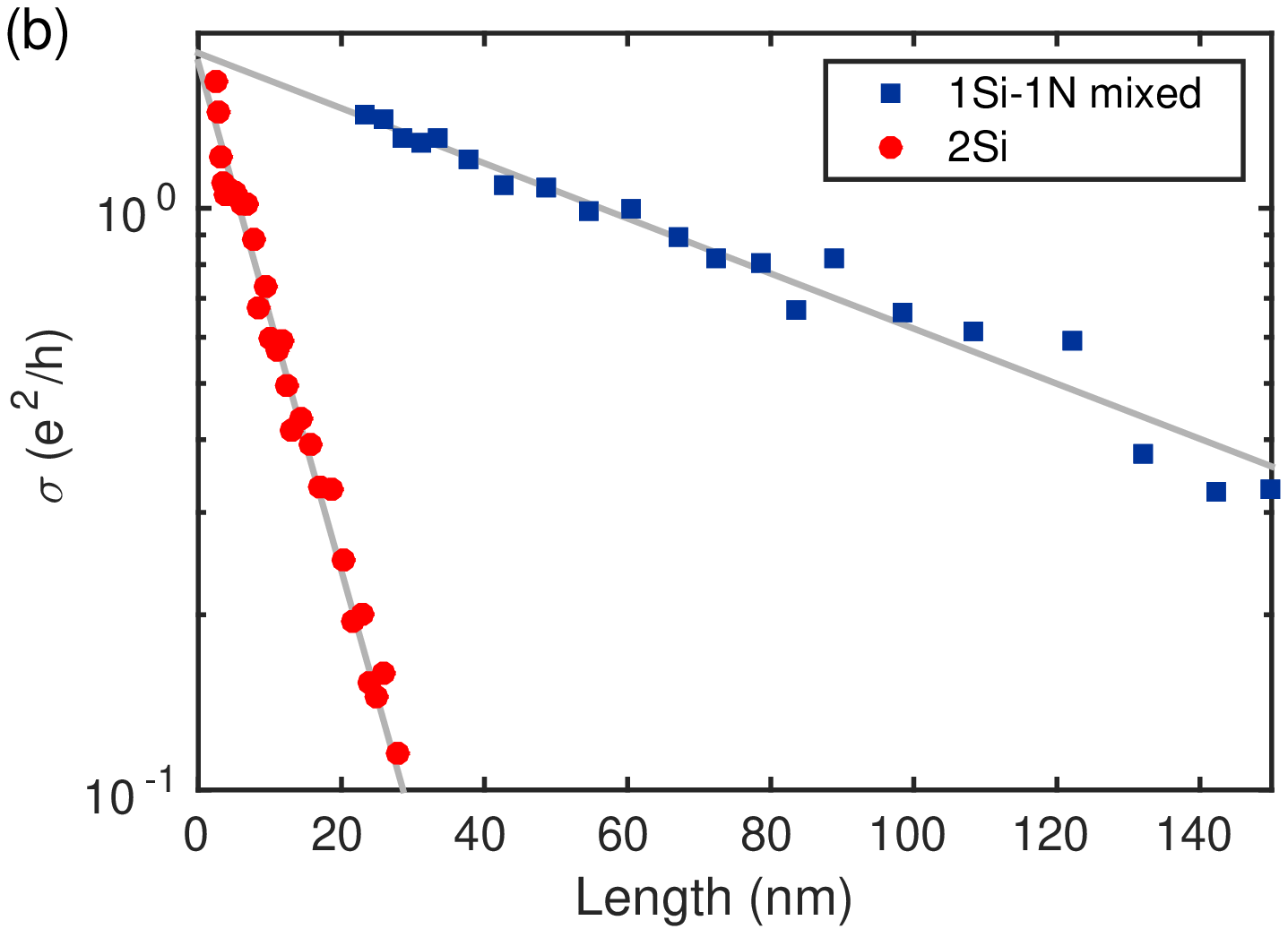}
\caption{(Color online). Conductivity as a function of (a) correlation time, and (b) length, at $E=0.0$ eV for 1Si and 1Si-1N, and at $E=0.15$ eV for 2Si, and at $E=0.3$ eV for 2Si-1N. The lines in (b) are exponential fits $\sigma \propto e^{-L/\xi}$ . \label{fig:localization}}
\end{figure}

The defects considered here break the graphene sublattice symmetry. Therefore pseudospin is not conserved, and backscattering is allowed. Localization effects are thus expected to occur at low temperatures in large systems with sufficiently large defect concentrations. The conductivities for the 1Si-N and 2Si defects at $E=0$ eV and at $E=0.15$ eV, respectively, shown in Fig.~\ref{fig:localization}(a), are small and decrease as a function of correlation time. This indicates the presence of strong localization around the charge neutrality point (CNP), where the semiclassical conductivities are close to the so-called ``minimum conductivity'' $\sigma_{\text{min}}=4e^2/\pi h$. Converting the correlation time to length \cite{Uppstu_2014}, and omitting the initial region of increasing conductivity that corresponds to the transition from ballistic to diffusive transport, we obtain the length dependence shown in Fig.~\ref{fig:localization}(b). Clearly, the conductivity decays exponentially as $\sigma \propto e^{-L/\xi}$, where the 2D localization lengths $\xi$ can be estimated to be $92$ nm for the 1Si-1N defects and $10$ nm for the 2Si defects at a defect concentration of $0.5\%$.

In systems with 1Si and 2Si-1N defects, the semiclassical conductivity is large at $E=0$ eV and at $E=0.3$ eV, respectively, as seen in Fig.~\ref{fig:localization}(a), which implies that the corresponding 2D localization lengths, which are exponentially proportional to the semiclassical conductivity \cite{Fan_2014_2}, are also large, presumably much larger than the phase coherence lengths in real systems. Therefore strong localization is not expected to occur. However, for the 2Si-1N defects we can see weak localization, characterized by the slight decrease of the conductivity as a function of the correlation time, which is most clear around $E=-0.1$ eV (not shown). At this point, the estimated 2D localization length is of the order of $1000$ nm at a $0.5\%$ defect concentration, and it would be even larger with a smaller defect concentration. In conclusion, the localization effects do not play an important role in systems with 1Si and 2Si-1N defects in real physical systems, whereas strong localization around the CNP is predominant for the 2Si and 1Si-1N defect types.

\section*{Conclusions}

By systematically optimizing the geometry of various defects in graphene containing silicon, nitrogen,
oxygen, and hydrogen atoms, we have identified the defect types with the
lowest formation energies. Namely, silicon and nitrogen prefer to fill a
monovacancy in graphene, whereas oxygen and hydrogen clearly favor to
stay on top of the graphene plane as adatoms. In general, various
impurities are effectively attracting each other to minimize the overall
distortion of the graphene backbone. In this sense, nitrogen, oxygen,
and hydrogen impurites and adatoms are trapped to silicon impurities.

To study the electronic properties of silicon and silicon-nitrogen
impurities, we have focused on defects where a silicon atom fills a
monovacancy (1Si) or a divacancy (2Si). Energetically the most
reasonable way to dope the 1Si and 2Si defects by nitrogen is to
substitute a nearest-neighbour carbon atom of the silicon atom with a
nitrogen atom, denoted as 1Si-1N and 2Si-1N. Such nitrogen doping
changes the electronic properties of the defects remarkably, most
notably resulting in defect-induced finite spin moments in the case of
1Si-1N defects. Each 1Si-1N defect has a spin moment of $1.0 \mu_B$ that
is mainly localized at the neighbourhood of the silicon atom. The spin
moments of the 1Si-1N defects interact with each other preferring
antiferromagnetic ordering. It also turns out that hydrogen adatoms
quench any finite spin moments of the silicon-nitrogen defects, further
enabling the use of such defects in spintronic devices.

We have derived tight-binding models to describe the 1Si, 2Si, 1Si-1N,
and 2Si-1N defects in graphene, and to evaluate electronic transport
properties in realistic systems with millions of atoms containing many
randomly placed impurities. The effective scattering cross sections have been
calculated to describe the intrinsic transport properties of systems
with each of the defect types. Systems with 1Si and 2Si-1N defects have
similar transport properties to pristine graphene, whereas the 1Si-1N
and 2Si defects scatter charge carriers heavily close to charge
neutrality. Furthermore, the spin-polarized 1Si-1N defects have strongly
spin-dependent electronic transport properties. Namely, there are
resonant impurity states localized at the defects in the majority spin
component, whereas the minority spin component is a semimetal with similar
properties to pristine graphene close to charge neutrality. Therefore the spin
transport properties can be controlled by applying an external field
that can tune the magnetic ordering of the defects. Furthermore, in
realistically sized graphene samples, the 2Si and 1Si-1N defects are
expected to cause strong localization, whereas systems with the
energetically preferred 1Si and 2Si-1N defects exhibit so large
localization lengths that localization effects are unlikely to be
observed experimentally.

Our results provide microscopic insights into the electronic structure
and transport of graphene with Si impurities and Si-N/O/H complex
defects. It is possible to extend graphene functionalities by
tuning its properties through controllable introduction of defects
either during the growth or by using post-synthesis methods.
Further exposure of silicon impurities to particular gases should give
rise to other defects. On the other hand, knowing the signatures of
particular types of defects in Si-doped graphene, one can use these 
systems in gas sensors \cite{Chen_2012,Niu_2013}, or to design and further improve graphene-based nanodevices.

\appendix

\begin{acknowledgments} We thank Morten Rish\o j Thomsen for useful
discussions. This research has been supported by the Academy of Finland
through its Centres of Excellence Program (project no. 251748) as well
as projects 263416 and 286279. We acknowledge the computational
resources provided by Aalto Science-IT project and Finland’s IT Center
for Science (CSC). \end{acknowledgments}

\bibliography{GraSiN}

\end{document}